\documentclass[12pt]{article}
\usepackage{amsmath,amssymb,amsthm,amsxtra,overpic,bbm,bm,epsfig,ulem}
\usepackage{color,ulem}
\usepackage{longtable,cite}
\def\thefootnote{\fnsymbol{footnote}}

\begin{document}

\vspace{0.2cm}

\begin{center}
{\large\bf Seesaw mirroring between light and heavy Majorana neutrinos with
the help of the $S^{}_3$ reflection symmetry}
\end{center}

\vspace{0.2cm}

\begin{center}
{\bf Zhi-zhong Xing$^{a, b}$}
and ~{\bf Di Zhang$^{a}$}\footnote{Email: zhangdi@ihep.ac.cn} \\
{$^a$Institute of High Energy Physics, and School of Physical
Sciences, \\ University of Chinese Academy of Sciences, Beijing 100049, China \\
$^b$Center for High Energy Physics, Peking University, Beijing
100080, China}
\end{center}

\vspace{1.5cm}

\begin{abstract}
In the canonical seesaw mechanism we require the relevant neutrino mass terms
to be invariant under the $S^{}_3$ charge-conjugation transformations of
left- and right-handed neutrino fields. Then both the Dirac mass matrix
$M^{}_{\rm D}$ and the right-handed neutrino mass matrix $M^{}_{\rm R}$
are well constrained, so is the effective light Majorana neutrino mass matrix
$M^{}_\nu$ via the seesaw formula. We
find that these mass matrices can be classified into 22 categories,
among which some textures respect the well-known $\mu$-$\tau$ permutation
or reflection symmetry and flavor democracy. It is also found that
there exist remarkable structural equalities or similarities
between $M^{}_\nu$ and $M^{}_{\rm R}$, reflecting a seesaw mirroring
relationship between light and heavy Majorana neutrinos. We calculate the
corresponding light neutrino masses and flavor mixing parameters as well as the
CP-violating asymmetries in decays of the lightest heavy Majorana neutrino,
and show that only the flavored leptogenesis mechanism is possible to work
for three categories of $M^{}_{\rm D}$ and $M^{}_{\rm R}$ in the $S^{}_3$
reflection symmetry limit.
\end{abstract}

\begin{flushleft}
\hspace{0.8cm} PACS number(s): 14.60.Pq, 11.30.Hv, 13.35.Hb.
\end{flushleft}

\newpage

\def\thefootnote{\arabic{footnote}}
\setcounter{footnote}{0}
\setcounter{table}{0}
\setcounter{equation}{0}
\setcounter{figure}{0}

\section{Introduction}

The experimental discoveries of neutrino oscillations \cite{PDG} have
confessedly demonstrated that the standard model (SM) of particle physics
is incomplete, because it cannot accommodate
and explain the finite but tiny neutrino masses and significant lepton
flavor mixing effects. The most canonical and popular way out is to introduce
three right-handed neutrino fields $N^{}_{\alpha \rm R}$
(for $\alpha = e, \mu, \tau$) and allow lepton number violation
\cite{SS,Yanagida1979,Gell-Mann1979,Glashow1980,Mohapatra1980},
with which the Yukawa interaction and a Majorana neutrino mass
term can be written as
\begin{eqnarray}
-{\cal L}^{}_{0} = \overline{\ell^{}_{\rm L}}
Y^{}_{\nu} \widetilde{H} N^{}_{\rm R} + \frac{1}{2} \overline{N^{\rm c}_{\rm R}}
M^{}_{\rm R} N^{}_{\rm R} + {\rm h.c.} \; ,
\end{eqnarray}
where $\widetilde{H} \equiv {\rm i}\sigma^{}_{2} H^{\ast}$ with $H$ being
the Higgs doublet of the SM, $\ell^{}_{\rm L}$ denotes the left-handed
lepton doublet column vector, $N^{}_{\rm R}$ represents the right-handed
neutrino column vector with the $N^{}_{\alpha \rm R}$ components, and
$N^{\rm c}_{\rm R} \equiv {\cal C} \overline{N^{}_{\rm R}}^{\rm T}$
with T denoting the transpose and $\cal C$ being the charge-conjugation
operator. After spontaneous electroweak symmetry breaking, Eq. (1.1) becomes
\begin{eqnarray}
-{\cal L}^{}_{\rm m} = \overline{\nu^{}_{\rm L}}
M^{}_{\rm D} N^{}_{\rm R} + \frac{1}{2} \overline{N^{\rm c}_{\rm R}}
M^{}_{\rm R} N^{}_{\rm R} + {\rm h.c.} \; ,
\end{eqnarray}
where $M^{}_{\rm D} \equiv Y^{}_{\nu}\langle H\rangle$ with
$\langle H\rangle \simeq 174$ GeV being the vacuum expectation value of
the Higgs field. The scale of $M^{}_{\rm R}$ can be much larger
than $\langle H\rangle$ because the right-handed neutrino fields are
the $SU(2)^{}_{\rm L} \times U(1)^{}_{\rm Y}$ singlets
and thus have nothing to do with electroweak symmetry breaking. In this
case one may integrate out the heavy degrees of freedom and then obtain an
effective mass term for the three light neutrinos:
\begin{eqnarray}
-{\cal L}^{}_{\nu} = \frac{1}{2} \overline{\nu^{}_{\rm L}}
M^{}_{\nu} \nu^{\rm c}_{\rm L} + {\rm h.c.} \; ,
\end{eqnarray}
where $\nu^{\rm c}_{\rm L} \equiv {\cal C} \overline{\nu^{}_{\rm L}}^{\rm T}$
is defined, and $M^{}_\nu = -M^{}_{\rm D} M^{-1}_{\rm R} M^{\rm T}_{\rm D}$
is the well-known seesaw formula
\cite{SS,Yanagida1979,Gell-Mann1979,Glashow1980,Mohapatra1980} in the
leading-order approximation, which naturally attributes the
smallness of the scale of $M^{}_\nu$ to the largeness of the scale of
$M^{}_{\rm R}$ as compared with the value of $\langle H\rangle$.

While the above seesaw relation can qualitatively explain why the masses
of three light Majorana neutrinos $m^{}_i$ (i.e., the eigenvalues of $M^{}_\nu$)
are strongly suppressed in magnitude,
it unfortunately has no quantitative prediction for
the values of $m^{}_i$ and flavor mixing parameters.
To reduce the number of unknown degrees of freedom and thus enhance
the predictability and testability of the seesaw mechanism, the structures of
$M^{}_{\rm D}$ and $M^{}_{\rm R}$ need to be specified with either some
empirical assumptions (e.g., texture zeros --- see Ref. \cite{FX2000} for a
review) or certain flavor symmetries (e.g., $A^{}_4$ and $S^{}_4$ symmetries ---
see Refs. \cite{Altarelli,Kobayashi,King} for recent reviews).
Since the observed pattern of the $3\times 3$
Pontecorvo-Maki-Nakagawa-Sakata (PMNS) neutrino
mixing matrix $V$ \cite{Maki,Pontecorvo}
exhibits an approximate $\mu$-$\tau$ permutation symmetry
(i.e., $|V^{}_{\mu i}| \simeq |V^{}_{\tau i}|$ for $i =1,2,3$), most of
the larger flavor symmetry groups considered for the neutrino sector
actually consist of a subgroup which allows $M^{}_\nu$ to respect the
$\mu$-$\tau$ flavor symmetry \cite{Zhao2016}.

Of course, building a realistic neutrino mass model based on a given flavor
symmetry is highly nontrivial because it is usually imperative to introduce
some hypothetical gauge-singlet scalar fields (i.e., the so-called flavon
fields) and make use of their vacuum expectation values to partly fix the
flavor structures of massive neutrinos and charged leptons. Hence the flavor
symmetry breaking is typically associated with many unknown parameters which
are normally put into a hidden dustbin in most of the model-building exercises,
since these new parameters are experimentally unaccessible for the time being.
The variety of such models makes it practically hard to judge which flavor
symmetry is closer to the truth \cite{Kobayashi2}.

In this situation one may follow a purely phenomenological way to
focus only on the mass terms of charged leptons and neutrinos and then
constrain their textures by means of certain flavor symmetries, so as to
predict an acceptable flavor mixing pattern which is consistent with
current neutrino oscillation data \cite{Altarelli,Kobayashi,King,Zhao2016}.
Although there is an obvious gap between such an approach and a real
neutrino mass model, the former can be regarded as a necessary or instructive step
towards the latter. Considering that the underlying flavor symmetry is most likely
to manifest itself at a high energy scale far above the electroweak scale,
the phenomenological approach under discussion actually fits the spirit of
the bottom-up approach of model building in particle physics.

Following the same phenomenological approach, here we are interested in exploring
the seesaw-induced relation between light and heavy Majorana neutrinos
with the help of possible $S^{}_3$ flavor symmetries. That is to say,
we consider the possibility of simultaneously constraining the textures
of $M^{}_{\rm D}$ and $M^{}_{\rm R}$ by requiring that ${\cal L}^{}_{\rm m}$
in Eq. (1.2) be invariant under the charge-conjugation
transformations $\nu^{}_{\rm L}
\leftrightarrow {\cal S}^{}_{\rm (L)} \nu^{\rm c}_{\rm L}$ and $N^{}_{\rm R}
\leftrightarrow {\cal S}^{}_{\rm (R)} N^{\rm c}_{\rm R}$, where
${\cal S}^{}_{\rm (L)}$ or ${\cal S}^{}_{\rm (R)}$
stands for an arbitrary element belonging to an
arbitrary subset of $S^{}_3$ group. In this way it is easy to show that such a
phenomenological requirement is equivalent to the constraints
$M^*_{\rm D} = {\cal S}^{\dagger}_{(\rm L)} M^{}_{\rm D} {\cal S}^{}_{(\rm R)}$
and $M^*_{\rm R} = {\cal S}^{}_{(\rm R)} M^{}_{\rm R} {\cal S}^{}_{(\rm R)}$,
and therefore the structures of $M^{}_{\rm D}$ and $M^{}_{\rm R}$ can be strongly
constrained. As a result, the structure of the light Majorana neutrino mass
matrix $M^{}_\nu$ can be partly determined via the seesaw
formula $M^{}_\nu = - M^{}_{\rm D} M^{-1}_{\rm R} M^{\rm T}_{\rm D}$,
leading to some intriguing predictions for the neutrino masses and
flavor mixing parameters. In comparison with the so-called $\mu$-$\tau$
reflection symmetry which has been used to directly constrain the
form of $M^{}_\nu$ \cite{Harrison}, our present method can be referred
to as the $S^{}_3$ reflection symmetry approach. Moreover, we find that
the obtained texture of $M^{}_\nu$ is either the same as or very similar
to that of $M^{}_{\rm R}$, a remarkable consequence of
our approach which is referred to as the seesaw mirroring relationship
between $M^{}_\nu$ and $M^{}_{\rm R}$. Along this line of thought, we
also examine which of the $S^{}_3$-constrained textures of
$M^{}_{\rm D}$ and $M^{}_{\rm R}$ can allow for CP violation in the
lepton-number-violating decays of the lightest heavy Majorana neutrino,
a necessary ingredient of the thermal leptogenesis mechanism \cite{FY}
which offers a natural explanation of the observed
baryon asymmetry of the Universe.

It is worth pointing out that the $S^{}_3$ reflection symmetry
approach under discussion is subject to the basis with the flavor
eigenstates of three charged leptons being the same as their mass
eigenstates (i.e., the charged-lepton mass matrix $M^{}_l$ is diagonal).
Such a basis choice is different from the conventional model building
exercises with the help of discrete flavor symmetries, in which the
charged-lepton fields usually transform together with the neutrino
fields under the given flavor groups \cite{Altarelli,Kobayashi,King}.
After spontaneous flavor symmetry breaking, the charged-lepton and
neutrino mass matrices are left with different residual symmetries.
The basis with $M^{}_l$ being diagonal can always be achieved by
choosing a suitable representation of the given symmetry group, but
it might not be convenient from the point of view of model building.
In the present work we simply assume $M^{}_l$ to be diagonal and
make the $S^{}_3$ reflection transformations only for the neutrino
sector. This simple treatment allows us to directly derive the PMNS neutrino
mixing matrix from the effective Majorana neutrino mass matrix $M^{}_\nu$
via the seesaw relation, with no concern about the charged-lepton sector.

The approach and main results of this paper are expected to be instructive and
useful for broadening our horizons in building realistic neutrino mass models
and understanding lepton flavor mixing and CP violation.
In fact, a lot of attention has been paid to applying the $S^{}_3$ flavor
symmetry to the quark and lepton sectors since the pioneering work
done in 1978 \cite{Pakvasa,Harari}, and in this connection remarkable progress
was made in 1996 and 1998 to predict quite large solar and atmospheric neutrino
mixing angles \cite{FX96,FX98,Tanimoto98}. Although some attempts have
been made in combining the seesaw mechanism and the $S^{}_3$ flavor symmetry
(see, e.g., Refs. \cite{Rodejohann,Kim,Mohapatra2006,XYZ,Ma2015,
Gomez-Izquierdo:2017rxi,Si:2017pdo,Mondragon,Garces:2018nar}),
our present work is different from them in several aspects:
\begin{itemize}
\item     We constrain the structures of $M^{}_{\rm D}$ and $M^{}_{\rm R}$ by
  dictating the two neutrino mass terms in Eq. (1.2) to be invariant under the
  $S^{}_3$ reflection (i.e., charge-conjugation) transformations
  $\nu^{}_{\rm L} \leftrightarrow {\cal S}^{}_{\rm (L)} \nu^{\rm c}_{\rm L}$
  and $N^{}_{\rm R} \leftrightarrow {\cal S}^{}_{\rm (R)} N^{\rm c}_{\rm R}$
  instead of the $S^{}_3$ permutation transformations
  $\nu^{}_{\rm L} \leftrightarrow {\cal S}^{}_{\rm (L)} \nu^{}_{\rm L}$ and
  $N^{}_{\rm R} \leftrightarrow {\cal S}^{}_{\rm (R)} N^{}_{\rm R}$. Such a new
  treatment makes sense because it is fully consistent with the spirit of the
  $\mu$-$\tau$ reflection symmetry --- a special case of the $S^{}_3$ reflection
  symmetry under discussion, in order to produce the experimentally favored
  results $\theta^{}_{23} = \pi/4$ and $\delta = 3\pi/2$ for the PMNS matrix
  $V$ in its standard parametrization form \cite{PDG}. In comparison, one will
  be left with $\delta = 0$ in the limit of the flavor democracy
  \cite{FX96,FX98,Tanimoto98} or $S^{}_3$ permutation symmetry \cite{Tanimoto1998yz,
  Tanimoto2000,Branco2001,Harrison2003aw,Jora2009gz,
  Jora2010at,Jora2012nw,Dev2011qy,Dev2012ns,Benaoum2013}.

\item    We carry out a systematic analysis of all the possible textures of
  $M^{}_{\rm D}$ and $M^{}_{\rm R}$ constrained by the $S^{}_3$ reflection symmetry,
  make a classification of them, and examine whether the resulting textures of
  $M^{}_\nu$ are seesaw-invariant or share the same flavor symmetry with
  $M^{}_{\rm R}$ and (or) $M^{}_{\rm D}$. Our results can therefore provide a very
  useful reference for further model-building exercises.

\item    We calculate the light neutrino masses, flavor mixing angles and CP-violting
  phases for each texture of $M^{}_\nu$ in the $S^{}_3$ refection symmetry limit, and
  examine whether the CP-violating asymmetries in decays of the lightest heavy
  Majorana neutrino are vanishing or not in the same limit.
  We find that in this case only
  flavored leptogenesis \cite{Barbieri:1999ma,Morozumi,Strumia} is possible
  to work for a few textures of $M^{}_{\rm D}$ and $M^{}_{\rm R}$.
\end{itemize}
As many other flavor symmetries, the $S^{}_3$ reflection symmetry
must be broken too, so as to make $M^{}_\nu$ fully fit current experimental data.
One may discuss such symmetry breaking effects by either taking account of the
renormalization-group evolution of $M^{}_\nu$ from the seesaw scale (where the
flavor symmetry is assumed to manifest itself) to the electroweak scale, or
introducing some explicit symmetry breaking terms into
$M^{}_{\rm D}$ and $M^{}_{\rm R}$ \cite{Zhao2016}.
A further work of this kind depends on more
technical details and empirical assumptions, and hence it is beyond the scope
of the present paper and will be done elsewhere as a follow-up.

The remaining parts of this paper are organized as follows. In section 2 we
first introduce the $S^{}_{3}$ reflection transformations for left-
and right-handed neutrino fields to constrain the structures of $M^{}_{\rm D}$
and $M^{}_{\rm R}$, and then determine the texture of $M^{}_{\nu}$
with the help of the seesaw formula. All the possibilities in this
connection are examined and classified. Section 3 is devoted to discussing
the phenomenological consequences of $M^{}_\nu$, where the light neutrino masses,
flavor mixing angles and CP-violating phases are calculated in a case-by-case way.
In section 4 we consider both unflavored and flavored leptogenesis mechanisms
and calculate the corresponding CP-violating asymmetries in decays of the lightest
heavy Majorana neutrino in the $S^{}_3$ reflection symmetry limit.
A summary of our approach and main results, together with some discussions
about extending $S^{}_3$ group to $A^{}_4$ group so as to illustrate the constrained
textures of neutrino mass matrices in a different way, is made in section 5.

\def\thefootnote{\arabic{footnote}}
\setcounter{footnote}{0}
\setcounter{table}{0}
\setcounter{equation}{0}
\setcounter{figure}{0}

\section{Applications of the $S^{}_{3}$ reflection symmetry}

\subsection{Textures of $M^{}_{\rm D}$ and $M^{}_{\rm R}$ under $S^{}_{3}$
reflection symmetry}

To begin with, we rewrite the mass terms in Eq. (1.2) in a more concise way
as follows:
\begin{eqnarray}
-{\cal L}^{}_{\rm m} = \frac{1}{2} \ \overline{\left(\begin{matrix}
\nu^{}_{\rm L} & N^{\rm c}_{\rm R} \end{matrix}\right)}
\left(\begin{matrix} {\bf 0} & M^{}_{\rm D} \cr
M^{\rm T}_{\rm D} & M^{}_{\rm R} \end{matrix}\right)
\left(\begin{matrix} \nu^{\rm c}_{\rm L} \cr N^{}_{\rm R} \end{matrix}
\right) + {\rm h.c.} \; .
\end{eqnarray}
To constrain flavor structures of the canonical seesaw mechanism, we
require the neutrino mass term in Eq. (2.1) to keep invariant
when $\nu^{}_{\rm L}$ and $N^{}_{\rm R}$ transform as
\begin{eqnarray}
\nu^{}_{\rm L} \leftrightarrow {\cal S}^{}_{({\rm L})} \nu^{\rm c}_{\rm L} \;,\quad
N^{}_{\rm R} \leftrightarrow {\cal S}^{}_{({\rm R})} N^{\rm c}_{\rm R} \;,
\end{eqnarray}
in which ${\cal S}^{}_{\rm L}$ or ${\cal S}^{}_{\rm R}$ denotes an arbitrary
element of $G$ --- a given subset of $S^{}_{3}$ group, and the
possibilities of both ${\cal S}^{}_{\rm L} = {\cal S}^{}_{\rm R}$ and
${\cal S}^{}_{\rm L} \neq {\cal S}^{}_{\rm R}$ are included.
It is worth pointing out that only the neutrino mass term ${\cal L}^{}_{\rm m}$
is dictated to be invariant under the transformations made in Eq. (2.2),
and hence the consequent $S^{}_3$ reflection symmetry is not a {\it real}
flavor symmetry for the whole Lagrangian of weak interactions. Instead,
it only works as an effective organizing principle to simplify and
constrain the structures of $M^{}_{\rm D}$ and $M^{}_{\rm R}$. Note
that the well-known $\mu$-$\tau$ reflection symmetry
and some other {\it working} flavor symmetries \cite{King,Zhao2016} were proposed
in the same spirit. If such a purely phenomenological approach turns out
to be compatible with current and future experimental data,
it may finally be embedded into a complete flavor
model of fermion masses based on a larger symmetry group.

Under the $S^{}_3$ reflection transformations given in Eq. (2.2), the $6\times 6$
neutrino mass matrix in Eq. (2.1) changes as follows:
\begin{eqnarray}
-{\cal L}^{\prime}_{\rm m} \hspace{-0.2cm}&=&\hspace{-0.2cm}
\frac{1}{2} \overline{\left(\begin{matrix} \nu^{\rm c}_{\rm L} & N^{}_{\rm R}
\end{matrix} \right)} \left(\begin{matrix} {\cal S}^{\dagger}_{({\rm L})} & \cr
& {\cal S}^{}_{({\rm R})} \end{matrix}\right)
\left(\begin{matrix} 0 & M^{}_{\rm D} \cr M^{\rm T}_{\rm D} & M^{}_{\rm R}
\end{matrix}\right) \left(\begin{matrix} {\cal S}^{\dagger}_{({\rm L})} & \cr
& {\cal S}^{}_{({\rm R})}\end{matrix}\right)
\left( \begin{matrix} \nu^{}_{\rm L} \cr N^{\rm c}_{\rm R} \end{matrix}\right)
\nonumber
\\
\hspace{-0.2cm}&+&\hspace{-0.2cm} \frac{1}{2} \overline{\left(\begin{matrix}
\nu^{}_{\rm L} & N^{\rm c}_{\rm R} \end{matrix} \right)} \left(\begin{matrix}
{\cal S}^{}_{({\rm L})} & \cr & {\cal S}^{\dagger}_{({\rm R})} \end{matrix}\right)
\left(\begin{matrix} 0 & M^{\ast}_{\rm D} \cr M^{\dagger}_{\rm D} &
M^{\dagger}_{\rm R}\end{matrix}\right) \left(\begin{matrix} {\cal S}^{}_{({\rm L})}
& \cr & {\cal S}^{\dagger}_{({\rm R})} \end{matrix}\right) \left( \begin{matrix}
\nu^{\rm c}_{\rm L} \cr N^{}_{\rm R} \end{matrix}\right) \;,
\end{eqnarray}
where the unitarity of ${\cal S}^{}_{\rm (L)}$ and ${\cal S}^{}_{\rm (R)}$
has been used. It becomes transparent that the neutrino mass terms will be
invariant (namely, ${\cal L}^{}_{\rm m} = {\cal L}^{\prime}_{\rm m}$)
if the whole neutrino mass matrix satisfies the condition
\begin{eqnarray}
\left(\begin{matrix} 0 & M^{}_{\rm D} \cr M^{\rm T}_{\rm D} & M^{}_{\rm R}
\end{matrix}\right) =  \left(\begin{matrix} 0 & {\cal S}^{}_{({\rm L})}
M^{\ast}_{\rm D} {\cal S}^{\dagger}_{({\rm R})} \cr {\cal S}^{\dagger}_{({\rm R})}
M^{\dagger}_{\rm D} {\cal S}^{}_{({\rm L})} & {\cal S}^{\dagger}_{({\rm R})}
M^{\dagger}_{\rm R} {\cal S}^{\dagger}_{({\rm R})} \end{matrix}\right) \; .
\end{eqnarray}
This in turn means that $M^{}_{\rm D}$ and $M^{}_{\rm R}$ should satisfy
the conditions
\begin{eqnarray}
M^{}_{\rm D} = {\cal S}^{}_{({\rm L})} M^{\ast}_{\rm D} {\cal S}^{\dagger}_{({\rm R)}}
\;, \quad M^{}_{\rm R} = {\cal S}^{\dagger}_{({\rm R)}} M^{\ast}_{\rm R}
{\cal S}^{\dagger}_{({\rm R})} \;.
\end{eqnarray}
Therefore, the $S^{}_3$ reflection symmetry imposed on the neutrino mass
terms in Eq. (2.1) allows us to constrain textures of the neutrino
mass matrices $M^{}_{\rm D}$ and $M^{}_{\rm R} $. Such a constraint can be
further transferred to the effective neutrino mass matrix $M^{}_{\nu}$
via the seesaw formula
\begin{eqnarray}
M^{}_\nu = -M^{}_{\rm D} M^{-1}_{\rm R} M^{\rm T}_{\rm D} \; ,
\end{eqnarray}
from which one may examine the structural similarity between $M^{}_\nu$
and $M^{}_{\rm R}$. In other words, it is possible to establish a seesaw
mirroring relationship between light and heavy Majorana neutrinos with
the help of the $S^{}_3$ reflection symmetry.

Explicitly, the three-dimensional unitary representations for six elements of
$S^{}_{3}$ group are
\begin{eqnarray}
S^{(123)}_{} &=& \left(\begin{matrix}
1 & 0 & 0 \cr 0 & 1 & 0 \cr 0 & 0 & 1
\end{matrix}\right) ,\;
S^{(231)}_{} = \left(\begin{matrix}
0 & 1 & 0 \cr 0 & 0 & 1 \cr 1 & 0 & 0
\end{matrix}\right) ,\;
S^{(312)}_{} = \left(\begin{matrix}
0 & 0 & 1 \cr 1 & 0 & 0 \cr 0 & 1 & 0
\end{matrix}\right) ,\;
\nonumber
\\
S^{(213)}_{} &=& \left(\begin{matrix}
0 & 1 & 0 \cr 1 & 0 & 0 \cr 0 & 0 & 1
\end{matrix}\right) ,\;
S^{(132)}_{} = \left(\begin{matrix}
1 & 0 & 0 \cr 0 & 0 & 1 \cr 0 & 1 & 0
\end{matrix}\right) ,\;
S^{(321)}_{} = \left(\begin{matrix}
0 & 0 & 1 \cr 0 & 1 & 0 \cr 1 & 0 & 0
\end{matrix}\right) .
\end{eqnarray}
These group elements can be categorized into three conjugacy classes:
${\cal C}^{}_{0} = \left\{ S^{(123)}_{} \right\} $,
${\cal C}^{}_{1} = \left\{ S^{(231)}_{} , S^{(312)}_{} \right\} $ and
${\cal C}^{}_{2} = \left\{ S^{(213)}_{} , S^{(132)}_{},
S^{(321)}_{} \right\} $. So $S^{}_3$ has one subgroup of order three,
$Z^{}_{3} = \left\{S^{(123)}_{}, S^{(231)}_{} , S^{(312)}_{}\right\}$,
as well as three subgroups of order two, $Z^{(12)}_{2} = \left\{S^{(123)}_{},
S^{(213)}_{}\right\}$, $Z^{(23)}_{2} = \left\{S^{(123)}_{}, S^{(132)}_{}
\right\}$ and $Z^{(31)}_{2} = \left\{S^{(123)}_{}, S^{(321)}_{}\right\}$.
Note that $S^{}_3$ group totally has $2^{3!} -1 = 63$ non-void subsets.
To characterize these subsets, we first reorder the elements of $S^{}_3$
group as
\begin{eqnarray}
G^{6}_{} = \left\{ S^{(123)}_{}, S^{(231)}_{}, S^{(312)}_{}, S^{(213)}_{},
S^{(132)}_{}, S^{(321)}_{}\right\} \;.
\end{eqnarray}
Then an arbitrary subset can be characterized by
$G^{n}_{i^{}_1 i^{}_2 \cdots i^{}_n}$
or $G^{n}_{\overline{i^{}_1 i^{}_2 \cdots i^{}_{6-n}}}$ (for
$n = 1, 2, \cdots, 6$), in which $n$ is the
number of elements in the subset, and $i^{}_1 i^{}_2 \cdots i^{}_n$
for $n \leq 3$ or $\overline{i^{}_1 i^{}_2 \cdots i^{}_{6-n}}$ for $n > 3$
is the index of different subsets
with the equal number of elements.
What is more, $i^{}_1, i^{}_2, \cdots, i^{}_n = 1, 2, \cdots, 6$ should
satisfy $i^{}_1 < i^{}_2 < \cdots < i^{}_n$. Note that we are making use of the
index $i^{}_1 i^{}_2 \cdots i^{}_n$ for
$n \leq 3$ which is a reordered sequence of the order numbers of elements
belonging to $G^{n}_{i^{}_1 i^{}_2 \cdots i^{}_n}$ in $G^{6}$, and the index
$\overline{i^{}_1 i^{}_2 \cdots i^{}_{6-n}}$ is used for $n > 3$
which is a reordered sequence of the order numbers of elements belonging
to the complement of $G^{n}_{\overline{i^{}_1 i^{}_2 \cdots i^{}_{6-n}}}$
with respect to $S^{}_{3}$ group in $G^{6}$. For illustration, let us
give several examples to make the notation issue clear:
$G^{1}_{2} = \left\{ S^{(231)}_{} \right\}$,
$G^{2}_{24} = \left\{ S^{(231)}_{}, S^{(213)}_{} \right\}$,
$G^{3}_{124} = \left\{ S^{(123)}_{}, S^{(231)}_{}, S^{(213)}_{} \right\}$,
$G^{4}_{\overline{36}} = \left\{ S^{(123)}_{}, S^{(231)}_{}, S^{(213)}_{},
S^{(132)}_{} \right\}$ and $G^{5}_{\overline{3}} = \left\{ S^{(123)}_{},
S^{(231)}_{}, S^{(213)}_{}, S^{(132)}_{}, S^{(321)}_{} \right\}$. It should
also be noted that
$G^{3}_{123} = Z^{}_{3}$, $G^{2}_{14} = Z^{(12)}_{2}$,
$G^{2}_{15} = Z^{(23)}_{2}$ and $G^{2}_{16} = Z^{(31)}_{2}$, and there are
totally $C^{n}_{6} = \frac{6!}{n! \left(6-n\right)!}$ different subsets
$ G^{n}_{i^{}_1 i^{}_2 \cdots i^{}_n}$ or
$G^{n}_{\overline{i^{}_1 i^{}_2 \cdots i^{}_{6-n}}}$ for a
given number $n$.

With the help of Eq. (2.5), we can now obtain all the possible structures of
$M^{}_{\rm D}$ and $M^{}_{\rm R}$ constrained by the subsets of $S^{}_3$ group.
Here we take set $G^{1}_{5}$, corresponding to ${\cal S}^{}_{\rm (L)} =
{\cal S}^{}_{\rm (R)} = S^{(132)}$, as a typical example to do some explicit
calculations. This case is particularly interesting because it actually works
like the $\mu$-$\tau$ reflection symmetry.
Since set $G^{1}_{5}$ contains only a single element (i.e., $S^{(132)}$),
the mass matrices $M^{}_{\rm D}$ and $M^{}_{\rm R}$ satisfy
\begin{eqnarray}
\left(\begin{matrix}
\langle M^{}_{\rm D} \rangle^{}_{11} & \langle M^{}_{\rm D} \rangle^{}_{12}
& \langle M^{}_{\rm D} \rangle^{}_{13} \cr
\langle M^{}_{\rm D} \rangle^{}_{21} & \langle M^{}_{\rm D} \rangle^{}_{22}
& \langle M^{}_{\rm D} \rangle^{}_{23} \cr
\langle M^{}_{\rm D} \rangle^{}_{31} & \langle M^{}_{\rm D} \rangle^{}_{32}
& \langle M^{}_{\rm D} \rangle^{}_{33}
\end{matrix}\right)
\hspace{-0.2cm}&=&\hspace{-0.2cm}
\left(\begin{matrix}
1 & 0 & 0 \cr 0 & 0 & 1 \cr 0 & 1 & 0
\end{matrix}\right)
\left(\begin{matrix}
\langle M^{}_{\rm D} \rangle^{\ast}_{11} &
\langle M^{}_{\rm D} \rangle^{\ast}_{12} &
\langle M^{}_{\rm D} \rangle^{\ast}_{13} \cr
\langle M^{}_{\rm D} \rangle^{\ast}_{21} &
\langle M^{}_{\rm D} \rangle^{\ast}_{22} &
\langle M^{}_{\rm D} \rangle^{\ast}_{23} \cr
\langle M^{}_{\rm D} \rangle^{\ast}_{31} &
\langle M^{}_{\rm D} \rangle^{\ast}_{32} &
\langle M^{}_{\rm D} \rangle^{\ast}_{33}
\end{matrix}\right)
\left(\begin{matrix}
1 & 0 & 0 \cr 0 & 0 & 1 \cr 0 & 1 & 0
\end{matrix}\right)
\nonumber
\\
\hspace{-0.2cm}&=&\hspace{-0.2cm}
\left(\begin{matrix}
\langle M^{}_{\rm D} \rangle^{\ast}_{11} &
\langle M^{}_{\rm D} \rangle^{\ast}_{13} &
\langle M^{}_{\rm D} \rangle^{\ast}_{12} \cr
\langle M^{}_{\rm D} \rangle^{\ast}_{31} &
\langle M^{}_{\rm D} \rangle^{\ast}_{33} &
\langle M^{}_{\rm D} \rangle^{\ast}_{32} \cr
\langle M^{}_{\rm D} \rangle^{\ast}_{21} &
\langle M^{}_{\rm D} \rangle^{\ast}_{23} &
\langle M^{}_{\rm D} \rangle^{\ast}_{22}
\end{matrix}\right) \; ,
\end{eqnarray}
and
\begin{eqnarray}
\left(\begin{matrix}
\langle M^{}_{\rm R} \rangle^{}_{11} &
\langle M^{}_{\rm R} \rangle^{}_{12} &
\langle M^{}_{\rm R} \rangle^{}_{13} \cr
\langle M^{}_{\rm R} \rangle^{}_{12} &
\langle M^{}_{\rm R} \rangle^{}_{22} &
\langle M^{}_{\rm R} \rangle^{}_{23} \cr
\langle M^{}_{\rm R} \rangle^{}_{13} &
\langle M^{}_{\rm R} \rangle^{}_{23} &
\langle M^{}_{\rm R} \rangle^{}_{33}
\end{matrix}\right)
\hspace{-0.2cm}&=&\hspace{-0.2cm}
\left(\begin{matrix}
1 & 0 & 0 \cr 0 & 0 & 1 \cr 0 & 1 & 0
\end{matrix}\right)
\left(\begin{matrix}
\langle M^{}_{\rm R} \rangle^{\ast}_{11} &
\langle M^{}_{\rm R} \rangle^{\ast}_{12} &
\langle M^{}_{\rm R} \rangle^{\ast}_{13} \cr
\langle M^{}_{\rm R} \rangle^{\ast}_{12} &
\langle M^{}_{\rm R} \rangle^{\ast}_{22} &
\langle M^{}_{\rm R} \rangle^{\ast}_{23} \cr
\langle M^{}_{\rm R} \rangle^{\ast}_{13} &
\langle M^{}_{\rm R} \rangle^{\ast}_{23} &
\langle M^{}_{\rm R} \rangle^{\ast}_{33}
\end{matrix}\right)
\left(\begin{matrix}
1 & 0 & 0 \cr 0 & 0 & 1 \cr 0 & 1 & 0
\end{matrix}\right)
\nonumber
\\
\hspace{-0.2cm}&=&\hspace{-0.2cm}
\left(\begin{matrix}
\langle M^{}_{\rm R} \rangle^{\ast}_{11} &
\langle M^{}_{\rm R} \rangle^{\ast}_{13} &
\langle M^{}_{\rm R} \rangle^{\ast}_{12} \cr
\langle M^{}_{\rm R} \rangle^{\ast}_{13} &
\langle M^{}_{\rm R} \rangle^{\ast}_{33} &
\langle M^{}_{\rm R} \rangle^{\ast}_{23} \cr
\langle M^{}_{\rm R} \rangle^{\ast}_{12} &
\langle M^{}_{\rm R} \rangle^{\ast}_{23} &
\langle M^{}_{\rm R} \rangle^{\ast}_{22}
\end{matrix}\right)\; .
\end{eqnarray}
As a result, we arrive at
\begin{eqnarray}
\langle M^{}_{\rm D} \rangle^{}_{11} \hspace{-0.2cm}&=&\hspace{-0.2cm}
\langle M^{}_{\rm D} \rangle^{\ast}_{11} \; , \quad
\langle M^{}_{\rm D} \rangle^{}_{12} =
\langle M^{}_{\rm D} \rangle^{\ast}_{13} \; ,
\nonumber
\\
\langle M^{}_{\rm D} \rangle^{}_{21} \hspace{-0.2cm}&=&\hspace{-0.2cm}
\langle M^{}_{\rm D} \rangle^{\ast}_{31} \; , \quad
\langle M^{}_{\rm D} \rangle^{}_{22} =
\langle M^{}_{\rm D} \rangle^{\ast}_{33} \; ,
\nonumber
\\
\langle M^{}_{\rm D} \rangle^{}_{23} \hspace{-0.2cm}&=&\hspace{-0.2cm}
\langle M^{}_{\rm D} \rangle^{\ast}_{32} \; ;
\end{eqnarray}
and
\begin{eqnarray}
\langle M^{}_{\rm R} \rangle^{}_{11} \hspace{-0.2cm}&=&\hspace{-0.2cm}
\langle M^{}_{\rm R} \rangle^{\ast}_{11} \; , \quad
\langle M^{}_{\rm R} \rangle^{}_{12} =
\langle M^{}_{\rm R} \rangle^{\ast}_{13} \; ,
\nonumber
\\
\langle M^{}_{\rm R} \rangle^{}_{22} \hspace{-0.2cm}&=&\hspace{-0.2cm}
\langle M^{}_{\rm R} \rangle^{\ast}_{33} \; , \quad
\langle M^{}_{\rm R} \rangle^{}_{23} =
\langle M^{}_{\rm R} \rangle^{\ast}_{23} \; .
\end{eqnarray}
According to Eqs. (2.11) and (2.12), the textures of $M^{}_{\rm D}$ and
$M^{}_{\rm R}$ can be parametrized as
\begin{eqnarray}
M^{}_{\rm D} = \left(\begin{matrix} A^{}_{\rm r} & B & B^{\ast}_{} \cr
E & C & D \cr E^{\ast}_{} & D^{\ast}_{} & C^{\ast}_{}
\end{matrix}\right) \;, \quad
M^{}_{\rm R} = \left(\begin{matrix} a^{}_{\rm r} & b & b^{\ast}_{} \cr
b & e & d^{}_{\rm r} \cr b^{\ast}_{} & d^{}_{\rm r} & e^{\ast}_{}
\end{matrix}\right) \;,
\end{eqnarray}
where the subscript ``${\rm r}$" means that this element is real. Taking
account of the seesaw formula in Eq. (2.6), we find that the effective
neutrino mass matrix $M^{}_\nu$ has the following texture:
\begin{eqnarray}
M^{}_{\nu} = \left(\begin{matrix} a^{\prime}_{\rm r} & b^{\prime}_{} &
b^{\prime \ast}_{} \cr b^{\prime}_{} & e^{\prime}_{} & d^{\prime}_{\rm r}
\cr b^{\prime \ast}_{} & d^{\prime}_{\rm r} & e^{\prime \ast}_{}
\end{matrix}\right) \;,
\end{eqnarray}
where $a^{\prime}_{\rm r}$, $b^{\prime}_{}$, $d^{\prime}_{\rm r}$
and $e^{\prime}_{}$ are explicitly given by
\begin{eqnarray}
a^{\prime}_{\rm r} \hspace{-0.2cm}&=&\hspace{-0.2cm}
- \frac{1}{\det{M^{}_{\rm R}}} \left\{ A^{2}_{\rm r} \left( |e|^{2}_{} -
d^{2}_{\rm r}\right) + 4A^{}_{\rm r} {\rm Re} \left[ B\left( d^{}_{\rm r}
b^{\ast}_{} - b e^{\ast}_{} \right)\right] + 2 {\rm Re} \left[ B^{2}_{}
\left( a^{}_{\rm r} e^{\ast}_{} - b^{\ast2}_{} \right)\right] \right.
\nonumber
\\
\hspace{-0.2cm}&&\hspace{-0.2cm} + \left. 2 |B|^{2}_{} \left( |b|^2
- a^{}_{\rm r}d \right) \right\} \;,
\nonumber
\\
b^{\prime}_{} \hspace{-0.2cm}&=&\hspace{-0.2cm} -
\frac{1}{\det{M^{}_{\rm R}}} \left\{ A^{}_{\rm r} E \left( |e|^{2}_{} - d^{2}_{\rm r}
\right) + 2E {\rm Re} \left[ B \left( d^{}_{\rm r}b^{\ast}_{} - b e^{\ast}_{}
\right)\right] + A^{}_{\rm r}C \left( d^{}_{\rm r}b^{\ast}_{} - b e^{\ast}_{} \right)
\right.
\nonumber
\\
\hspace{-0.2cm}&&\hspace{-0.2cm} + BC \left( a^{}_{\rm r} e^{\ast}_{} -b^{\ast2}_{}
\right) + \left( B^{\ast}_{}C + BD \right)\left( |b|^{2}_{} - ad^{}_{\rm r} \right)
+ A^{}_{\rm r}D \left( bd^{}_{\rm r} - b^{\ast}_{}e \right)
\nonumber
\\
\hspace{-0.2cm}&&\hspace{-0.2cm} + \left. B^{\ast}_{}D \left( a^{}_{\rm r}e - b^{2}_{}
\right) \right\} \;,
\nonumber
\\
e^{\prime}_{} \hspace{-0.2cm}&=&\hspace{-0.2cm}
- \frac{1}{\det{M^{}_{\rm R}}} \left[ E^{2}_{} \left( |e|^{2}_{} - d^{2}_{\rm r}
\right) + 2EC \left( d^{}_{\rm r}b^{\ast}_{} - be^{\ast}_{} \right) +
2ED \left( bd^{}_{\rm r} - b^{\ast}_{}e \right) \right.
\nonumber
\\
\hspace{-0.2cm}&&\hspace{-0.2cm} + \left.  C^{2}_{} \left( a^{}_{\rm r}
e^{\ast}_{}
 - b^{\ast2}_{} \right) + 2CD \left( |b|^{2}_{} - a^{}_{\rm r} d^{}_{\rm r} \right)
 + D^{2}_{} \left( a^{}_{\rm r}e - b^{2}_{}\right) \right]\;,
\nonumber
\\
d^{\prime}_{\rm r} \hspace{-0.2cm}&=&\hspace{-0.2cm}
- \frac{1}{\det{M^{}_{\rm R}}} \left\{ |E|^{2}_{} \left( |e|^{2}_{} - d^{2}_{\rm r}
\right) + 2 {\rm Re} \left[ \left( E^{\ast}_{}C + ED^{\ast} \right) \left(
d^{}_{\rm r}b^{\ast}_{} - b e^{\ast}_{} \right)\right]
\right.
\nonumber
\\
\hspace{-0.2cm}&&\hspace{-0.2cm} + \left. \left( |C|^{2}_{} + |D|^{2}_{} \right)
\left( |b|^{2}_{} - a^{}_{\rm r} d^{}_{\rm r} \right) + 2 {\rm Re}
\left[ CD^{\ast}_{} \left( a^{}_{\rm r} e^{\ast}_{} - b^{\ast2}_{} \right)\right]
\right\} \;,
\end{eqnarray}
with $\det{M^{}_{\rm R}} = a^{}_{\rm r}|e|^{2}_{} + 2|b|^{2}_{}d^{}_{\rm r} -
a^{}_{\rm r} d^{2}_{\rm r} - 2 {\rm Re} \left( b^{2}_{} e^{\ast}_{} \right)$.
We see that $M^{}_{\nu}$ and $M^{}_{\rm R}$ have the same structure respecting
the $\mu$-$\tau$ reflection symmetry, and therefore there exists an interesting
seesaw mirroring relationship between light and heavy Majorana neutrinos.

The other possibilities can be similarly discussed by repeating the
above procedure with either sets $G^n$ corresponding to ${\cal S}^{}_{\rm (L)} =
{\cal S}^{}_{\rm (R)}$ or sets $G^n_{\rm L} \times G^n_{\rm R}$ including
both ${\cal S}^{}_{\rm (L)} = {\cal S}^{}_{\rm (R)}$ and
${\cal S}^{}_{\rm (L)} \neq {\cal S}^{}_{\rm (R)}$ options.
In Table 2.1 we list and classify the textures of $M^{}_{\rm D}$, $M^{}_{\rm R}$
and $M^{}_\nu$ corresponding to all the possible sets under consideration.
For the sake of simplicity, the explicit relations between the parameters
of $M^{}_\nu$ and those of $M^{}_{\rm D}$ and $M^{}_{\rm R}$ have been
omitted from Table 2.1.
\begin{table}[h!]
\centering
\newcommand{\tabincell}[2]{\begin{tabular}{@{}#1@{}}#2\end{tabular}}
\caption{All the possible structures of
$M^{}_{\rm D}$ and $M^{}_{\rm R}$ constrained by sets $G^{n}$ or
$G^{n}_{\rm L} \times G^{n}_{\rm R}$ in the $S^{}_3$ reflection
symmetry limit, and the consequent structures of $M^{}_\nu$ via
the seesaw formula. The subscript ``${\rm r}$" of a given matrix
element means that this element is real.}
\renewcommand\arraystretch{1.09}
\begin{longtable}{ccccc}
\hline\hline
Cases & $M^{}_{\rm D}$ & $M^{}_{\rm R}$ & $M^{}_{\nu}$ & Sets
\\
\hline
$A^{}_1$ & $\left(\begin{matrix} A^{}_{\rm r} & B & B^{\ast}_{} \cr
E & C & D \cr E^{\ast}_{} & D^{\ast}_{} & C^{\ast}_{}
\end{matrix}\right)$ & $\left(\begin{matrix} a^{}_{\rm r} & b &
b^{\ast}_{} \cr b & e & d^{}_{\rm r} \cr b^{\ast}_{} & d^{}_{\rm r}
& e^{\ast}_{} \end{matrix}\right)$ & $\left(\begin{matrix}
a^{\prime}_{\rm r} & b^{\prime}_{} & b^{\prime \ast}_{} \cr
b^{\prime}_{} & e^{\prime}_{} & d^{\prime}_{\rm r} \cr
b^{\prime \ast}_{} & d^{\prime}_{\rm r} & e^{\prime \ast}_{}
\end{matrix}\right)$ &
$G^{1}_{5}$, $G^{1}_{5{\rm L}} \times G^{1}_{5{\rm R}}$
\\
$A^{}_2$ & $\left(\begin{matrix} C & D & E \cr D^{\ast}_{} & C^{\ast}_{} &
E^{\ast}_{} \cr B & B^{\ast}_{} & A^{}_{\rm r} \end{matrix}\right)$ &
$\left(\begin{matrix} e & d^{}_{\rm r} & b \cr d^{}_{\rm r} &
e^{\ast}_{} & b^{\ast}_{} \cr b & b^{\ast}_{} & a^{}_{\rm r}
\end{matrix}\right)$ & $\left(\begin{matrix} e^{\prime}_{} &
d^{\prime}_{\rm r} &  b^{\prime}_{} \cr d^{\prime}_{\rm r} &
e^{\prime \ast}_{} & b^{\prime \ast}_{} \cr b^{\prime}_{} &
b^{\prime \ast}_{} & a^{\prime}_{\rm r} \end{matrix}\right)$ &
$G^{1}_{4}$, $G^{1}_{4{\rm L}} \times G^{1}_{4{\rm R}}$
\\
$A^{}_3$ & $\left(\begin{matrix} C^{\ast}_{} & E^{\ast}_{} & D^{\ast}_{}
\cr  B^{\ast}_{} & A^{}_{\rm r} & B \cr D & E & C \end{matrix}\right)$
& $\left(\begin{matrix} e^{\ast}_{} & b^{\ast}_{} & d^{}_{\rm r} \cr
b^{\ast}_{} & a^{}_{\rm r} & b \cr d^{}_{\rm r} & b & e
\end{matrix}\right)$ & $\left(\begin{matrix} e^{\prime \ast}_{} &
b^{\prime \ast}_{} &  d^{\prime}_{\rm r} \cr b^{\prime \ast}_{} &
a^{\prime}_{\rm r} & b^{\prime}_{} \cr d^{\prime}_{\rm r} &
b^{\prime}_{} & e^{\prime}_{} \end{matrix}\right)$ &
$G^{1}_{6}$, $G^{1}_{6{\rm L}} \times G^{1}_{6{\rm R}}$
\\
$B^{}_1$ & $\left(\begin{matrix} A^{}_{\rm r} & B^{}_{\rm r} & B^{}_{\rm r}
\cr E^{}_{\rm r} & C^{}_{\rm r} & D^{}_{\rm r} \cr E^{}_{\rm r} &
D^{}_{\rm r} & C^{}_{\rm r} \end{matrix}\right)$ &
$\left(\begin{matrix} a^{}_{\rm r} & b^{}_{\rm r} & b^{}_{\rm r} \cr
b^{}_{\rm r} & e^{}_{\rm r} & d^{}_{\rm r} \cr b^{}_{\rm r} &
d^{}_{\rm r} & e^{}_{\rm r} \end{matrix}\right)$ & $\left(\begin{matrix}
a^{\prime}_{\rm r} & b^{\prime}_{\rm r} & b^{\prime}_{\rm r} \cr
b^{\prime}_{\rm r} & e^{\prime}_{\rm r} & d^{\prime}_{\rm r} \cr
b^{\prime}_{\rm r} & d^{\prime}_{\rm r} & e^{\prime}_{\rm r}
\end{matrix}\right)$ & $ G^{2}_{15}$
\\
$B^{}_2$ & $\left(\begin{matrix} C^{}_{\rm r} & D^{}_{\rm r} & E^{}_{\rm r}
\cr D^{}_{\rm r} & C^{}_{\rm r} & E^{}_{\rm r} \cr B^{}_{\rm r} &
B^{}_{\rm r} & A^{}_{\rm r} \end{matrix}\right)$ & $\left(\begin{matrix}
e^{}_{\rm r} & d^{}_{\rm r} & b^{}_{\rm r} \cr d^{}_{\rm r} &
e^{}_{\rm r} & b^{}_{\rm r} \cr b^{}_{\rm r} & b^{}_{\rm r} &
a^{}_{\rm r} \end{matrix}\right)$ & $\left(\begin{matrix}
e^{\prime}_{\rm r} & d^{\prime}_{\rm r} &  b^{\prime}_{\rm r} \cr
d^{\prime}_{\rm r} & e^{\prime}_{\rm r} & b^{\prime}_{\rm r} \cr
b^{\prime}_{\rm r} & b^{\prime}_{\rm r} & a^{\prime}_{\rm r}
\end{matrix}\right)$ & $G^{2}_{14}$
\\
$B^{}_3$ & $\left(\begin{matrix} C^{}_{\rm r} & E^{}_{\rm r} &
D^{}_{\rm r} \cr  B^{}_{\rm r} & A^{}_{\rm r} & B^{}_{\rm r}
\cr D^{}_{\rm r} & E^{}_{\rm r} & C^{}_{\rm r} \end{matrix}
\right)$ & $\left(\begin{matrix} e^{}_{\rm r} & b^{}_{\rm r} &
d^{}_{\rm r} \cr b^{}_{\rm r} & a^{}_{\rm r} & b^{}_{\rm r}
\cr d^{}_{\rm r} & b^{}_{\rm r} & e^{}_{\rm r} \end{matrix}
\right)$ & $\left(\begin{matrix} e^{\prime}_{\rm r} &
b^{\prime}_{\rm r} &  d^{\prime}_{\rm r} \cr b^{\prime}_{\rm r}
& a^{\prime}_{\rm r} & b^{\prime}_{\rm r} \cr d^{\prime}_{\rm r}
& b^{\prime}_{\rm r} & e^{\prime}_{\rm r} \end{matrix}\right)$
& $G^{2}_{16}$
\\
$C$ & $\left(\begin{matrix} A^{}_{\rm r} & B & B^{\ast}_{} \cr
B^{\ast}_{} & A^{}_{\rm r} & B \cr B & B^{\ast}_{} & A^{}_{\rm r}
\end{matrix}\right)$ & $\left(\begin{matrix} a^{}_{\rm r} &
b^{}_{\rm r} & b^{}_{\rm r} \cr b^{}_{\rm r} & a^{}_{\rm r} &
b^{}_{\rm r} \cr b^{}_{\rm r} & b^{}_{\rm r} & a^{}_{\rm r}
\end{matrix}\right)$ & $\left(\begin{matrix} a^{\prime}_{\rm r}
& b^{\prime}_{\rm r} &  b^{\prime}_{\rm r} \cr b^{\prime}_{\rm r}
& a^{\prime}_{\rm r} & b^{\prime}_{\rm r} \cr b^{\prime}_{\rm r}
& b^{\prime}_{\rm r} & a^{\prime}_{\rm r} \end{matrix}\right)$ &
\tabincell{c}{$G^{2}_{45}$, $G^{2}_{46}$ \\
$G^{2}_{56}$, $G^{3}_{456}$}
\\
$D$ & $\left(\begin{matrix} A^{}_{\rm r} & B^{}_{\rm r} &
B^{}_{\rm r} \cr  B^{}_{\rm r} & A^{}_{\rm r} & B^{}_{\rm r}
\cr B^{}_{\rm r} & B^{}_{\rm r} & A^{}_{\rm r} \end{matrix}
\right)$ & $\left(\begin{matrix} a^{}_{\rm r} & b^{}_{\rm r} &
b^{}_{\rm r} \cr b^{}_{\rm r} & a^{}_{\rm r} & b^{}_{\rm r} \cr
b^{}_{\rm r} & b^{}_{\rm r} & a^{}_{\rm r} \end{matrix}\right)$
& $\left(\begin{matrix} a^{\prime}_{\rm r} & b^{\prime}_{\rm r}
&  b^{\prime}_{\rm r} \cr b^{\prime}_{\rm r} & a^{\prime}_{\rm r}
& b^{\prime}_{\rm r} \cr b^{\prime}_{\rm r} & b^{\prime}_{\rm r}
& a^{\prime}_{\rm r} \end{matrix}\right)$ &
\tabincell{c}{$G^{3}_{145}$, $G^{3}_{146}$ \\
$G^{3}_{156}$, $G^{4}_{\overline{23}}$}
\\
$E^{}_1$ & $\left(\begin{matrix} A^{}_{\rm r} & B^{}_{\rm r} &
B^{}_{\rm r} \cr  B^{}_{\rm r} & A^{}_{\rm r} & B^{}_{\rm r}
\cr B^{}_{\rm r} & B^{}_{\rm r} & A^{}_{\rm r} \end{matrix}
\right)$ & $\left(\begin{matrix} a^{}_{\rm r} & b^{}_{\rm r}
& b^{}_{\rm r} \cr b^{}_{\rm r} & b^{}_{\rm r} & a^{}_{\rm r}
\cr b^{}_{\rm r} & a^{}_{\rm r} & b^{}_{\rm r} \end{matrix}
\right)$ & $\left(\begin{matrix} a^{\prime}_{\rm r} &
b^{\prime}_{\rm r} &  b^{\prime}_{\rm r} \cr b^{\prime}_{\rm r}
& b^{\prime}_{\rm r} & a^{\prime}_{\rm r} \cr b^{\prime}_{\rm r}
& a^{\prime}_{\rm r} & b^{\prime}_{\rm r} \end{matrix}\right)$
& \tabincell{c}{$G^{2}_{25}$, $G^{2}_{35}$, $G^{3}_{125}$ \\
$G^{3}_{135}$, $G^{3}_{235}$, $G^{4}_{\overline{46}}$}
\\
$E^{}_2$ & $\left(\begin{matrix} A^{}_{\rm r} & B^{}_{\rm r} &
B^{}_{\rm r} \cr  B^{}_{\rm r} & A^{}_{\rm r} & B^{}_{\rm r}
\cr B^{}_{\rm r} & B^{}_{\rm r} & A^{}_{\rm r} \end{matrix}
\right)$ & $\left(\begin{matrix} b^{}_{\rm r} & a^{}_{\rm r}
& b^{}_{\rm r} \cr a^{}_{\rm r} & b^{}_{\rm r} & b^{}_{\rm r}
\cr b^{}_{\rm r} & b^{}_{\rm r} & a^{}_{\rm r} \end{matrix}
\right)$ & $\left(\begin{matrix} b^{\prime}_{\rm r} &
a^{\prime}_{\rm r} &  b^{\prime}_{\rm r} \cr a^{\prime}_{\rm r}
& b^{\prime}_{\rm r} & b^{\prime}_{\rm r} \cr b^{\prime}_{\rm r}
& b^{\prime}_{\rm r} & a^{\prime}_{\rm r} \end{matrix}\right)$ &
\tabincell{c}{$G^{2}_{24}$, $G^{2}_{34}$, $G^{3}_{124}$ \\
$G^{3}_{134}$, $G^{3}_{234}$, $G^{4}_{\overline{56}}$}
\\
$E^{}_3$ & $\left(\begin{matrix} A^{}_{\rm r} & B^{}_{\rm r} &
B^{}_{\rm r} \cr  B^{}_{\rm r} & A^{}_{\rm r} & B^{}_{\rm r}
\cr B^{}_{\rm r} & B^{}_{\rm r} & A^{}_{\rm r} \end{matrix}
\right)$ & $\left(\begin{matrix} b^{}_{\rm r} & b^{}_{\rm r}
& a^{}_{\rm r} \cr b^{}_{\rm r} & a^{}_{\rm r} & b^{}_{\rm r}
\cr a^{}_{\rm r} & b^{}_{\rm r} & b^{}_{\rm r} \end{matrix}
\right)$ & $\left(\begin{matrix} b^{\prime}_{\rm r} &
b^{\prime}_{\rm r} &  a^{\prime}_{\rm r} \cr b^{\prime}_{\rm r}
& a^{\prime}_{\rm r} & b^{\prime}_{\rm r} \cr a^{\prime}_{\rm r}
& b^{\prime}_{\rm r} & b^{\prime}_{\rm r} \end{matrix}\right)$ &
\tabincell{c}{$G^{2}_{26}$, $G^{2}_{36}$, $G^{3}_{126}$ \\
$G^{3}_{136}$, $G^{3}_{236}$, $G^{4}_{\overline{45}}$}
\\
\hline\hline
\end{longtable}
\end{table}
\begin{table}[h!]
\centering
\newcommand{\tabincell}[2]{\begin{tabular}{@{}#1@{}}#2\end{tabular}}
{Table 2.1: Continued}
\vspace{0.25cm}
\renewcommand\arraystretch{1.1}
\begin{longtable}{ccccc}
\hline\hline
Cases & $M^{}_{\rm D}$ & $M^{}_{\rm R}$ & $M^{}_{\nu}$ & Sets
\\
\hline
$F$ & $\left(\begin{matrix} A^{}_{\rm r} & B^{}_{\rm r} &
C^{}_{\rm r} \cr  C^{}_{\rm r} & A^{}_{\rm r} & B^{}_{\rm r} \cr
B^{}_{\rm r} & C^{}_{\rm r} & A^{}_{\rm r} \end{matrix}\right)$ &
$\left(\begin{matrix} a^{}_{\rm r} & b^{}_{\rm r} & e^{}_{\rm r}
\cr b^{}_{\rm r} & e^{}_{\rm r} & a^{}_{\rm r} \cr e^{}_{\rm r} &
a^{}_{\rm r} & b^{}_{\rm r} \end{matrix}\right)$ & $\left(
\begin{matrix} a^{\prime}_{\rm r} & b^{\prime}_{\rm r} &
e^{\prime}_{\rm r} \cr b^{\prime}_{\rm r} & e^{\prime}_{\rm r} &
a^{\prime}_{\rm r} \cr e^{\prime}_{\rm r} & a^{\prime}_{\rm r} &
b^{\prime}_{\rm r} \end{matrix}\right)$ & \tabincell{c}{$G^{1}_{2}
$, $G^{1}_{3}$, $G^{2}_{12}$
\\
$G^{2}_{13}$, $G^{2}_{23}$, $G^{3}_{123}$
\\
$G^{1}_{2{\rm L}} \times G^{1}_{2{\rm R}} $, $G^{1}_{3{\rm L}}
\times G^{1}_{3{\rm R}} $
}
\\
$H^{}_1$ & $\left(\begin{matrix} A^{}_{\rm r} & B^{}_{\rm r} &
B^{}_{\rm r} \cr  D^{}_{\rm r} & C^{}_{\rm r} & C^{}_{\rm r}
\cr D^{}_{\rm r} & C^{}_{\rm r} & C^{}_{\rm r} \end{matrix}
\right)$ & $\left(\begin{matrix} a^{}_{\rm r} & b^{}_{\rm r}
& b^{}_{\rm r} \cr b^{}_{\rm r} & e^{}_{\rm r} & d^{}_{\rm r}
\cr b^{}_{\rm r} & d^{}_{\rm r} & e^{}_{\rm r} \end{matrix}
\right)$ & $\left(\begin{matrix} a^{\prime}_{\rm r} &
b^{\prime}_{\rm r} &  b^{\prime}_{\rm r} \cr b^{\prime}_{\rm r}
& e^{\prime}_{\rm r} & e^{\prime}_{\rm r} \cr b^{\prime}_{\rm r}
& e^{\prime}_{\rm r} & e^{\prime}_{\rm r} \end{matrix}\right)$
& $G^{2}_{15{\rm L}} \times G^{2}_{15{\rm R}} $
\\
$H^{}_2$ & $\left(\begin{matrix} C^{}_{\rm r} & C^{}_{\rm r} &
D^{}_{\rm r} \cr  C^{}_{\rm r} & C^{}_{\rm r} & D^{}_{\rm r}
\cr B^{}_{\rm r} & B^{}_{\rm r} & A^{}_{\rm r} \end{matrix}
\right)$ & $\left(\begin{matrix} e^{}_{\rm r} & d^{}_{\rm r}
& b^{}_{\rm r} \cr d^{}_{\rm r} & e^{}_{\rm r} & b^{}_{\rm r}
\cr b^{}_{\rm r} & b^{}_{\rm r} & a^{}_{\rm r} \end{matrix}
\right)$ & $\left(\begin{matrix} e^{\prime}_{\rm r} &
e^{\prime}_{\rm r} &  b^{\prime}_{\rm r} \cr e^{\prime}_{\rm r}
& e^{\prime}_{\rm r} & b^{\prime}_{\rm r} \cr b^{\prime}_{\rm r}
& b^{\prime}_{\rm r} & a^{\prime}_{\rm r} \end{matrix}
\right)$ & $G^{2}_{14{\rm L}} \times G^{2}_{14{\rm R}} $
\\
$H^{}_3$ & $\left(\begin{matrix} C^{}_{\rm r} & D^{}_{\rm r} &
C^{}_{\rm r} \cr  B^{}_{\rm r} & A^{}_{\rm r} & B^{}_{\rm r}
\cr C^{}_{\rm r} & D^{}_{\rm r} & C^{}_{\rm r} \end{matrix}
\right)$ & $\left(\begin{matrix} e^{}_{\rm r} & b^{}_{\rm r}
& d^{}_{\rm r} \cr b^{}_{\rm r} & a^{}_{\rm r} & b^{}_{\rm r}
\cr d^{}_{\rm r} & b^{}_{\rm r} & e^{}_{\rm r} \end{matrix}
\right)$ & $\left(\begin{matrix} e^{\prime}_{\rm r} &
b^{\prime}_{\rm r} &  e^{\prime}_{\rm r} \cr b^{\prime}_{\rm r}
& a^{\prime}_{\rm r} & b^{\prime}_{\rm r} \cr e^{\prime}_{\rm r}
& b^{\prime}_{\rm r} & e^{\prime}_{\rm r} \end{matrix}\right)$
& $G^{2}_{16{\rm L}} \times G^{2}_{16{\rm R}} $
\\
$I^{}_1$ & $ A^{}_{\rm r} \left(\begin{matrix} 1 & 1 & 1 \cr  1 &
1 & 1 \cr 1 & 1 & 1 \end{matrix}\right)$ & $\left(\begin{matrix}
a^{}_{\rm r} & b^{}_{\rm r} & b^{}_{\rm r} \cr b^{}_{\rm r} &
b^{}_{\rm r} & a^{}_{\rm r} \cr b^{}_{\rm r} & a^{}_{\rm r} &
b^{}_{\rm r} \end{matrix}\right)$ & $ a^{\prime}_{\rm r} \left(
\begin{matrix} 1 & 1 &  1 \cr 1 & 1 & 1 \cr 1 & 1 & 1 \end{matrix}
\right)$ & \tabincell{c}{$G^{2}_{25 {\rm L}} \times G^{2}_{25
{\rm R}} $, $G^{2}_{35 {\rm L}} \times G^{2}_{35 {\rm R}} $
\\
$G^{3}_{125 {\rm L}} \times G^{3}_{125 {\rm R}} $ ,
$G^{3}_{135 {\rm L}} \times G^{3}_{135 {\rm R}} $
\\
$G^{3}_{235 {\rm L}} \times G^{3}_{235 {\rm R}} $,
$G^{4}_{\overline{46} {\rm L}} \times G^{4}_{\overline{46}
{\rm R}} $}
\\
$I^{}_2$ & $ A^{}_{\rm r} \left(\begin{matrix} 1 & 1 & 1 \cr  1
& 1 & 1 \cr 1 & 1 & 1 \end{matrix}\right)$ &
$\left(\begin{matrix} b^{}_{\rm r} & a^{}_{\rm r} &
b^{}_{\rm r} \cr a^{}_{\rm r} & b^{}_{\rm r} & b^{}_{\rm r}
\cr b^{}_{\rm r} & b^{}_{\rm r} & a^{}_{\rm r} \end{matrix}
\right)$ & $ a^{\prime}_{\rm r} \left(\begin{matrix} 1 & 1
&  1 \cr 1 & 1 & 1 \cr 1 & 1 & 1 \end{matrix}\right)$ &
\tabincell{c}{$G^{2}_{24 {\rm L}} \times G^{2}_{24 {\rm R}} $,
$G^{2}_{34 {\rm L}} \times G^{2}_{34 {\rm R}} $
\\
$G^{3}_{124 {\rm L}} \times G^{3}_{124 {\rm R}} $ ,
$G^{3}_{134 {\rm L}} \times G^{3}_{134 {\rm R}} $
\\
$G^{3}_{234 {\rm L}} \times G^{3}_{234 {\rm R}} $,
$G^{4}_{\overline{56} {\rm L}} \times G^{4}_{\overline{56}
{\rm R}} $}
\\
$I^{}_3$ & $ A^{}_{\rm r} \left(\begin{matrix} 1 & 1 & 1 \cr
1 & 1 & 1 \cr 1 & 1 & 1 \end{matrix}\right)$ &
$\left(\begin{matrix} b^{}_{\rm r} & b^{}_{\rm r} & a^{}_{\rm r}
\cr b^{}_{\rm r} & a^{}_{\rm r} & b^{}_{\rm r} \cr a^{}_{\rm r}
& b^{}_{\rm r} & b^{}_{\rm r} \end{matrix}\right)$ &
$ a^{\prime}_{\rm r} \left(\begin{matrix} 1 & 1 &  1 \cr 1 & 1
& 1 \cr 1 & 1 & 1 \end{matrix}\right)$ & \tabincell{c}{
$G^{2}_{26 {\rm L}} \times G^{2}_{26 {\rm R}} $,
$G^{2}_{36 {\rm L}} \times G^{2}_{36 {\rm R}} $
\\
$G^{3}_{126 {\rm L}} \times G^{3}_{126 {\rm R}} $ ,
$G^{3}_{136 {\rm L}} \times G^{3}_{136 {\rm R}} $
\\
$G^{3}_{236 {\rm L}} \times G^{3}_{236 {\rm R}} $,
$G^{4}_{\overline{45} {\rm L}} \times G^{4}_{\overline{45}
{\rm R}} $}
\\
$J$ & $ A^{}_{\rm r} \left(\begin{matrix} 1 & 1 & 1 \cr
1 & 1 & 1 \cr 1 & 1 & 1 \end{matrix}\right)$ &
$\left(\begin{matrix} a^{}_{\rm r} & b^{}_{\rm r} &
e^{}_{\rm r} \cr b^{}_{\rm r} & e^{}_{\rm r} & a^{}_{\rm r}
\cr e^{}_{\rm r} & a^{}_{\rm r} & b^{}_{\rm r} \end{matrix}
\right)$ & $ a^{\prime}_{\rm r} \left(\begin{matrix} 1 & 1
&  1 \cr 1 & 1 & 1 \cr 1 & 1 & 1 \end{matrix}\right)$ &
\tabincell{c}{$G^{2}_{12 {\rm L}} \times G^{2}_{12 {\rm R}}
$, $G^{2}_{13 {\rm L}} \times G^{2}_{13 {\rm R}} $
\\
$G^{2}_{23 {\rm L}} \times G^{2}_{23 {\rm R}} $ ,
$G^{3}_{123 {\rm L}} \times G^{3}_{123 {\rm R}} $}
\\
$K$ & $ A^{}_{\rm r} \left(\begin{matrix} 1 & 1 & 1 \cr
1 & 1 & 1 \cr 1 & 1 & 1 \end{matrix}\right)$ &
$\left(\begin{matrix} a^{}_{\rm r} & b^{}_{\rm r} &
b^{}_{\rm r} \cr b^{}_{\rm r} & a^{}_{\rm r} & b^{}_{\rm r}
\cr b^{}_{\rm r} & b^{}_{\rm r} & a^{}_{\rm r} \end{matrix}
\right)$ & $ a^{\prime}_{\rm r} \left(\begin{matrix} 1 & 1
&  1 \cr 1 & 1 & 1 \cr 1 & 1 & 1 \end{matrix}\right)$ &
\tabincell{c}{$G^{2}_{45 {\rm L}} \times G^{2}_{45 {\rm R}}
$, $G^{2}_{46 {\rm L}} \times G^{2}_{46 {\rm R}} $
\\
$G^{2}_{56 {\rm L}} \times G^{2}_{56 {\rm R}} $ ,
$G^{3}_{145 {\rm L}} \times G^{3}_{145 {\rm R}} $
\\
$G^{3}_{146 {\rm L}} \times G^{3}_{146 {\rm R}} $ ,
$G^{3}_{156 {\rm L}} \times G^{3}_{156 {\rm R}} $
\\
$G^{3}_{456 {\rm L}} \times G^{3}_{456 {\rm R}} $ ,
$G^{4}_{\overline{23} {\rm L}} \times G^{4}_{\overline{23}
{\rm R}} $}
\\
$L$ & $\left(\begin{matrix} a^{}_{\rm r} & b^{}_{\rm r} &
b^{}_{\rm r} \cr b^{}_{\rm r} & a^{}_{\rm r} & b^{}_{\rm r}
\cr b^{}_{\rm r} & b^{}_{\rm r} & a^{}_{\rm r} \end{matrix}
\right)$ &  $ A^{}_{\rm r} \left(\begin{matrix} 1 & 1 & 1
\cr  1 & 1 & 1 \cr 1 & 1 & 1 \end{matrix}\right)$ & --------
& \tabincell{c}{All $G^{n}_{i^{}_1 i^{}_2 \cdots i^{}_n }$ or $G^{n}_{
\overline{i^{}_1 i^{}_2 \cdots i^{}_{6-n}}}$
\\
that are not listed above}
\\
$N$ &  $ A^{}_{\rm r} \left(\begin{matrix} 1 & 1 & 1 \cr
1 & 1 & 1 \cr 1 & 1 & 1 \end{matrix}\right)$ &  $ A^{}_{\rm r}
\left(\begin{matrix} 1 & 1 & 1 \cr  1 & 1 & 1 \cr 1 & 1 & 1
\end{matrix}\right)$ & -------- & \tabincell{c}{
$G^{n}_{i^{}_1 i^{}_2 \cdots i^{}_n {\rm L}} \times
G^{n}_{i^{}_1 i^{}_2 \cdots i^{}_n {\rm R}}
$ or
\\
$G^{n}_{\overline{i^{}_1 i^{}_2 \cdots i^{}_{6-n}} {\rm L}} \times G^{n}_{
\overline{i^{}_1 i^{}_2 \cdots i^{}_{6-n}} {\rm R}}$
\\
that are not listed above}
\\
\hline\hline
\end{longtable}
\end{table}

\subsection{The seesaw mirroring structure of $M^{}_\nu$}

Table 2.1 provides a classification of all the possible structures
of $M^{}_\nu$ in accordance with those of $M^{}_{\rm D}$ and $M^{}_{\rm R}$.
For each category of $M^{}_\nu$, its structure is the same as or similar to
the structure of $M^{}_{\rm R}$ or $M^{}_{\rm D}$, reflecting the seesaw
mirroring feature that we have stressed.

The classification is certainly based on Eqs. (2.5) and (2.6). If the textures
of $M^{}_{\rm D}$, $M^{}_{\rm R}$ and $M^{}_\nu$ constrained by different sets,
such as $G^{n}_{i^{}_{1} i^{}_{2} \cdots i^{}_{n}}$
(or $G^{n}_{i^{}_{1} i^{}_{2} \cdots i^{}_{n} {\rm L}} \times
G^{n}_{i^{}_{1} i^{}_{2} \cdots i^{}_{n} {\rm R}}$) or
$G^{n}_{\overline{i^{}_{1} i^{}_{2} \cdots i^{}_{6-n}}}$
(or $G^{n}_{\overline{i^{}_{1} i^{}_{2} \cdots i^{}_{6-n}} {\rm L}} \times
G^{n}_{\overline{i^{}_{1} i^{}_{2} \cdots i^{}_{6-n}} {\rm R}}$), are all
the same, then they will be sorted into one group. In this way we are totally
left with 22 categories of distinctive structures of the mass matrices,
as listed in Table 2.1.
Note that the mass matrices belonging to categories $A^{}_1$, $A^{}_2$ and
$A^{}_3$ are actually correlated with each other via a transformation associated
with $S^{(231)}_{}$ and $S^{{(312)}}_{}$. To be specific,
\begin{eqnarray}
M^{(A^{}_2)}_{\rm D} \hspace{-0.2cm}&=&\hspace{-0.2cm}
S^{(231)} M^{(A^{}_1)}_{\rm D} S^{(312)} \;,
\quad M^{(A^{}_3)}_{\rm D} = S^{(312)} M^{(A^{}_1)}_{\rm D} S^{(231)} \;,
\nonumber \\
M^{(A^{}_2)}_{\rm R} \hspace{-0.2cm}&=&\hspace{-0.2cm}
S^{(231)} M^{(A^{}_1)}_{\rm R} S^{(312)} \;,
\quad M^{(A^{}_3)}_{\rm R} = S^{(312)} M^{(A^{}_1)}_{\rm R} S^{(231)} \;,
\nonumber \\
M^{(A^{}_2)}_{\nu} \hspace{-0.2cm}&=&\hspace{-0.2cm}
S^{(231)} M^{(A^{}_1)}_{\nu} S^{(312)} \;,
\quad M^{(A^{}_3)}_{\nu} = S^{(312)} M^{(A^{}_1)}_{\nu} S^{(231)} \; .
\end{eqnarray}
We find that the same correlations exist for $M^{}_{\rm D}$, $M^{}_{\rm R}$
and $M^{}_\nu$ in categories $B^{}_{i}$, $E^{}_{i}$, $H^{}_{i}$ and $I^{}_{i}$ (for
$i = 1, 2, 3$). In fact, Eq. (2.16) for categories $A^{}_i$ and similar
relations of this kind for other categories can be understood
as follows.
\begin{enumerate}
\item     The corresponding sets in categories $X^{}_{1}$, $X^{}_{2}$ and
$X^{}_{3}$ (for $X = A$, $B$, $E$, $H$ or $I$) contain $S^{(132)}$,
$S^{(213)}$ and $S^{(321)}$, respectively. The other possible elements
(i.e., $S^{(123)}$, $S^{(231)}$, $S^{(312)}$) contained by $X^{}_1$
are simultaneously contained by $X^{}_2$ and $X^{}_3$.

\item     The three-dimensional representation of $S^{}_{3}$ group in Eq. (2.7)
is a unitary representation, and hence ${\cal S}^{\dagger}_{\rm (L)} =
{\cal S}^{-1}_{\rm (L)}$ and ${\cal S}^{\dagger}_{\rm (R)} =
{\cal S}^{-1}_{\rm (R)}$ hold.

\item     Since $S^{(132)}$, $S^{(213)}$ and $S^{(321)}$ belong to one
conjugacy class ${\cal C}^{}_{2}$, they can be connected with one another
by one element of $S^{}_{3}$ group. Namely,
$S^{(213)} = S^{(231)} S^{(132)} [S^{(231)}]^{-1}$,
$S^{(321)} = S^{(312)} S^{(132)} [S^{(312)}]^{-1}$ and
$S^{(213)} = S^{(312)} S^{(321)} [S^{(312)}]^{-1}$.

\item     The conjugacy class ${\cal C}^{}_{1}$ containing elements
$S^{(231)}$ and $S^{(312)}$ is a self reciprocal class, and the subgroup
$Z^{}_{3}$ is an Abelian group. As a result,
$S^{(231)} = [S^{(312)}]^{-1}$, $S^{(312)} = [S^{(231)}]^{-1}$,
$[S^{(231)}]^{2} = S^{(312)}$ and $[S^{(312)}]^{2} = S^{(231)}$ hold,
and the three elements of $Z^{}_{3}$ commute with one another
(i.e., $\left[S^{(123)} , S^{(231)}\right]=\left[S^{(123)} ,
S^{(312)}\right]=\left[S^{(312)} , S^{(231)}\right]=0$).
\end{enumerate}
These properties, together with Eqs. (2.5) and (2.6), allow us to easily obtain
Eq. (2.16) and other similar relations. Such relations are very helpful in
the sense that once the result for one case is achieved, the results for the
other two cases can be conveniently figured out in no need of
repeating the relevant calculations.

The notation $G^{n}_{\rm L} \times G^{n}_{\rm R}$ in Table 2.1
means that the left-handed fields $\nu^{}_{\rm L}$ and the
right-handed fields $N^{}_{\rm R}$ can transform with different elements of $G^{n}$,
corresponding to ${\cal S}^{}_{\rm L}$ for $\nu^{}_{\rm L}$ and ${\cal S}^{}_{\rm R}$
for $N^{}_{\rm R}$ shown in Eq. (2.2), where ${\cal S}^{}_{\rm L}$ and
${\cal S}^{}_{\rm R}$ can be either identical or different
\footnote{Without invoking any confusion, we have omitted the subscript
$i^{}_1 i^{}_2 \cdots i^{}_n$ (or $\overline{i^{}_1 i^{}_2 \cdots i^{}_{6-n}}$)
of set $G^{n}_{i^{}_1 i^{}_2 \cdots i^{}_n}$
(or $G^{n}_{\overline{i^{}_1 i^{}_2 \cdots i^{}_{6-n}}}$) here and
hereafter for the sake of simplicity.}.
Note that sets $G^{1}_{1}$ and $G^{1}_{1{\rm L}} \times G^{1}_{1{\rm R}}$
are trivial in the sense that they only restrict all the elements of a given
mass matrix to be real. That is why for categories $L$ and $N$ listed in
Table 2.1 the corresponding sets do not include $G^{1}_{1}$ and
$G^{1}_{1{\rm L}} \times G^{1}_{1{\rm R}}$. In these two cases the heavy
Majorana neutrino mass matrix $M^{}_{\rm R}$ has a democracy texture
of rank one, and thus its determinant is zero, making the seesaw formula
in Eq. (2.6) does not work anymore.

It is obvious that if a set contains element $S^{(123)}$, then $M^{}_{\rm D}$,
$M^{}_{\rm R}$ and $M^{}_\nu$ will all be real. Eq. (2.5) tells us
that the structure of $M^{}_{\rm D}$ is constrained by both ${\cal S}^{}_{\rm L}$
and ${\cal S}^{}_{\rm R}$, and that of $M^{}_{\rm R}$ is constrained only by
${\cal S}^{}_{\rm R}$. As a result, $M^{}_{\rm D}$ is constrained more strictly
in the case associated with $G^{n}_{\rm L} \times G^{n}_{\rm R}$
than in the case associated with $G^{n}$, but the constraints on $M^{}_{\rm R}$
in these two situations are the same. In fact, $G^{1}$ and
$G^{1}_{\rm L} \times G^{1}_{\rm R}$ are identical and thus lead to
the same textures for relevant mass matrices.

Of course, the structure of $M^{}_{\nu}$ is in general different from that
of $M^{}_{\rm R}$. But as shown in Table 2.1, $M^{}_{\nu}$
and $M^{}_{\rm R}$ {\it do} share the same texture for categories $A^{}_i$ to $F$,
in which the structures of relevant mass matrices are dominated by sets
$G^{n}$ and $G^{1}_{\rm L} \times G^{1}_{\rm R}$.
As for categories $H^{}_i$ to $K$, in which sets $G^{n}_{\rm L}
\times G^{n}_{\rm R}$ (for $n = 2, 3, \cdots, 6$) dominate,
the structure of $M^{}_{\nu}$ is quite similar to that of $M^{}_{\rm R}$
or to a combination of the structures of $M^{}_{\rm D}$ and $M^{}_{\rm R}$.
In these cases $M^{}_{\rm D}$ is constrained more strictly than $M^{}_{\rm R}$,
and hence it possesses a much simpler texture which dominates
the texture pattern of $M^{}_\nu$ via the seesaw formula
in Eq. (2.6). Especially in categories $I^{}_i$ to $K$, the mass matrices
$M^{}_{\nu}$ and $M^{}_{\rm D}$ exactly share the same democracy texture.
To characterize the relationship between light and heavy Majorana neutrinos
in the seesaw framework under consideration, we refer to the structural equality or
similarity between $M^{}_{\nu}$ and $M^{}_{\rm R}$ as a seesaw mirroring
relationship.

Another thing that deserves attention is that the mass matrices constrained by any
one of $S^{(123)}$, $S^{(231)}$ and $S^{(312)}$ must be real. In other words,
$M^{}_{\rm D}$ and $M^{}_{\rm R}$ will be real if the corresponding set in a
given category contains one of the above three elements. Only categories
$A^{}_1$, $A^{}_2$, $A^{}_3$ and $C$, in which $S^{(123)}$, $S^{(231)}$ and
$S^{(312)}$ are not involved, give rise to complex $M^{}_{\rm D}$ and
$M^{}_{\rm R}$. Among them, only categories $A^{}_1$, $A^{}_2$ and $A^{}_3$
allow us to obtain the complex textures of $M^{}_{\nu}$ via the seesaw formula.
This observation means that in the $S^{}_3$ reflection symmetry limit there are
only four possibilities to accommodate CP violation in the lepton-number-violating
decays of heavy Majorana neutrinos, and only three possibilities to accommodate
CP violation in the effective light neutrino mass matrix $M^{}_\nu$.

At this point it is also worth mentioning that $S^{}_3$ is not a symmetry of
the Lagrangian in the neutrino sector. Although we have considered the subgroups
and subsets of $S^{}_3$ group, they are mainly used as a tool to constrain and
classify possible structures of the neutrino mass matrices. From the phenomenological
point of view, our strategy is expected to be helpful for understanding the neutrino
flavor structures under $S^{}_3$ symmetry and providing a reference about which
larger group should be introduced and which representations should be determined
when doing a realistic model-building exercise. We admit that a larger flavor
symmetry group may not have a direct connection with $S^{}_3$, but the latter
is likely to play an indirect but suggestive role in bridging an underlying flavor
symmetry and a phenomenologically favored pattern of $M^{}_\nu$. Since $S^{}_3$
is so simple and instructive in reflecting the possible interchange among three
flavor families, it should be qualified as a good bottom-up example in probing what
is behind tiny neutrino masses and significant flavor mixing effects.

\setcounter{table}{0}
\setcounter{equation}{0}
\setcounter{figure}{0}

\section{Neutrino masses and flavor mixing patterns}

Now we proceed to calculate the light neutrino masses and flavor mixing
parameters for each of the textures of $M^{}_\nu$ listed in Table 2.1
in the basis where the flavor eigenstates of three charged leptons are
identical with their mass eigenstates. Although some of the flavor mixing
patterns derived from $M^{}_\nu$ in the $S^{}_3$ reflection symmetry limit
are expected to be far away from the observed pattern of the PMNS matrix,
it remains instructive to see their salient features from a phenomenological
point of view.

Since $M^{}_{\nu}$ is symmetric, it can be diagonalized by a unitary
transformation matrix $V$ as follows:
$V^{\dagger}_{} M^{}_{\nu} V^{\ast}_{} = \widehat{ M^{}_{\nu}}$, where
$\widehat{M^{}_{\nu}} \equiv {\rm Diag} \left\{ m^{}_{1}, m^{}_{2}, m^{}_{3}
\right\}$ with $m^{}_i$ (for $i=1,2,3$) being the neutrino masses. In the
chosen flavor basis $V$ is just the PMNS matrix which describes
the effects of neutrino mixing and CP violation, and its
standard parametrization form is
\begin{eqnarray}
V = P^{}_{l} \left(\begin{matrix}
c^{}_{12} c^{}_{13} & s^{}_{12} c^{}_{13} & s^{}_{13} e^{-{\rm i}\delta}_{} \cr
-s^{}_{12} c^{}_{23} - c^{}_{12} s^{}_{13} s^{}_{23} e^{{\rm i}\delta}_{} &
c^{}_{12} c^{}_{23} - s^{}_{12} s^{}_{13} s^{}_{23} e^{{\rm i}\delta}_{} &
c^{}_{13} s^{}_{23} \cr s^{}_{12} s^{}_{23} - c^{}_{12} s^{}_{13}
c^{}_{23} e^{{\rm i}\delta}_{} & -c^{}_{12} s^{}_{23} -s^{}_{12} s^{}_{13}
c^{}_{23} e^{{\rm i}\delta}_{} & c^{}_{13} c^{}_{23}
\end{matrix}\right) P^{}_{\nu} \;,
\end{eqnarray}
where $c^{}_{ij} = \cos{\theta^{}_{ij}}$ and
$s^{}_{ij} =  \sin{\theta^{}_{ij}}$ with $ij = 12, 13, 23$,
$P^{}_{l} = {\rm Diag} \left\{e^{{\rm i}\phi^{}_{e}}, e^{{\rm i}\phi^{}_{\mu}},
e^{{\rm i}\phi^{}_{\tau}} \right\}$ contains three unphysical phases which
can be absorbed by rephasing the charged-lepton fields, and
$P^{}_{\nu} = {\rm Diag} \left\{ e^{{\rm i}\rho}, e^{{\rm i}\sigma}, 1 \right\}$
contains two physical Majorana phases. Therefore, a diagonalization of
the effective Majorana neutrino mass matrix $M^{}_\nu$ allows us to determine
three neutrino masses $m^{}_i$, three flavor mixing angles $\theta^{}_{ij}$
and three CP-violating phases $\delta$, $\rho$ and $\sigma$. In the
following we do such exercises by examining all the textures of $M^{}_\nu$
listed in Table 2.1.

\subsection{Categories $A^{}_i$}

In category $A^{}_{1}$ the light neutrino mass matrix $M^{}_{\nu}$ satisfies
the $\mu$-$\tau$ reflection symmetry, which naturally predicts the
phenomenologically favored results $\theta^{}_{23} = \pi/4$
and $\delta = - \pi/2$ \cite{Zhao2016,Harrison}. It is therefore interesting
to reproduce this texture from the canonical seesaw mechanism in the
$S^{}_3$ reflection symmetry limit. To be specific, the $\mu$-$\tau$
reflection symmetry structure of $M^{}_\nu$ in this case leads us to
\begin{eqnarray}
\hspace{-0.2cm}&&\hspace{-0.2cm} \theta^{A^{}_{1}}_{23} = \frac{\pi}{4} \; ,
\quad \delta^{A^{}_{1}}_{} = \pm \frac{\pi}{2} \;, \quad \rho^{A^{}_{1}} ,
\sigma^{A^{}_{1}}_{} = 0 \; {\rm or} \; \frac{\pi}{2} \;,
\nonumber
\\
\hspace{-0.2cm}&&\hspace{-0.2cm} \phi^{A^{}_{1}}_{e} = 0 \;{\rm or}\;
\frac{\pi}{2} \; , \quad \phi^{A^{}_{1}}_{\mu} + \phi^{A^{}_{1}}_{\tau} =
2 \phi^{A^{}_{1}}_{e} \pm \pi \; .
\end{eqnarray}
In addition, the other two flavor mixing angles and the three neutrino masses
in category $A^{}_1$ can be expressed as follows:
\begin{eqnarray}
\hspace{-0.2cm}&&\hspace{-0.2cm} \tan{\theta^{A^{}_{1}}_{13}} = \left|
\frac{{\rm Im} (e^{\prime\prime}_{})}
{ \sqrt{2} {\rm Re} (b^{\prime\prime}_{})} \right| \;,
\nonumber
\\
\hspace{-0.2cm}&&\hspace{-0.2cm} \tan{2\theta^{A^{}_{1}}_{12}} =
\frac{ 2 \sqrt{2} \cos{2\theta^{A^{}_{1}}_{13}
{\rm Re} (b^{\prime\prime}_{}) }}{ c^{A^{}_{1}}_{13} \left\{ \left[
{\rm Re} (e^{\prime\prime}_{}) -d^{\prime\prime}_{} \right]
\cos{2\theta^{A^{}_{1}}_{13}} - \left[ {\rm Re} (e^{\prime\prime}_{})
+ d^{\prime\prime}_{} \right] s^{A^{}_{1}2}_{13} - a^{\prime\prime}_{}
c^{A^{}_{1}2}_{13} \right\} } \;,
\nonumber
\\
\hspace{-0.2cm}&&\hspace{-0.2cm}
m^{}_{1} = \left| - d^{\prime\prime}_{}
- \frac{\sqrt{2} {\rm Re} (b^{\prime\prime}_{}) }{c^{A^{}_{1}}_{13}
\sin{2\theta^{A^{}_{1}}_{12}}} + \frac{ \left[ a^{\prime\prime}_{}
+ {\rm Re} (e^{\prime\prime}_{}) + d^{\prime\prime}_{} \right]
c^{A^{}_{1}2}_{13} }{ 2 \cos{2\theta^{A^{}_{1}}_{13}} } \right|  \;,
\nonumber
\\
\hspace{-0.2cm}&&\hspace{-0.2cm}
m^{}_{2} = \left| - d^{\prime\prime}_{}
+ \frac{\sqrt{2} {\rm Re} (b^{\prime\prime}_{}) }{c^{A^{}_{1}}_{13}
\sin{2\theta^{A^{}_{1}}_{12}}} + \frac{ \left[ a^{\prime\prime}_{}
+ {\rm Re} (e^{\prime\prime}_{}) + d^{\prime\prime}_{} \right]
c^{A^{}_{1}2}_{13} }{ 2 \cos{2\theta^{A^{}_{1}}_{13}} } \right| \;,
\nonumber
\\
\hspace{-0.2cm}&&\hspace{-0.2cm}
m^{}_{3} = \frac{ a^{\prime\prime}_{} s^{A^{}_{1}2}_{13} +
\left[ {\rm Re} (e^{\prime\prime}_{}) + d^{\prime\prime}_{} \right]
c^{A^{}_{1}2}_{13} }{\cos{2\theta^{A^{}_{1}}_{13}}} \;,
\end{eqnarray}
where $a^{\prime\prime}_{} = a^{\prime}_{\rm r}
\exp \left(-2{\rm i}\phi^{A^{}_{1}}_{e}\right)$, $b^{\prime\prime}_{}
= b^{\prime}_{} \exp \left[-{\rm i} \left( \phi^{A^{}_{1}}_{e} +
\phi^{A^{}_{1}}_{\mu}\right)\right]$, $e^{\prime\prime}_{} = e^{\prime}_{}
\exp \left(-2{\rm i} \phi^{A^{}_{1}}_{\mu}\right)$ and $d^{\prime\prime}_{}
= d^{\prime}_{\rm r} \exp \left[-{\rm i} \left( \phi^{A^{}_{1}}_{\mu} +
\phi^{A^{}_{1}}_{\tau} \right)\right]$.

For categories $A^{}_{2}$ and $A^{}_{3}$, the corresponding textures of
$M^{}_\nu$ are related to that in category $A^{}_1$ via Eq. (2.16).
One may therefore choose the same order of three mass eigenvalues and
then establish similar correlations among the three PMNS matrices of
categories $A^{}_1$, $A^{}_2$ and $A^{}_3$ with the help of Eq. (2.16):
\begin{eqnarray}
V^{A^{}_{2}}_{} = S^{(231)}_{} V^{A^{}_{1}}_{} \;, \quad
V^{A^{}_{3}}_{} = S^{(312)}_{} V^{A^{}_{1}}_{} \;.
\end{eqnarray}
As a consequence, the relevant flavor mixing parameters in categories
$A^{}_{2}$ and $A^{}_{3}$ can be respectively related to those of
category $A^{}_{1}$ as follows:
\begin{eqnarray}
\hspace{-0.2cm}&&\hspace{-0.2cm}
\tan{\theta^{A^{}_{2}}_{23}} = \frac{1}{\sqrt{2}
\tan{\theta^{A^{}_{1}}_{13}}} \;,
\nonumber
\\
\hspace{-0.2cm}&&\hspace{-0.2cm}
\sin{\theta^{A^{}_{2}}_{13}} = \frac{1}{\sqrt{2}}
\cos{\theta^{A^{}_{1}}_{13}} \;,
\nonumber
\\
\hspace{-0.2cm}&&\hspace{-0.2cm}
\cos{2\theta^{A^{}_{2}}_{12}} = - \frac{ 1 - \sin^{2}{
\theta^{A^{}_{1}}_{13}} }{ 1 + \sin^{2}{\theta^{A^{}_{1}}_{13}} }
\cos{2\theta^{A^{}_{1}}_{12}} \;,
\nonumber
\\
\hspace{-0.2cm}&&\hspace{-0.2cm}
\sin{\delta^{A^{}_{2}}} = \frac{ \cos{\theta^{A^{}_{1}}_{12}}
\sin{\theta^{A^{}_{1}}_{12}} \cos^2{\theta^{A^{}_{1}}_{13}}
\sin{\theta^{A^{}_{1}}_{13}} }{ 2 \cos{\theta^{A^{}_{2}}_{12}}
\sin{\theta^{A^{}_{2}}_{12}} \cos^2{\theta^{A^{}_{2}}_{13}}
\sin{\theta^{A^{}_{2}}_{13}} \cos{\theta^{A^{}_{2}}_{23}}
\sin{\theta^{A^{}_{2}}_{23}} } \sin{\delta^{A^{}_{1}}} \;,
\nonumber
\\
\hspace{-0.2cm}&&\hspace{-0.2cm}
\phi^{A^{}_{2}}_{e} = \phi^{A^{}_{1}}_{\mu} + \delta^{A^{}_{2}}_{} \;,\;
\phi^{A^{}_{2}}_{\mu} = \phi^{A^{}_{1}}_{\tau} \;,\;
\phi^{A^{}_{2}}_{\tau} = \phi^{A^{}_{1}}_{e} - \delta^{A^{}_{1}}_{} \;,
\nonumber
\\
\hspace{-0.2cm}&&\hspace{-0.2cm}
\rho^{A^{}_{2}}_{} = \varphi^{}_{1} + \rho^{A^{}_{1}}_{} -
\delta^{A^{}_{2}}_{} \;,\; \sigma^{A^{}_{2}}_{} = \varphi^{}_{2}
+ \sigma^{A^{}_{1}}_{} -  \delta^{A^{}_{2}}_{} \; ;
\end{eqnarray}
and
\begin{eqnarray}
\hspace{-0.2cm}&&\hspace{-0.2cm}
\tan{\theta^{A^{}_{3}}_{23}} =
\sqrt{2} \tan{\theta^{A^{}_{1}}_{13}} \;,
\nonumber
\\
\hspace{-0.2cm}&&\hspace{-0.2cm}
\sin{\theta^{A^{}_{3}}_{13}} =
\frac{1}{\sqrt{2}} \cos{\theta^{A^{}_{1}}_{13}} \;,
\nonumber
\\
\hspace{-0.2cm}&&\hspace{-0.2cm}
\cos{2\theta^{A^{}_{3}}_{12}} = - \frac{ 1 - \sin^{2}{
\theta^{A^{}_{1}}_{13}} }{ 1 + \sin^{2}{\theta^{A^{}_{1}}_{13}} }
\cos{2\theta^{A^{}_{1}}_{12}} \;,
\nonumber
\\
\hspace{-0.2cm}&&\hspace{-0.2cm}
\sin{\delta^{A^{}_{3}}} = \frac{ \cos{\theta^{A^{}_{1}}_{12}}
\sin{\theta^{A^{}_{1}}_{12}} \cos^2{\theta^{A^{}_{1}}_{13}}
\sin{\theta^{A^{}_{1}}_{13}} }{ 2 \cos{\theta^{A^{}_{3}}_{12}}
\sin{\theta^{A^{}_{3}}_{12}} \cos^2{\theta^{A^{}_{3}}_{13}}
\sin{\theta^{A^{}_{3}}_{13}} \cos{\theta^{A^{}_{3}}_{23}}
\sin{\theta^{A^{}_{3}}_{23}} } \sin{\delta^{A^{}_{1}}} \;,
\nonumber
\\
\hspace{-0.2cm}&&\hspace{-0.2cm}
\phi^{A^{}_{3}}_{e} = \phi^{A^{}_{1}}_{\tau} + \delta^{A^{}_{3}}_{} \;,\;
\phi^{A^{}_{3}}_{\mu} = \phi^{A^{}_{1}}_{e} - \delta^{A^{}_{1}}_{} \;,\;
\phi^{A^{}_{3}}_{\tau} = \phi^{A^{}_{1}}_{\mu} \;,
\nonumber
\\
\hspace{-0.2cm}&&\hspace{-0.2cm}
\rho^{A^{}_{3}}_{} =\pi -  \varphi^{}_{1} + \rho^{A^{}_{1}}_{} -
\delta^{A^{}_{3}}_{} \;,\; \sigma^{A^{}_{3}}_{} = \pi - \varphi^{}_{2} +
\sigma^{A^{}_{1}}_{} -  \delta^{A^{}_{3}}_{} \;,
\end{eqnarray}
where $\sin{\varphi^{}_{1}} = \mp
\cos{\theta^{A^{}_{1}}_{12}} \sin{\theta^{A^{}_{1}}_{13}}/
\sqrt{ 1- \cos^2{\theta^{A^{}_{1}}_{12}} \cos^2{\theta^{A^{}_{1}}_{13}}}$,
$\cos{\varphi^{}_{1}} = - \sin{\theta^{A^{}_{1}}_{12}}/
\sqrt{ 1- \cos^2{\theta^{A^{}_{1}}_{12}} \cos^2{\theta^{A^{}_{1}}_{13}}}$,
$\sin{\varphi^{}_{2}} = \mp \sin{\theta^{A^{}_{1}}_{12}}
\sin{\theta^{A^{}_{1}}_{13}}/\sqrt{ 1-\sin^2{\theta^{A^{}_{1}}_{12}}
\cos^2{\theta^{A^{}_{1}}_{13}}}$
and $\cos{\varphi^{}_{2}} = \cos{\theta^{A^{}_{1}}_{12}}/
\sqrt{ 1- \sin^2{\theta^{A^{}_{1}}_{12}} \cos^2{\theta^{A^{}_{1}}_{13}}}$
with the ``$\mp$" signs corresponding to $\delta^{A^{}_1} = \pm \pi/2$.
The analytical results of three neutrino masses in these two categories
are formally the same as those given in Eq. (3.3), but of course the
relevant flavor mixing parameters need to be substituted with the ones
obtained in Eq. (3.5) or (3.6). It is obvious that none of the flavor
mixing angles and CP-violating phases in categories $A^{}_{2}$ and
$A^{}_{3}$ take special values, and this simply means that the standard
parametrization of $V$ is not the best choice for these two cases. One
may therefore consider to choose another parametrization of $V$ which
can automatically reveal the $S^{}_3$ reflection symmetry hidden
in $M^{}_\nu$ in categories $A^{}_2$ and $A^{}_3$.

\subsection{Categories $B^{}_i$}

In category $B^{}_{1}$ the structure of $M^{}_\nu$ possesses the
$\mu$-$\tau$ permutation symmetry
\footnote{That is, the light Majorana neutrino mass term is invariant under the
permutation transformations $\nu^{}_e \leftrightarrow \nu^{}_e$ and
$\nu^{}_\mu \leftrightarrow \nu^{}_\tau$, and thus the structure
of $M^{}_\nu$ gets constrained.},
which naturally predicts
$\theta^{}_{13} =0$ and $\theta^{}_{23} = \pi/4$ in the standard
parametrization of $V$ \cite{PS,Ma2001,Lam,Grimus}. The whole
pattern of $V$ in this case is found to be
\begin{eqnarray}
V^{B^{}_{1}}_{} =
\left(\begin{matrix}
\displaystyle \mp \frac{  \left( - a^{\prime}_{\rm r} + e^{\prime}_{\rm r}
+d^{\prime}_{\rm r} + \sqrt{\Delta}\right)}
{ \sqrt{8 b^{\prime 2}_{\rm r} + \left( - a^{\prime}_{\rm r}
+ e^{\prime}_{\rm r} + d^{\prime}_{\rm r} + \sqrt{\Delta}
\right)^{2}_{}} } & \displaystyle \pm \frac{ \left( a^{\prime}_{\rm r}
- e^{\prime}_{\rm r} - d^{\prime}_{\rm r} + \sqrt{\Delta}
\right)}{ \sqrt{8 b^{\prime 2}_{\rm r}
+ \left( a^{\prime}_{\rm r} - e^{\prime}_{\rm r}
- d^{\prime}_{\rm r} + \sqrt{\Delta} \right)^{2}_{}} } & 0 \cr
\displaystyle \frac{2 | b^{\prime}_{\rm r}| }{ \sqrt{8 b^{\prime 2}_{\rm r}
+ \left( - a^{\prime}_{\rm r} + e^{\prime}_{\rm r}
+ d^{\prime}_{\rm r} + \sqrt{\Delta} \right)^{2}_{}} } &
\displaystyle \frac{2 | b^{\prime}_{\rm r}| }{ \sqrt{8 b^{\prime 2}_{\rm r}
+ \left(  a^{\prime}_{\rm r} - e^{\prime}_{\rm r}
- d^{\prime}_{\rm r} + \sqrt{\Delta} \right)^{2}_{}} } &
\displaystyle -\frac{1}{\sqrt{2}} \cr
\displaystyle \frac{2 | b^{\prime}_{\rm r}| }
{ \sqrt{8 b^{\prime 2}_{\rm r} + \left( - a^{\prime}_{\rm r}
+ e^{\prime}_{\rm r} + d^{\prime}_{\rm r} + \sqrt{\Delta}
\right)^{2}_{}} } & \displaystyle \frac{2 | b^{\prime}_{\rm r}| }{
\sqrt{8 b^{\prime 2}_{\rm r} + \left(  a^{\prime}_{\rm r}
- e^{\prime}_{\rm r} - d^{\prime}_{\rm r} + \sqrt{\Delta}
\right)^{2}_{}} } & \displaystyle \frac{1}{\sqrt{2}}
\end{matrix}\right) \;
\end{eqnarray}
with two sign options corresponding to the positive or negative sign of
$b^{\prime}_{\rm r}$. It is easy to see that $\left| V^{B^{}_{1}}_{\mu i}
\right| = \left| V^{B^{}_{1}}_{\tau i} \right|$ holds (for $i=1,2,3$), a
clear reflection of the $\mu$-$\tau$ permutation symmetry. Note,
however, that the structure of $M^{B^{}_{1}}_{\nu}$ shown in Table 2.1
is just a particular example which respects the $\mu$-$\tau$
permutation symmetry but has no complex elements.
A general form of $M^{}_\nu$ in the $\mu$-$\tau$ permutation
symmetry limit must have a structure similar to $M^{B^{}_{1}}_{\nu}$,
but it should contain some complex elements \cite{Zhao2016}.

Given the texture of $M^{B^{}_{1}}_{\nu}$ in Table 2.1 and the pattern
of $V^{B^{}_1}$ in Eq. (3.7), it is straightforward to calculate
the neutrino masses and flavor mixing parameters. We obtain
\begin{eqnarray}
m^{}_{1} \hspace{-0.2cm}&=&\hspace{-0.2cm} \frac{1}{2}
\left( a^{\prime}_{\rm r} + e^{\prime}_{\rm r} +d^{\prime}_{\rm r}
- \sqrt{\Delta} \right) \;,
\nonumber
\\
m^{}_{2} \hspace{-0.2cm}&=&\hspace{-0.2cm} \frac{1}{2}
\left( a^{\prime}_{\rm r} + e^{\prime}_{\rm r} +d^{\prime}_{\rm r}
+ \sqrt{\Delta} \right) \;,
\nonumber
\\
m^{}_{3} \hspace{-0.2cm}&=&\hspace{-0.2cm}
e^{\prime}_{\rm r} - d^{\prime}_{\rm r}  \; ;
\end{eqnarray}
and
\begin{eqnarray}
\hspace{-0.2cm}&&\hspace{-0.2cm}
\theta^{B^{}_{1}}_{23} = \frac{\pi}{4} \;, \quad
\theta^{B^{}_{1}}_{13} = 0 \;, \quad
\theta^{B^{}_{1}}_{12} = \arccos{\left( \frac{2\sqrt{2}
|b^{\prime}_{\rm r}|}{\sqrt{8 b^{\prime 2}_{\rm r} +
\left( a^{\prime}_{\rm r} - e^{\prime}_{\rm r}
- d^{\prime}_{\rm r} + \sqrt{\Delta} \right)^{2}_{} }} \right)} \;,
\nonumber
\\
\hspace{-0.2cm}&&\hspace{-0.2cm}
\delta^{B^{}_{1}}_{} \in \left[0, 2\pi\right) \;, \quad
\rho^{B^{}_{1}}_{} = 0 \;, \quad \sigma^{B^{}_{1}}_{}  = \pi  \;, \quad
\phi^{B^{}_{1}}_{e} = 0\;{\rm or}\; \pi \;, \quad
\phi^{B^{}_{1}}_{\mu} = \pi \;, \quad \phi^{B^{}_{1}}_{\tau} = 0 \;,
\hspace{1cm}
\end{eqnarray}
where $\Delta = 8 b^{\prime 2}_{\rm r} + \left( - a^{\prime}_{\rm r}
+ e^{\prime}_{\rm r} + d^{\prime}_{\rm r} \right)^{2}_{} $. Note that
$m^{}_{1}$, $m^{}_{2}$ or $ m^{}_{3}$ in Eq. (3.8) may be negative,
but a minus sign can always be absorbed into three unphysical phases
and two Majorana phases.

Analogous to categories $A^{}_1$, $A^{}_2$ and $A^{}_3$,
the neutrino masses and flavor mixing parameters of $B^{}_1$,
$B^{}_2$ and $B^{}_3$ are also correlated with one another via
Eq. (2.16). In fact, the three neutrino masses for both categories
$B^{}_2$ and $B^{}_3$ are the same as those given by Eq. (3.8),
and the flavor mixing parameters in these two cases are found to be
\begin{eqnarray}
\hspace{-0.2cm}&&\hspace{-0.2cm}
\theta^{B^{}_{2}}_{23} = \frac{\pi}{2} \;, \quad
\theta^{B^{}_{2}}_{13} = \frac{\pi}{4} \;, \quad
\theta^{B^{}_{2}}_{12} = \arccos{\left( \frac{2\sqrt{2}
|b^{\prime}_{\rm r}|}{\sqrt{8 b^{\prime 2}_{\rm r} +
\left( - a^{\prime}_{\rm r} + e^{\prime}_{\rm r} +
d^{\prime}_{\rm r} + \sqrt{\Delta} \right)^{2}_{} }}
\right)} \;,
\nonumber
\\
\hspace{-0.2cm}&&\hspace{-0.2cm}
\rho^{B^{}_{2}}_{} =  \sigma^{B^{}_{2}}_{}
= \pi - \delta^{B^{}_{2}}_{} = - \phi^{B^{}_{2}}_{e}
=  -  \phi^{B^{}_{2}}_{\tau} \;{\rm or} \; \pi - \phi^{B^{}_{2}}_{\tau}  \;, \quad
\phi^{B^{}_{2}}_{\mu} = 0 \; ;
\end{eqnarray}
and
\begin{eqnarray}
\hspace{-0.2cm}&&\hspace{-0.2cm}\theta^{B^{}_{3}}_{23} = 0 \;,\;
\theta^{B^{}_{3}}_{13} = \frac{\pi}{4} \;, \quad
\theta^{B^{}_{3}}_{12} = \arccos{\left( \frac{2\sqrt{2}
|b^{\prime}_{\rm r}|}{\sqrt{8 b^{\prime 2}_{\rm r} +
\left( - a^{\prime}_{\rm r} + e^{\prime}_{\rm r} +
d^{\prime}_{\rm r} + \sqrt{\Delta} \right)^{2}_{} }}
\right)} \;,
\nonumber
\\
\hspace{-0.2cm}&&\hspace{-0.2cm}
\rho^{B^{}_{3}}_{} =  \sigma^{B^{}_{3}}_{}
= - \delta^{B^{}_{3}}_{} = - \phi^{B^{}_{3}}_{e}
= -  \phi^{B^{}_{3}}_{\mu} \; {\rm or} \; -\pi - \phi^{B^{}_{3}}_{\mu} \;, \quad
\phi^{B^{}_{3}}_{\tau} = \pi \;,
\end{eqnarray}
respectively. Needless to say, these two possibilities are strongly
disfavored by current neutrino oscillation data \cite{BF2018}.

\subsection{Categories $C$ and $D$}

The structures of $M^{}_\nu$ in categories $C$ and $D$ are exactly the
same, and thus their phenomenological consequences are also the same.
In particular, they lead us to the well-known tri-bimaximal neutrino
mixing pattern \cite{TBM,Xing2002,He:2003rm},
\begin{eqnarray}
V= \left(\begin{matrix} \displaystyle -\frac{2}{\sqrt{6}} &
\displaystyle \frac{1}{\sqrt{3}}
& 0 \cr \displaystyle \frac{1}{\sqrt{6}} & \displaystyle \frac{1}{\sqrt{3}}
& \displaystyle \frac{1}{\sqrt{2}} \cr \displaystyle \frac{1}{\sqrt{6}}
& \displaystyle \frac{1}{\sqrt{3}} & \displaystyle -\frac{1}{\sqrt{2}}
\end{matrix}\right) \;.
\end{eqnarray}
To be explicit, the results for neutrino masses and flavor mixing parameters
are
\begin{eqnarray}
m^{}_{1} = a^{\prime}_{\rm r} - b^{\prime}_{\rm r} \;, \quad
m^{}_{2} = a^{\prime}_{\rm r} + 2b^{\prime}_{\rm r}\;, \quad
m^{}_{3} = a^{\prime}_{\rm r} - b^{\prime}_{\rm r}  \;,
\end{eqnarray}
and
\begin{eqnarray}
\hspace{-0.2cm}&&\hspace{-0.2cm}
\theta^{}_{23} = \frac{\pi}{4} \;, \quad
\theta^{}_{13} = 0 \;, \quad
\theta^{}_{12} = \arctan{\left( \frac{1}{\sqrt{2}} \right)} \;,
\nonumber
\\
\hspace{-0.2cm}&&\hspace{-0.2cm}
\delta \in \left[0, 2\pi\right) \;, \quad
\rho= \pi \;, \quad \sigma = 0  \;, \quad \phi^{}_{e} = 0  \;, \quad
\phi^{}_{\mu} = 0 \;, \quad \phi^{}_{\tau} = \pi \;.
\end{eqnarray}
So far many model-building exercises have been done in this connection
to introduce small perturbations to $M^{}_\nu$ and consequently small
corrections to $V$ in Eq. (3.12), so as to
arrive at a better fit of current experimental data
\cite{Jora2009gz,Jora2010at,Jora2012nw,Dev2011qy,Dev2012ns,Benaoum2013}.

\subsection{Categories $E^{}_{i}$}

In category $E^{}_{1}$ the PMNS matrix reads
\begin{eqnarray}
V^{E^{}_{1}}_{} = \left(\begin{matrix} \displaystyle \frac{2}{\sqrt{6}} &
\displaystyle - \frac{1}{\sqrt{3}} & 0 \cr
\displaystyle - \frac{1}{\sqrt{6}} & \displaystyle - \frac{1}{\sqrt{3}} &
\displaystyle \frac{{\rm i}}{\sqrt{2}} \cr \displaystyle - \frac{1}{\sqrt{6}}
& \displaystyle - \frac{1}{\sqrt{3}} & \displaystyle
-\frac{{\rm i}}{\sqrt{2}}  \end{matrix}\right) \; ,
\end{eqnarray}
which is also the tri-bimaximal flavor mixing pattern with some trivial phases.
A straightforward calculation allows us to obtain the neutrino masses for
category $E^{}_1$:
\begin{eqnarray}
m^{}_{1} = a^{\prime}_{\rm r} - b^{\prime}_{\rm r} \;, \quad
m^{}_{2} = a^{\prime}_{\rm r} + 2b^{\prime}_{\rm r}\;, \quad
m^{}_{3} = a^{\prime}_{\rm r} - b^{\prime}_{\rm r}  \; ,
\end{eqnarray}
and the same result is true for categories $E^{}_2$ and $E^{}_3$.
The corresponding flavor mixing parameters are found to be
\begin{eqnarray}
\hspace{-0.2cm}&&\hspace{-0.2cm}
\theta^{E^{}_{1}}_{23} = \frac{\pi}{4} \;, \quad
\theta^{E^{}_{1}}_{13} = 0 \;, \quad
\theta^{E^{}_{1}}_{12} = \arctan{\left(\frac{1}{ {\sqrt{2}}} \right)} \;,
\nonumber
\\
\hspace{-0.2cm}&&\hspace{-0.2cm}
\delta^{E^{}_{1}}_{} \in \left[0, 2\pi \right) \;, \quad
\rho^{E^{}_{1}}_{} = \frac{3\pi}{2} \;, \quad
\sigma^{E^{}_{1}}_{} = \frac{\pi}{2} \;,
\nonumber
\\
\hspace{-0.2cm}&&\hspace{-0.2cm}
\phi^{E^{}_{1}}_{e} = \frac{\pi}{2} \;, \quad
\phi^{E^{}_{1}}_{\mu} = \frac{\pi}{2} \;, \quad
\phi^{E^{}_{1}}_{\tau} = \frac{3\pi}{2} \;
\end{eqnarray}
in category $E^{}_1$;
\begin{eqnarray}
\hspace{-0.2cm}&&\hspace{-0.2cm}
\theta^{E^{}_{2}}_{23} = \frac{\pi}{2} \;, \quad
\theta^{E^{}_{2}}_{13} = \frac{\pi}{4} \;, \quad
\theta^{E^{}_{2}}_{12} = \arctan{\left( {\sqrt{2}} \right)} \;,
\nonumber
\\
\hspace{-0.2cm}&&\hspace{-0.2cm}
\rho^{E^{}_{2}}_{} = \sigma^{E^{}_{2}}_{} = \frac{\pi}{2}
- \delta^{E^{}_{2}}_{} = \pi - \phi^{E^{}_{2}}_{e} =
- \phi^{E^{}_{2}}_{\tau} \;, \quad \phi^{E^{}_{2}}_{\mu} =
\frac{3\pi}{2} \;
\end{eqnarray}
in category $E^{}_2$; and
\begin{eqnarray}
\hspace{-0.2cm}&&\hspace{-0.2cm}
\theta^{E^{}_{3}}_{23} = 0 \;, \quad
\theta^{E^{}_{3}}_{13} = \frac{\pi}{4} \;, \quad
\theta^{E^{}_{3}}_{12} = \arctan{\left( {\sqrt{2}} \right)} \;,
\nonumber
\\
\hspace{-0.2cm}&&\hspace{-0.2cm}
\rho^{E^{}_{3}}_{} = \sigma^{E^{}_{3}}_{} = \frac{3\pi}{2}
- \delta^{E^{}_{3}}_{} = \pi - \phi^{E^{}_{3}}_{e} = \pi
- \phi^{E^{}_{3}}_{\mu} \;, \quad \phi^{E^{}_{3}}_{\tau} =
\frac{\pi}{2} \;
\end{eqnarray}
in category $E^{}_3$. One can see that the latter two cases are
strongly disfavored by current neutrino oscillation data \cite{BF2018}.

\subsection{Category $F$}

In this category of $M^{}_\nu$ the PMNS matrix is given by
\begin{eqnarray}
V = \left(\begin{matrix}
\displaystyle {\rm i} \frac{ - a^{\prime}_{\rm r} + b^{\prime}_{\rm r} + \lambda }
{ a^{\prime}_{\rm r} - e^{\prime}_{\rm r}} \sqrt{
\frac{ -2b^{\prime}_{\rm r} + a^{\prime}_{\rm r} +
e^{\prime}_{\rm r} + 2\lambda }{6\lambda}}  & \displaystyle
\frac{ - a^{\prime}_{\rm r} + b^{\prime}_{\rm r} - \lambda }
{ a^{\prime}_{\rm r} - e^{\prime}_{\rm r}} \sqrt{
\frac{ 2b^{\prime}_{\rm r} - a^{\prime}_{\rm r}
- e^{\prime}_{\rm r} + 2\lambda }{6\lambda}} & \displaystyle \frac{1}{\sqrt{3}}
\cr \displaystyle {\rm i} \frac{ - b^{\prime}_{\rm r} + e^{\prime}_{\rm r} - \lambda }
{ a^{\prime}_{\rm r} - e^{\prime}_{\rm r}} \sqrt{
\frac{ -2b^{\prime}_{\rm r} + a^{\prime}_{\rm r} +
e^{\prime}_{\rm r} + 2\lambda }{6\lambda}}  & \displaystyle
\frac{ - b^{\prime}_{\rm r} + e^{\prime}_{\rm r} + \lambda }
{ a^{\prime}_{\rm r} - e^{\prime}_{\rm r}} \sqrt{
\frac{ 2b^{\prime}_{\rm r} - a^{\prime}_{\rm r} -
e^{\prime}_{\rm r} + 2\lambda }{6\lambda}} & \displaystyle \frac{1}{\sqrt{3}}
\cr \displaystyle {\rm i} \sqrt{\frac{ -2b^{\prime}_{\rm r} + a^{\prime}_{\rm r} +
e^{\prime}_{\rm r} + 2\lambda }{6\lambda}}  & \displaystyle \sqrt{
\frac{ 2b^{\prime}_{\rm r} - a^{\prime}_{\rm r}
- e^{\prime}_{\rm r} + 2\lambda }{6\lambda}} & \displaystyle \frac{1}{\sqrt{3}}
\cr \end{matrix}\right) \; \hspace{0.3cm}
\end{eqnarray}
with $\lambda = \sqrt{a^{\prime2}_{\rm r} + b^{\prime2}_{\rm r}
+ e^{\prime2}_{\rm r} - a^{\prime}_{\rm r}b^{\prime}_{\rm r}
- a^{\prime}_{\rm r}e^{\prime}_{\rm r} - b^{\prime}_{\rm r}
e^{\prime}_{\rm r}}$. The masses of three light neutrinos are
\begin{eqnarray}
m^{}_{1} = m^{}_{2} = \lambda \;, \quad
m^{}_{3} = a^{\prime}_{\rm r} +  b^{\prime}_{\rm r} + e^{\prime}_{\rm r} \;,
\end{eqnarray}
and the corresponding flavor mixing parameters are found to be
\begin{eqnarray}
\hspace{-0.2cm}&&\hspace{-0.2cm}
\theta^{}_{23} = \frac{\pi}{4} \;, \quad
\theta^{}_{13} = \arccos{\left( \frac{2}{\sqrt{6}} \right)} \;,
\nonumber
\\
\hspace{-0.2cm}&&\hspace{-0.2cm}
\theta^{}_{12} = \arctan{\left(\frac{a - b + \lambda}{a-b-\lambda}
\sqrt{\frac{ 2b^{\prime}_{\rm r} - a^{\prime}_{\rm r} - e^{\prime}_{\rm r}
+ 2\lambda }{-2b^{\prime}_{\rm r} + a^{\prime}_{\rm r} + e^{\prime}_{\rm r}
+ 2\lambda}} \right) }\;,
\nonumber
\\
\hspace{-0.2cm}&&\hspace{-0.2cm}
\delta = 0 \; {\rm or} \; \pi \;,
\quad \rho = \frac{\pi}{2} - \delta \;,
\quad \sigma = \phi^{}_{e} = \delta \;, \quad
\phi^{}_{\mu} = \phi^{}_{\tau} = 0 \; .
\end{eqnarray}
This case turns out to be strongly disfavored by current experimental data.

\subsection{Categories $H^{}_{i}$}

For categories $H^{}_1$, $H^{}_2$ and $H^{}_3$, the corresponding
neutrino masses and flavor mixing parameters can easily be obtained
from categories $B^{}_1$, $B^{}_2$ and $B^{}_3$
by taking $d^{\prime}_{\rm r} = e^{\prime}_{\rm r}$.
In this way one is left with $m^{}_{3} = 0$,
corresponding to the inverted neutrino mass hierarchy.
Here let us focus on the normal neutrino mass hierarchy. The PMNS matrix
is found to be
\begin{eqnarray}
V^{H^{}_{1}}_{} = \left(\begin{matrix}
0 & \displaystyle \mp \frac{ \left( - a^{\prime}_{\rm r} + 2e^{\prime}_{\rm r} +
\sqrt{\Delta^{\prime}_{}} \right)}{
\sqrt{8 b^{\prime 2}_{\rm r} + \left( - a^{\prime}_{\rm r} +
2e^{\prime}_{\rm r} + \sqrt{\Delta^{\prime}_{}} \right)^{2}_{}} }
& \displaystyle \pm \frac{ \left( a^{\prime}_{\rm r} - 2e^{\prime}_{\rm r} +
\sqrt{\Delta^{\prime}_{}} \right)}{
\sqrt{8 b^{\prime 2}_{\rm r} + \left( a^{\prime}_{\rm r} -
2e^{\prime}_{\rm r} + \sqrt{\Delta^{\prime}_{}} \right)^{2}_{}} }
\cr  \displaystyle -\frac{1}{\sqrt{2}} & \displaystyle
\frac{2 | b^{\prime}_{\rm r}| }
{ \sqrt{8 b^{\prime 2}_{\rm r} + \left( - a^{\prime}_{\rm r} +
2e^{\prime}_{\rm r} + \sqrt{\Delta^{\prime}_{}} \right)^{2}_{}} }
& \displaystyle \frac{2 | b^{\prime}_{\rm r}| }{ \sqrt{8 b^{\prime 2}_{\rm r} +
\left(  a^{\prime}_{\rm r} - 2e^{\prime}_{\rm r} +
\sqrt{\Delta^{\prime}_{}} \right)^{2}_{}} } \cr  \displaystyle \frac{1}{\sqrt{2}}
& \displaystyle \frac{2 | b^{\prime}_{\rm r}| }{ \sqrt{8 b^{\prime 2}_{\rm r} +
\left( - a^{\prime}_{\rm r} + 2e^{\prime}_{\rm r} +
\sqrt{\Delta^{\prime}_{}} \right)^{2}_{}} } & \displaystyle \frac{2
| b^{\prime}_{\rm r}| }{ \sqrt{8 b^{\prime 2}_{\rm r} +
\left(  a^{\prime}_{\rm r} - 2e^{\prime}_{\rm r} +
\sqrt{\Delta^{\prime}_{}} \right)^{2}_{}} }  \end{matrix}\right) \; ,
\hspace{0.5cm}
\end{eqnarray}
and the neutrino masses are given by
\begin{eqnarray}
m^{}_{1} \hspace{-0.2cm}&=&\hspace{-0.2cm} 0 \;,
\nonumber
\\
m^{}_{2} \hspace{-0.2cm}&=&\hspace{-0.2cm}
\frac{1}{2} \left( a^{\prime}_{\rm r} + 2e^{\prime}_{\rm r} -
\sqrt{\Delta^{\prime}_{}} \right) \;,
\nonumber
\\
m^{}_{3} \hspace{-0.2cm}&=&\hspace{-0.2cm}
\frac{1}{2} \left( a^{\prime}_{\rm r} + 2e^{\prime}_{\rm r} +
\sqrt{\Delta^{\prime}_{}} \right) \;,
\end{eqnarray}
where $\Delta ^{\prime}_{} = 8 b^{\prime 2}_{\rm r} +
\left( - a^{\prime}_{\rm r} + 2e^{\prime}_{\rm r}
\right)^{2}_{}$. The results of $m^{}_i$ in Eq. (3.24) are also valid
for categories $H^{}_2$ and $H^{}_3$. To be explicit, the flavor mixing
parameters in these three cases are obtained below:
\begin{eqnarray}
\hspace{-0.2cm}&&\hspace{-0.2cm}
\theta^{H^{}_{1}}_{23} = \frac{\pi}{4} \;, \quad
\theta^{H^{}_{1}}_{12} = \frac{\pi}{2} \;, \quad
\theta^{H^{}_{1}}_{13} = \arccos{\left( \frac{2\sqrt{2}
|b^{\prime}_{\rm r}|}{\sqrt{8 b^{\prime 2}_{\rm r} +
\left( a^{\prime}_{\rm r} - 2e^{\prime}_{\rm r} +
\sqrt{\Delta^{\prime}_{}} \right)^{2}_{} }} \right)} \;,
\nonumber
\\
\hspace{-0.2cm}&&\hspace{-0.2cm}
\delta^{H^{}_{1}}_{} = \pi - \sigma^{H^{}_{1}}_{} =
\phi^{H^{}_{1}}_{e} \; {\rm or } \; \phi^{H^{}_{1}}_{e} + \pi  \;,
\quad \rho^{H^{}_{1}}_{} =
\phi^{H^{}_{1}}_{\mu} = \phi^{H^{}_{1}}_{\tau} = 0 \;
\end{eqnarray}
for category $H^{}_1$;
\begin{eqnarray}
\hspace{-0.2cm}&&\hspace{-0.2cm}
\theta^{H^{}_{2}}_{23} = \arctan{\left( \frac{2|b^{\prime}_{\rm r}|}
{a^{\prime}_{\rm r} - 2e^{\prime}_{\rm r} + \sqrt{\Delta^{\prime}_{}}}
\right)} \;,
\nonumber
\\
\hspace{-0.2cm}&&\hspace{-0.2cm}
\theta^{H^{}_{2}}_{12} = \arctan{\left( \frac{2\sqrt{2}
|b^{\prime}_{\rm r}|}{\sqrt{8 b^{\prime 2}_{\rm r} +
\left( - a^{\prime}_{\rm r} + 2e^{\prime}_{\rm r} +
\sqrt{\Delta^{\prime}_{}} \right)^{2}_{} }} \right)} \;,
\nonumber
\\
\hspace{-0.2cm}&&\hspace{-0.2cm}
\theta^{H^{}_{2}}_{13} = \arcsin{\left( \frac{ 2|b^{\prime}_{\rm r}|}
{\sqrt{8 b^{\prime 2}_{\rm r} + \left( a^{\prime}_{\rm r}
- 2e^{\prime}_{\rm r} + \sqrt{\Delta^{\prime}_{}} \right)^{2}_{} }}
\right)} \;,
\nonumber
\\
\hspace{-0.2cm}&&\hspace{-0.2cm}
\delta^{H^{}_{2}}_{} = \sigma^{ H^{}_{2} }_{} = \phi^{H^{}_{2}}_{e}
= \phi^{H^{}_{2}}_{\mu} = 0 \;, \quad
\rho^{H^{}_{2}}_{} = \pi \;, \quad
\phi^{H^{}_{2}}_{\tau} = 0 \;{\rm or}\; -\pi \;
\end{eqnarray}
for category $H^{}_2$; and
\begin{eqnarray}
\hspace{-0.2cm}&&\hspace{-0.2cm}
\theta^{H^{}_{3}}_{23} = \arctan{\left( \frac{a^{\prime}_{\rm r} -
2e^{\prime}_{\rm r} + \sqrt{\Delta^{\prime}_{}}}{2|b^{\prime}_{\rm r}|}
\right)} \;,
\nonumber
\\
\hspace{-0.2cm}&&\hspace{-0.2cm}
\theta^{H^{}_{3}}_{12} = \arctan{\left( \frac{2\sqrt{2}
|b^{\prime}_{\rm r}|}{\sqrt{8 b^{\prime 2}_{\rm r} +
\left( - a^{\prime}_{\rm r} + 2e^{\prime}_{\rm r} +
\sqrt{\Delta^{\prime}_{}} \right)^{2}_{} }} \right)} \;,
\nonumber
\\
\hspace{-0.2cm}&&\hspace{-0.2cm}
\theta^{H^{}_{3}}_{13} = \arcsin{\left( \frac{ 2|b^{\prime}_{\rm r}|}
{\sqrt{8 b^{\prime 2}_{\rm r} + \left( a^{\prime}_{\rm r}
- 2e^{\prime}_{\rm r} + \sqrt{\Delta^{\prime}_{}} \right)^{2}_{} }}
\right)} \;,
\nonumber
\\
\hspace{-0.2cm}&&\hspace{-0.2cm}
\delta^{H^{}_{3}}_{} = \rho^{H^{}_{3}}_{} = \sigma^{ H^{}_{3} }_{}
= \phi^{H^{}_{3}}_{e} = \pi  \;, \quad
\phi^{H^{}_{3}}_{\tau}  = 0 \;, \quad
\phi^{H^{}_{3}}_{\mu} = 0 \;{\rm or}\; -\pi \;
\end{eqnarray}
for category $H^{}_3$, respectively.
Note that the flavor mixing angles in the latter two cases satisfy the
relations $\tan{ \theta^{ H^{}_{2} }_{12} } \tan{ \theta^{H^{}_{2}}_{23}}
= \sin{\theta^{H^{}_{2}}_{13}}$ and $\tan{ \theta^{ H^{}_{3} }_{12} }
= \tan{ \theta^{H^{}_{3}}_{23}} \sin{\theta^{H^{}_{3}}_{13}}$.

\subsection{Categories $I^{}_{i}$, $J$ and $K$}

In these five categories the textures of $M^{}_\nu$ are all democratic, and
thus the corresponding PMNS matrix is of the form
\begin{eqnarray}
V = \left(\begin{matrix}
\displaystyle \frac{1}{\sqrt{2}} & \displaystyle \frac{1}{\sqrt{6}} &
\displaystyle \frac{1}{\sqrt{3}} \cr
\displaystyle -\frac{1}{\sqrt{2}} & \displaystyle \frac{1}{\sqrt{6}} &
\displaystyle \frac{1}{\sqrt{3}} \cr
0 & \displaystyle -\frac{2}{\sqrt{6}} & \displaystyle
\frac{1}{\sqrt{3}} \end{matrix}\right)\;,
\end{eqnarray}
corresponding to a special neutrino mass spectrum with
$m^{}_{1} = m^{}_{2} = 0$ and $m^{}_{3} = 3 a^{\prime}_{\rm r}$.
To be explicit, the pattern of $V$ in Eq. (3.28) leads us to
\begin{eqnarray}
\hspace{-0.2cm}&&\hspace{-0.2cm}
\theta^{}_{23} = \frac{\pi}{4} \;, \quad
\theta^{}_{13} = \arccos{ \left( \frac{2}{\sqrt{6}} \right)} \;, \quad
\theta^{}_{12} = \frac{\pi}{6} \;,
\nonumber
\\
\hspace{-0.2cm}&&\hspace{-0.2cm}
\delta = \rho = \sigma = \phi^{}_{e}
= \phi^{}_{\mu} = \phi^{}_{\tau} = 0 \; ,
\end{eqnarray}
which are strongly disfavored by current neutrino oscillation data.

\setcounter{footnote}{0}
\setcounter{table}{0}
\setcounter{equation}{0}
\setcounter{figure}{0}

\section{Leptogenesis in the $S^{}_3$ symmetry limit}

Now we examine whether the leptogenesis mechanism \cite{FY,Davidson},
which can provide a natural way to account for the observed
matter-antimatter asymmetry of the Universe \cite{Sakharov}, works or
not in the $S^{}_3$ reflection symmetry limit under discussion.
According to this mechanism, the lepton-number-violating, CP-violating and
out-of-equilibrium decays of heavy Majorana neutrinos
$N^{}_{i}$ may result in a lepton-antilepton asymmetry in the early Universe,
and the latter can subsequently be converted to the wanted baryon-antibaryon
asymmetry through the $B$-$L$ conserving sphaleron process
\cite{Klinkhamer,Buchmuller}.
Here what we are concerned with are the CP-violating asymmetries
between the decay modes $N^{}_{i} \to \ell^{}_{\alpha} + H$
and their CP-conjugate processes $N^{}_{i} \to \bar{\ell}^{}_{\alpha} +
\overline{H} $, usually denoted as $\epsilon^{}_{i\alpha}$
(for $\alpha = e, \mu, \tau$ and $i= 1, 2, 3$), because they will
finally determine the strength of baryogenesis via leptogenesis.
Assuming that the masses of three heavy Majorana neutrinos are hierarchical
(i.e., $M^{}_{1} \ll M^{}_{2} < M^{}_{3}$), it has been shown that only the
CP-violating asymmetries $\epsilon^{}_{1\alpha}$ survive and contribute
to the lepton-antilepton asymmetry. In this case the expression of
$\epsilon^{}_{1\alpha}$ is given by \cite{Morozumi,XZ2011}
\begin{eqnarray}
\epsilon^{}_{1\alpha} \hspace{-0.2cm}&=&\hspace{-0.2cm}  \frac{\Gamma
\left( N^{}_{i} \rightarrow \ell^{}_{\alpha} + H \right) - \Gamma
\left( N^{}_{i} \rightarrow \bar{\ell}^{}_{\alpha} + \overline{H} \right) }
{\sum\limits_\alpha \left[ \Gamma \left( N^{}_{i} \rightarrow \ell^{}_{\alpha}
+ H \right) + \Gamma \left( N^{}_{i} \rightarrow \bar{\ell}^{}_{\alpha} +
\overline{H} \right) \right]}
\nonumber
\\
\hspace{-0.2cm}&=&\hspace{-0.2cm} \frac{1}{8\pi v^2
\left( \widetilde{M}^{\dagger}_{\rm D} \widetilde{M}^{}_{\rm D}
\right)^{}_{11} } \sum^{}_{j \neq 1} \left\{ {\rm Im} \left[ \left(
\widetilde{M}^{\ast}_{\rm D} \right)^{}_{\alpha 1}
\left( \widetilde{M}^{}_{\rm D} \right)^{}_{\alpha j} \left(
\widetilde{M}^{\dagger}_{\rm D} \widetilde{M}^{}_{\rm D}\right)^{}_{1j}
\right] \times \mathcal{F} \left( \frac{M^{2}_{j}}{M^{2}_{1}} \right)
\right.
\nonumber
\\
\hspace{-0.2cm}& &\hspace{-0.2cm} + \left. {\rm Im} \left[ \left(
\widetilde{M}^{\ast}_{\rm D} \right)^{}_{\alpha 1} \left(
\widetilde{M}^{}_{\rm D} \right)^{}_{\alpha j} \left(
\widetilde{M}^{\dagger}_{\rm D} \widetilde{M}^{}_{\rm D}\right)^{\ast}_{1j}
\right] \times \mathcal{G} \left( \frac{M^{2}_{j}}{M^{2}_{1}} \right)
\right\} \;,
\end{eqnarray}
where $\widetilde{M}^{}_{\rm D} = M^{}_{\rm D} U^{\ast}_{\rm R} $
with $U^{}_{\rm R}$ being the unitary matrix used to diagonalize
$M^{}_{\rm R}$ (i.e., $U^{\dagger}_{\rm R} M^{}_{\rm R} U^{\ast}_{\rm R} =
\widehat{M}^{}_N = {\rm Diag} \left\{ M^{}_{1}, M^{}_{2}, M^{}_{3}
\right\}$), $M^{}_{i}$ (for $i=1, 2, 3$) stands for the mass of the
heavy Majorana neutrino $N^{}_{i}$,
$v \equiv \langle H \rangle \approx 174 {\rm GeV}$ is the vacuum expectation
value of the Higgs field, ${\cal F} \left( x \right) $ and
${\cal G} \left( x \right) $ are the loop functions defined as
${\cal F}\left(x\right) = \sqrt{x} \left\{ (2-x)/(1-x) + (1+x)
\ln{[x/(1+x)]} \right\}$ and ${\cal G} (x) = 1/(1-x)$, respectively.
If all the interactions in the period of leptogenesis are blind to lepton
flavors, then only the total CP-violating asymmetry $\epsilon^{}_{1}$
is relevant,
\begin{eqnarray}
\epsilon^{}_{1} = \sum^{}_{\alpha} \epsilon^{}_{1 \alpha} =
\frac{1}{8\pi v^2 \left( \widetilde{M}^{\dagger}_{\rm D}
\widetilde{M}^{}_{\rm D}\right)^{}_{11} } \sum^{}_{j \neq 1} {\rm Im}
\left[ \left( \widetilde{M}^{\dagger}_{\rm D} \widetilde{M}^{}_{\rm D}
\right)^{2}_{1j} \right] \times \mathcal{F}
\left( \frac{M^{2}_{j}}{M^{2}_{1}} \right) \;.
\end{eqnarray}
In the literature $\epsilon^{}_1$ and $\epsilon^{}_{1\alpha}$ correspond
to the so-called ``unflavored" and ``flavored" leptogenesis. In view of
Table 2.1, it is obvious that only categories $A^{}_i$ (for $i=1,2,3$)
and $C$ are likely to lead us to nonzero $\epsilon{}_{1}$ or
$\epsilon^{}_{1 \alpha}$, and thus we are going to calculate
them in the following.

\subsection{Unflavored leptogenesis}

\subsubsection{Categories $A^{}_{i}$}

Let us first consider category $A^{}_{1}$, and then turn to categories
$A^{}_{2}$ and $A^{}_{3}$. In category $A^{}_{1}$ the three mass matrices
all respect the $\mu$-$\tau$ reflection symmetry, so it is easy to
calculate their corresponding mass eigenvalues and flavor mixing parameters.
To be more specific, the unitary matrix $U^{}_{\rm R}$ used to diagonalize
$M^{}_{\rm R}$ can be decomposed as
$ U^{}_{\rm R} = P^{\rm R}_{1} \widetilde{U}^{\rm R}_{} P^{\rm R}_{2} $, where
$\widetilde{U}^{\rm R}_{} = O^{}_{23} \tilde{O}^{}_{13} O^{}_{12}$ is of
the same form as the standard parameterization shown in Eq. (3.1), and
$P^{\rm R}_{1} = {\rm Diag}
\left\{ e^{{\rm i}\phi^{\rm R}_{1}}_{}, e^{{\rm i}\phi^{\rm R}_{2}}_{},
e^{{\rm i}\phi^{\rm R}_{3}}_{} \right\}$ and $P^{\rm R}_{2} = {\rm Diag}
\left\{ e^{{\rm i}\rho^{\rm R}_{}}_{}, e^{{\rm i}\sigma^{\rm R}_{}}_{}, 1 \right\}$
are the diagonal phase matrices. Then we obtain
\begin{eqnarray}
\hspace{-0.2cm}&&\hspace{-0.2cm} \theta^{\rm R}_{23} = \frac{\pi}{4} \; \quad
\delta^{\rm R}_{} = \pm \frac{\pi}{2} \; , \quad
\rho^{\rm R}_{}, \sigma^{\rm R}_{} = 0 ~ {\rm or} ~ \frac{\pi}{2} \; ,
\nonumber
\\
\hspace{-0.2cm}&&\hspace{-0.2cm} \phi^{\rm R}_{1} = 0
~ {\rm or} ~ \frac{\pi}{2} \; ,
\quad \phi^{\rm R}_{2} + \phi^{\rm R}_{3} = 2 \phi^{\rm R}_{1} \pm \pi \;.
\end{eqnarray}
One can see that all the phase parameters take very special values.

We proceed to calculate the elements $\left(\widetilde{M}^{\dagger}_{\rm D}
\widetilde{M}^{}_{\rm D} \right)^{}_{1j}$ which appear in Eq. (4.2). The
Hermitian matrix $\widetilde{M}^{\dagger}_{\rm D} \widetilde{M}^{}_{\rm D}$
can be rewritten as
\begin{eqnarray}
\widetilde{M}^{\dagger}_{\rm D} \widetilde{M}^{}_{\rm D} =
U^{\rm T}_{\rm R} M^{\dagger}_{\rm D} M^{}_{\rm D} U^{\ast}_{\rm R} =
U^{\rm T}_{\rm R} U^{}_{23} U^{\dagger}_{23} M^{\dagger}_{\rm D}
M^{}_{\rm D} U^{}_{23} U^{\dagger}_{23} U^{\ast}_{\rm R} =
U^{\prime \rm T}_{\rm R} H U^{\prime \ast}_{\rm R} \;,
\end{eqnarray}
in which
\begin{eqnarray}
U^{}_{23} = \left(\begin{matrix}
1 & 0 & 0 \cr 0 & \displaystyle \frac{{\rm i}}{\sqrt{2}} &
\displaystyle \frac{1}{\sqrt{2}} \cr
0 & \displaystyle \frac{-{\rm i}}{\sqrt{2}} &
\displaystyle \frac{1}{\sqrt{2}} \end{matrix}\right) \; ,
\end{eqnarray}
and
\begin{eqnarray}
H = U^{\dagger}_{23} M^{\dagger}_{\rm D} M^{}_{\rm D} U^{}_{23} =
\left(\begin{matrix} A^{\prime}_{} & B^{\prime}_{} & C^{\prime}_{} \cr
B^{\prime}_{} & E^{\prime}_{} & D^{\prime}_{} \cr
C^{\prime}_{} & D^{\prime}_{} & F^{\prime}_{}\end{matrix}\right)
\end{eqnarray}
is a real symmetric matrix whose elements are given by
\begin{eqnarray}
A^{\prime}_{} \hspace{-0.2cm}&=&\hspace{-0.2cm}
A^{2}_{\rm r} + 2|E|^{2}_{} \;,
\nonumber
\\
B^{\prime}_{} \hspace{-0.2cm}&=&\hspace{-0.2cm}
- \sqrt{2} {\rm Im} \left( A^{}_{\rm r}B + E^{\ast}_{}C
+ ED^{\ast}_{} \right) \;,
\nonumber
\\
C^{\prime}_{} \hspace{-0.2cm}&=&\hspace{-0.2cm}
\sqrt{2} {\rm Re} \left( A^{}_{\rm r}B + E^{\ast}_{}C
+ ED^{\ast}_{} \right) \;,
\nonumber
\\
D^{\prime}_{} \hspace{-0.2cm}&=&\hspace{-0.2cm}
- {\rm Im} \left( B^{2}_{} + 2CD^{\ast}_{} \right) \;,
\nonumber
\\
E^{\prime}_{} \hspace{-0.2cm}&=&\hspace{-0.2cm}
|B|^{2}_{} + |C|^{2}_{} + |D|^{2}_{} - {\rm Re} \left( B^{2}_{}
+ 2CD^{\ast}_{} \right) \;,
\nonumber
\\
F^{\prime}_{} \hspace{-0.2cm}&=&\hspace{-0.2cm}
|B|^{2}_{} + |C|^{2}_{} + |D|^{2}_{} + {\rm Re} \left( B^{2}_{}
+ 2CD^{\ast}_{} \right) \;,
\end{eqnarray}
and finally
\begin{eqnarray}
U^{\prime}_{\rm R} = U^{T}_{23} U^{}_{\rm R} =
\left(\begin{matrix} \eta & 0 & 0 \cr 0 & {\rm i}x & {\rm i}y \cr 0 & y & x
\end{matrix}\right)
\left(\begin{matrix}
c^{\rm R}_{13}c^{\rm R}_{12} & c^{\rm R}_{13}s^{\rm R}_{12} &
s^{\rm R}_{13}e^{-{\rm i}\delta^{\rm R}}_{} \cr -s^{\rm R}_{12} &
c^{\rm R}_{12} & 0 \cr
-s^{\rm R}_{13}c^{\rm R}_{12} e^{{\rm i}\delta^{\rm R}}_{} &
-s^{\rm R}_{13}s^{\rm R}_{12}e^{{\rm i}\delta^{\rm R}}_{} &
c^{\rm R}_{13}
\end{matrix}\right)
\left(\begin{matrix}
e^{{\rm i}\rho^{\rm R}_{}}_{} & 0 & 0 \cr 0 & e^{{\rm i}\sigma^{\rm R}_{}}_{}
& 0 \cr 0 & 0 & 1
\end{matrix}\right)
\end{eqnarray}
with $c^{\rm R}_{ij} = \cos{\theta^{\rm R}_{ij}}$,
$s^{\rm R}_{ij} = \sin{\theta^{\rm R}_{ij}}$, and
$x = {\rm i} \sin{\phi^{\rm R}_{2}}$ and $y = \cos{\phi^{\rm R}_{2}}$
for $\eta = 1$ (i.e., $\phi^{\rm R}_{1} = 0$) or $x = \cos{\phi^{\rm R}_{2}}$ and
$y = {\rm i} \sin{\phi^{\rm R}_{2}}$ for $\eta = {\rm i}$ (i.e.,
$\phi^{\rm R}_{1} = \pi/2$). With the help of Eqs. (4.6) and (4.8),
Eq. (4.4) can be expressed as
\begin{eqnarray}
\left(\widetilde{M}^{\dagger}_{\rm D} \widetilde{M}^{}_{\rm D}
\right)^{}_{1j} = \sum^{3}_{k=1}  W^{}_{k} \left(\begin{matrix}
c^{\rm R}_{13}c^{\rm R}_{12} & c^{\rm R}_{13}s^{\rm R}_{12} &
s^{\rm R}_{13}e^{{\rm i}\delta^{\rm R}}_{} \cr -s^{\rm R}_{12} &
c^{\rm R}_{12} & 0 \cr
-s^{\rm R}_{13}c^{\rm R}_{12} e^{-{\rm i}\delta^{\rm R}}_{} &
-s^{\rm R}_{13}s^{\rm R}_{12}e^{-{\rm i}\delta^{\rm R}}_{} &
c^{\rm R}_{13}
\end{matrix}\right)^{}_{\hspace{-0.15cm} kj}
P^{}_{jj} \;,
\end{eqnarray}
in which $P = {\rm Diag} \left\{1 , e^{{\rm i}(\rho^{\rm R}_{}
- \sigma^{\rm R}_{})}_{}, e^{{\rm i}\rho^{\rm R}_{}} \right\}$ and
\begin{eqnarray}
W^{}_{k}= \left[
\left(\begin{matrix}
c^{\rm R}_{13}c^{\rm R}_{12} & -s^{\rm R}_{12} &
-s^{\rm R}_{13}c^{\rm R}_{12} e^{{\rm i}\delta^{\rm R}}_{} \cr
c^{\rm R}_{13}s^{\rm R}_{12} & c^{\rm R}_{12} &
-s^{\rm R}_{13}s^{\rm R}_{12}e^{{\rm i}\delta^{\rm R}}_{} \cr
s^{\rm R}_{13}e^{-{\rm i}\delta^{\rm R}}_{} & 0 & c^{\rm R}_{13}
\end{matrix}\right)
\left(\begin{matrix}
\eta & 0 & 0 \cr 0 & {\rm i}x & y \cr 0 & {\rm i}y & x
\end{matrix}\right) H
\left(\begin{matrix}
\eta^{\ast}_{} & 0 & 0 \cr 0 & -{\rm i}x^{\ast}_{} & -{\rm i}y^{\ast}_{} \cr
0 & y^{\ast}_{} & x^{\ast}_{}
\end{matrix}\right)\right]^{}_{1k} \;.
\end{eqnarray}
Concretely,
\begin{eqnarray}
W^{}_{1} \hspace{-0.2cm}&=&\hspace{-0.2cm} A^{\prime}_{} c^{\rm R}_{12}
c^{\rm R}_{13} -  p s^{}_{12} - q s^{\rm R}_{13} c^{\rm R}_{12}
e^{{\rm i}\delta^{\rm R}_{}}_{} \;,
\nonumber
\\
W^{}_{2} \hspace{-0.2cm}&=&\hspace{-0.2cm} p^{\ast}_{} c^{\rm R}_{12}
c^{\rm R}_{13}  -  t^{}_{1} s^{}_{12} - r s^{\rm R}_{13} c^{\rm R}_{12}
e^{{\rm i}\delta^{\rm R}_{}}_{} \;,
\nonumber
\\
W^{}_{3} \hspace{-0.2cm}&=&\hspace{-0.2cm} q^{\ast}_{} c^{\rm R}_{12}
c^{\rm R}_{13}  - r^{\ast}_{} s^{}_{12} - t^{}_{2} s^{\rm R}_{13}
c^{\rm R}_{12} e^{{\rm i}\delta^{\rm R}_{}}_{} \;,
\end{eqnarray}
where
\begin{eqnarray}
p \hspace{-0.2cm}&=&\hspace{-0.2cm}
\eta^{\ast}_{} ({\rm i}xB^{\prime}_{} + yC^{\prime}_{}) \;,
\nonumber
\\
q \hspace{-0.2cm}&=&\hspace{-0.2cm}
\eta^{\ast}_{} ({\rm i}yB^{\prime}_{} + xC^{\prime}_{}) \;,
\nonumber
\\
t^{}_{1} \hspace{-0.2cm}&=&\hspace{-0.2cm}
|x|^{2}_{}E^{\prime}_{} + |y|^{2}_{}F^{\prime}_{}
- 2 D^{\prime}_{} {\rm Im} \left( xy^{\ast}_{} \right) \;,
\nonumber
\\
t^{}_{2} \hspace{-0.2cm}&=&\hspace{-0.2cm}
|x|^{2}_{}F^{\prime}_{} + |y|^{2}_{}E^{\prime}_{} +
2 D^{\prime}_{} {\rm Im} \left( xy^{\ast}_{} \right) \;,
\nonumber
\\
r \hspace{-0.2cm}&=&\hspace{-0.2cm}
x^{\ast}_{}yE^{\prime}_{} + xy^{\ast}_{}F^{\prime}_{}
- {\rm i} D^{\prime}_{} \left( |x|^2 - |y|^2 \right) \;,
\end{eqnarray}
with $t^{}_{1}$ and $t^{}_{2}$ being real. Taking account
of Eqs. (4.9)---(4.12), we obtain
\begin{eqnarray}
\left(\widetilde{M}^{\dagger}_{\rm D} \widetilde{M}^{}_{\rm D}
\right)^{}_{11} \hspace{-0.2cm}&=&\hspace{-0.2cm}
A^{\prime 2}_{} c^{\rm R2}_{12} c^{\rm R2}_{13} - 2 s^{\rm R}_{12}
c^{\rm R}_{12} c^{\rm R}_{13} {\rm Re} (p) - 2{\rm i}e^{{\rm i}\delta^{R}_{}}_{}
s^{\rm R}_{13} c^{\rm R}_{13} c^{\rm R2}_{12} {\rm Im} (q)
\nonumber
\\
\hspace{-0.2cm}&&\hspace{-0.2cm} + 2{\rm i}e^{{\rm i}\delta^{R}_{}}_{} s^{\rm R}_{13}
s^{\rm R}_{12} c^{\rm R}_{12} {\rm Im} (r) + t^{}_{1} s^{\rm R2}_{12} +
t^{}_{2} s^{\rm R2}_{13} c^{\rm R2}_{12} \;,
\nonumber
\\
\left(\widetilde{M}^{\dagger}_{\rm D} \widetilde{M}^{}_{\rm D}
\right)^{}_{12} \hspace{-0.2cm}&=&\hspace{-0.2cm} e^{{\rm i}
\left(\rho^{\rm R}_{} - \sigma^{\rm R}_{} \right)} \left[
A^{\prime}_{} s^{\rm R}_{12} c^{\rm R}_{12} c^{\rm R2}_{13} -
2s^{\rm R2}_{12} c^{\rm R}_{13} {\rm Re} (p) + p^{\ast}_{}
c^{\rm R}_{13} \right.
\nonumber
\\
\hspace{-0.2cm}&&\hspace{-0.2cm} - 2{\rm i}e^{{\rm i}\delta^{R}_{}}_{} s^{\rm R}_{13}
c^{\rm R}_{13} s^{\rm R}_{12} c^{\rm R}_{12} {\rm Im} (q) +
2{\rm i}e^{{\rm i}\delta^{R}_{}}_{} s^{\rm R}_{13} s^{\rm R2}_{12} {\rm Im} (r)
\nonumber
\\
\hspace{-0.2cm}&&\hspace{-0.24cm} \left. - r e^{{\rm i}\delta^{R}_{}}_{}
s^{\rm R}_{13} - t^{}_{1} s^{\rm R}_{12} c^{\rm R}_{12} + t^{}_{2}
s^{\rm R2}_{13} s^{\rm R}_{12} c^{\rm R}_{12} \right] \;,
\nonumber
\\
\left(\widetilde{M}^{\dagger}_{\rm D} \widetilde{M}^{}_{\rm D}
\right)^{}_{13} \hspace{-0.2cm}&=&\hspace{-0.2cm}
e^{{\rm i}\rho^{\rm R}_{}} \left[ A^{\prime}_{} e^{{\rm i}\delta^{R}_{}}_{}
s^{\rm R}_{13} c^{\rm R}_{13} c^{\rm R2}_{12}
-p e^{{\rm i}\delta^{R}_{}}_{} s^{\rm R}_{13} c^{\rm R}_{12}
- t^{}_{2} e^{{\rm i}\delta^{R}_{}}_{} s^{\rm R}_{13} c^{\rm R}_{13}
c^{\rm R}_{12} \right.
\nonumber
\\
\hspace{-0.2cm}&&\hspace{-0.24cm} \left. - r^{\ast}_{} s^{\rm R}_{12}
c^{\rm R}_{13} + q c^{\rm R}_{12} - 2{\rm i} c^{\rm R}_{12} c^{\rm R2}_{13}
{\rm Im} (q) \right] \;.
\end{eqnarray}
In view of Eqs. (4.3) and (4.12) together with the definitions of $\eta$, $x$
and $y$, it is apparent that $p$, $t^{}_{1}$ and $t^{}_{2}$ are real;
$q$ and $r$ are purely imaginary; ${\rm i}e^{{\rm i}\delta^{R}_{}}_{}
= \pm 1$, $e^{{\rm i}\left(\rho^{\rm R}_{} - \sigma^{\rm R}_{} \right)}
= 1 $ or $\pm {\rm i}$ and $e^{{\rm i}\rho^{\rm R}_{}} = 1$ or $\rm i$.
We are therefore left with
\begin{eqnarray}
{\rm Im} \left[ \left(\widetilde{M}^{\dagger}_{\rm D}
\widetilde{M}^{}_{\rm D}\right)^{2}_{12} \right] = {\rm Im}
\left[ \left(\widetilde{M}^{\dagger}_{\rm D} \widetilde{M}^{}_{\rm D}
\right)^{2}_{13} \right] = 0 \;, \quad \epsilon^{}_{1} = 0 \;.
\end{eqnarray}
In other words, there is no CP violation at all in $N^{}_1$ decays
for category $A^{}_1$.

If the three heavy Majorana neutrinos have the same mass hierarchy in
categories $A^{}_{1}$, $A^{}_2$ and $A^{}_3$, then the expressions of
three eigenvalues of $M^{A^{}_{i}}_{\rm R}$ are of the same form,
and therefore Eq. (2.16) leads us to
\begin{eqnarray}
U^{A^{}_{2}}_{\rm R} = S^{(231)}_{} U^{A^{}_{1}}_{\rm R} \;,  \quad
U^{A^{}_{3}}_{\rm R} = S^{(312)}_{} U^{A^{}_{1}}_{\rm R} \;.
\end{eqnarray}
With the help of Eqs. (2.16) and (4.15), we find
\begin{eqnarray}
\widetilde{M}^{A^{}_{2}\dagger}_{\rm D} \widetilde{M}^{A^{}_{2}}_{\rm D}
\hspace{-0.2cm}&=&\hspace{-0.2cm} U^{A^{}_{2}T}_{\rm R} M^{A^{}_{2}
\dagger}_{\rm D} M^{A^{}_{2}}_{\rm D} U^{A^{}_{2}\ast}_{\rm R}
\nonumber
\\
\hspace{-0.2cm}&=&\hspace{-0.2cm} U^{A^{}_{1}T}_{\rm R} S^{(312)}_{}
S^{(231)}_{} M^{A^{}_{1} \dagger}_{\rm D} S^{(312)}_{} S^{(231)}_{}
M^{A^{}_{1}}_{\rm D} S^{(312)}_{} S^{(231)}_{} U^{A^{}_{1}\ast}_{\rm R}
\nonumber
\\
\hspace{-0.2cm}&=&\hspace{-0.2cm} U^{A^{}_{1}T}_{\rm R}
M^{A^{}_{1} \dagger}_{\rm D} M^{A^{}_{1}}_{\rm D}
U^{A^{}_{1}\ast}_{\rm R}
\nonumber
\\
\hspace{-0.2cm}&=&\hspace{-0.2cm} \widetilde{M}^{A^{}_{1}\dagger}_{\rm D}
\widetilde{M}^{A^{}_{1}}_{\rm D} \;,
\nonumber
\\
\widetilde{M}^{A^{}_{3}\dagger}_{\rm D} \widetilde{M}^{A^{}_{3}}_{\rm D}
\hspace{-0.2cm}&=&\hspace{-0.2cm} U^{A^{}_{3}T}_{\rm R}
M^{A^{}_{3} \dagger}_{\rm D} M^{A^{}_{3}}_{\rm D} U^{A^{}_{3}\ast}_{\rm R}
\nonumber
\\
\hspace{-0.2cm}&=&\hspace{-0.2cm} U^{A^{}_{1}T}_{\rm R} S^{(231)}_{}
S^{(312)}_{} M^{A^{}_{1} \dagger}_{\rm D} S^{(231)}_{} S^{(312)}_{}
M^{A^{}_{1}}_{\rm D} S^{(231)}_{} S^{(312)}_{} U^{A^{}_{1}\ast}_{\rm R}
\nonumber
\\
\hspace{-0.2cm}&=&\hspace{-0.2cm} U^{A^{}_{1}T}_{\rm R}
M^{A^{}_{1} \dagger}_{\rm D} M^{A^{}_{1}}_{\rm D} U^{A^{}_{1}\ast}_{\rm R}
\nonumber
\\
\hspace{-0.2cm}&=&\hspace{-0.2cm} \widetilde{M}^{A^{}_{1}\dagger}_{\rm D}
 \widetilde{M}^{A^{}_{1}}_{\rm D} \;.
\end{eqnarray}
This result in turn means
\begin{eqnarray}
\epsilon^{A^{}_{2}}_{1} = \epsilon^{A^{}_{3}}_{1} =
\epsilon^{A^{}_{1}}_{1} = 0 \;.
\end{eqnarray}
We conclude that in the $S^{}_3$ reflection symmetry limit there is no
way to realize unflavored leptogenesis for categories $A^{}_i$.
This conclusion will change when the lepton flavor effects are taken into
account.

\subsubsection{Category $C$}

In this case the three eigenvalues of $M^{}_{\rm R}$ are given by
$a^{}_{\rm r} - b^{}_{\rm r}$, $a^{}_{\rm r} - b^{}_{\rm r}$ and
$a^{}_{\rm r} + 2 b^{}_{\rm r}$, respectively. For simplicity, let us
assume $ a^{}_{\rm r} > 0$ and $b^{}_{\rm r} < 0$, such that
$M^{}_{1} = a^{}_{\rm r} + 2 b^{}_{\rm r} \ll
M^{}_{2} = M^{}_{3} = a^{}_{\rm r} - b^{}_{\rm r}$ can be
satisfied. The corresponding unitary matrix
$U^{}_{\rm R}$ is
\begin{eqnarray}
U^{}_{\rm R} = \left(\begin{matrix}
\displaystyle \frac{1}{\sqrt{3}} & \displaystyle - \frac{2}{\sqrt{6}} & 0 \cr
\displaystyle \frac{1}{\sqrt{3}} &
\displaystyle \frac{1}{\sqrt{6}} & \displaystyle  \frac{1}{\sqrt{2}} \cr
\displaystyle \frac{1}{\sqrt{3}} &
\displaystyle \frac{1}{\sqrt{6}} & \displaystyle - \frac{1}{\sqrt{2}}
\end{matrix}\right) \;.
\end{eqnarray}
Consequently,
\begin{eqnarray}
\widetilde{M}^{}_{\rm D} = M^{}_{\rm D} U^{\ast}_{\rm R} =
\left(\begin{matrix}
\displaystyle \frac{1}{\sqrt{3}} \left( A^{}_{\rm r} + 2 {\rm Re} B \right) &
\displaystyle - \frac{1}{\sqrt{6}} \left( 2A^{}_{\rm r} - 2 {\rm Re} B \right)
& \displaystyle \sqrt{2} \ {\rm i} \ {\rm Im} B \cr
\displaystyle \frac{1}{\sqrt{3}} \left( A^{}_{\rm r}
+ 2 {\rm Re} B \right) &  \displaystyle \frac{1}{\sqrt{6}} \left( A^{}_{\rm r}
-2B^{\ast}_{} + B \right) & \displaystyle \frac{1}{\sqrt{2}} \left( A^{}_{\rm r}
- B \right) \cr \displaystyle \frac{1}{\sqrt{3}} \left( A^{}_{\rm r} + 2 {\rm Re}
B \right) & \displaystyle \frac{1}{\sqrt{6}} \left( A^{}_{\rm r} + B^{\ast}_{} -
2B \right) & \displaystyle - \frac{1}{\sqrt{2}} \left( A^{}_{\rm r} - B^{\ast}_{}
\right) \end{matrix}\right)\;.
\end{eqnarray}
Then it is straightforward for us to obtain
\begin{eqnarray}
\left(\widetilde{M}^{\dagger}_{\rm D} \widetilde{M}^{}_{\rm D}
\right)^{}_{11} = \left( A^{}_{\rm r} + 2{\rm Re} B \right)^2 \;, \quad
\left(\widetilde{M}^{\dagger}_{\rm D} \widetilde{M}^{}_{\rm D}
\right)^{}_{12} = 0 \;, \quad
\left(\widetilde{M}^{\dagger}_{\rm D} \widetilde{M}^{}_{\rm D}
\right)^{}_{13} = 0  \; .
\end{eqnarray}
As a result,
\begin{eqnarray}
\epsilon^{}_{1\alpha} = 0 \;, \quad
\epsilon^{}_{1} = \sum^{}_{\alpha} \epsilon^{}_{1\alpha} = 0 \; ,
\end{eqnarray}
where $\alpha$ runs over $e$, $\mu$ and $\tau$. Therefore, there is no way for
both unflavored and flavored leptogenesis to work in category $C$.

\subsection{Flavored leptogenesis}

In the unflavored leptogenesis case as discussed above,
the Yukawa interactions of charged leptons are not taken into
account, since the equilibrium temperature of heavy Majorana
neutrinos is assumed to be high enough that such interactions
cannot distinguish one lepton flavor from another. In other words,
all the relevant Yukawa interactions are blind to lepton flavors.
When the equilibrium temperature is lower, however, it is possible that
the Yukawa interactions of charged leptons become faster than the (inverse)
decays of $N^{}_{i}$ or equivalently comparable to the expansion rate
of the Universe. In this case the flavor effects must be taken into
consideration \cite{Barbieri:1999ma,Morozumi}.

Here we focus on the possibility that the equilibrium temperature $T$
lies in the range $ 10^{9} \; {\rm GeV} < T < 10^{12} \;
{\rm GeV}$, in which the $\tau$ lepton can be in thermal equilibrium
and thus are distinguishable from the $e$ and $\mu$ flavors. In this case
both the CP-violating asymmetries and washout effects involving the $\tau$ flavor
should be treated separately \cite{Nardi:2006fx,Abada:2006fw}.
It is then possible to achieve successful leptogenesis provided
$\epsilon^{}_{1\alpha} \neq 0$ holds, even though the total CP-violating asymmetry
$\epsilon^{}_1$ is vanishing or vanishingly small.

We have shown in Eq. (4.21) that both $\epsilon^{}_{1\alpha}$ and
$\epsilon^{}_1$ are vanishing in category $C$, and thus it is impossible
to realize either unflavored or flavored leptogenesis in this case in
the $S^{}_3$ reflection symmetry limit. In the following we calculate
the flavor-dependent CP-violating asymmetries $\epsilon^{}_{1\alpha}$
for categories $A^{}_i$ by using Eq. (4.1), to examine whether flavored
leptogenesis has a chance to work or not in this case.

Given $M^{}_{\rm D}$ and $U^{}_{\rm R}$ in category $A^{}_1$, a lengthy but
straightforward calculation leads us to
\begin{eqnarray}
\left(\widetilde{M}^{\ast}_{\rm D}\right)^{}_{e1}
\left(\widetilde{M}^{}_{\rm D}\right)^{}_{e2}
\hspace{-0.2cm}&=&\hspace{-0.2cm}
e^{{\rm i}\left( \rho^{\rm R}_{} - \sigma^{\rm R}_{} \right)}_{}
\left\{ A^{2}_{\rm r} s^{\rm R}_{12} c^{\rm R}_{12} c^{\rm R2}_{13}
+ \sqrt{2}A^{}_{\rm r} c^{\rm R}_{13} \cos{2 \theta^{\rm R}_{12}}
{\rm Re} \left( \eta^{\ast}_{} B e^{ -{\rm i}\phi^{\rm R}_{2}}_{} \right)
\right.
\nonumber
\\
\hspace{-0.2cm}&&\hspace{-0.2cm}
+ 2 \ {\rm i} \ e^{{\rm i}\delta^{\rm R}_{}}_{} s^{\rm R}_{13} \cos{2\theta^{\rm R}_{12}}
{\rm Re} \left( \eta^{\ast}_{} B e^{ -{\rm i}\phi^{\rm R}_{2}}_{} \right)
{\rm Im} \left( \eta^{\ast}_{} B e^{ -{\rm i}\phi^{\rm R}_{2}}_{} \right)
\nonumber
\\
\hspace{-0.2cm}&&\hspace{-0.2cm} + 2 s^{\rm R2}_{13} s^{\rm R}_{12}
c^{\rm R}_{12} \left[ {\rm Im} \left( \eta^{\ast}_{} B e^{ -{\rm i}\phi^{\rm R}_{2}}_{}
\right) \right]^{2}_{} - 2 s^{\rm R}_{12} c^{\rm R}_{12} \left[ {\rm Re}
\left( \eta^{\ast}_{} B e^{ -{\rm i}\phi^{\rm R}_{2}}_{} \right) \right]^{2}_{}
\nonumber
\\
\hspace{-0.2cm}&&\hspace{-0.24cm}
\left.  + 2\sqrt{2} \ {\rm i} \ e^{{\rm i}\delta^{\rm R}_{}}_{} A^{}_{\rm r} s^{\rm R}_{13}
c^{\rm R}_{13} s^{\rm R}_{12} c^{\rm R}_{12} {\rm Im}
\left( \eta^{\ast}_{} B e^{ - {\rm i}\phi^{\rm R}_{2}}_{} \right) \right\} \;,
\nonumber
\\
\left(\widetilde{M}^{\ast}_{\rm D}\right)^{}_{e1}
\left(\widetilde{M}^{}_{\rm D}\right)^{}_{e3}
\hspace{-0.2cm}&=&\hspace{-0.2cm}
e^{ {\rm i} \rho^{\rm R}_{} }_{} \left\{ A^{2}_{\rm r} s^{\rm R}_{13}
c^{\rm R}_{13} c^{\rm R}_{12} e^{ {\rm i} \delta^{\rm R}_{}}_{} -
2e^{{\rm i}\delta^{\rm R}_{}}_{} s^{\rm R}_{13} c^{\rm R}_{13} c^{\rm R}_{12}
\left[ {\rm Im} \left( \eta^{\ast}_{} B e^{ -{\rm i}\phi^{\rm R}_{2}}_{} \right)
\right]^{2}_{}
\right.
\nonumber
\\
\hspace{-0.2cm}&&\hspace{-0.2cm}
- 2 \ {\rm i} \ s^{\rm R}_{12} c^{\rm R}_{13} {\rm Re} \left( \eta^{\ast}_{} B
e^{ - {\rm i}\phi^{\rm R}_{2}}_{} \right) {\rm Im} \left( \eta^{\ast}_{} B
e^{ - {\rm i}\phi^{\rm R}_{2}}_{} \right)
\nonumber
\\
\hspace{-0.2cm}&&\hspace{-0.2cm}
+ \sqrt{2} \ {\rm i} \ A^{}_{\rm r} \cos{2\theta^{\rm R}_{13}} c^{\rm R}_{12} {\rm Im}
\left( \eta^{\ast}_{} B e^{ -{\rm i}\phi^{\rm R}_{2}}_{} \right)
\nonumber
\\
\hspace{-0.2cm}&&\hspace{-0.24cm}
\left. - \sqrt{2} \ A^{}_{\rm r} s^{\rm R}_{13}
s^{\rm R}_{12} e^{{\rm i}\delta^{\rm R}_{}}_{} {\rm Re} \left( \eta^{\ast}_{}
B e^{ -{\rm i}\phi^{\rm R}_{2}}_{} \right)  \right\} \;,
\end{eqnarray}
and
\begin{eqnarray}
\left(\widetilde{M}^{\ast}_{\rm D}\right)^{}_{\mu1}
\left(\widetilde{M}^{}_{\rm D}\right)^{}_{\mu2}
\hspace{-0.2cm}&=&\hspace{-0.2cm}
e^{{\rm i}\left( \rho^{\rm R}_{} - \sigma^{\rm R}_{} \right)}_{}
\left\{ |E|^{2}_{} s^{\rm R}_{12} c^{\rm R}_{12} c^{\rm R2}_{13}
- \frac{1}{2} |z^{}_{1} + z^{}_{2}|^{2}_{} s^{\rm R}_{12} c^{\rm R}_{12}
+ \frac{1}{2} |z^{}_{1} - z^{}_{2}|^{2}_{} s^{\rm R2}_{13} s^{\rm R}_{12}
c^{\rm R}_{12} \right.
\nonumber
\\
\hspace{-0.2cm}&&\hspace{-0.2cm}
-\sqrt{2} \ s^{\rm R2}_{12} c^{\rm R}_{13} {\rm Re} \left[ E^{\ast}_{}
\left( z^{}_{1} + z^{}_{2} \right) \right] + \sqrt{2} \ {\rm i} \
e^{{\rm i}\delta^{\rm R}_{}}_{} s^{\rm R}_{13} c^{\rm R}_{13} s^{\rm R}_{12}
c^{\rm R}_{12} {\rm Im} \left[ E^{\ast}_{} \left( z^{}_{1} - z^{}_{2}
\right) \right]
\nonumber
\\
\hspace{-0.2cm}&&\hspace{-0.2cm}
- {\rm i} \ e^{{\rm i}\delta^{\rm R}_{}}_{} s^{\rm R}_{13}
\cos{2\theta^{\rm R}_{12}} {\rm Im} \left( z^{\ast}_{1} z^{}_{2} \right)
+ \frac{1}{\sqrt{2}} \ c^{\rm R}_{13} E^{\ast}_{} \left( z^{}_{1} + z^{}_{2}
\right)
\nonumber
\\
\hspace{-0.2cm}&&\hspace{-0.24cm}
\left. - \frac{1}{2} s^{\rm R}_{13} e^{{\rm i}\delta^{\rm R}_{}}_{}
\left( |C|^{2}_{} - |D|^{2}_{} \right)  \right\} \;,
\nonumber
\\
\left(\widetilde{M}^{\ast}_{\rm D}\right)^{}_{\mu1}
\left(\widetilde{M}^{}_{\rm D}\right)^{}_{\mu3}
\hspace{-0.2cm}&=&\hspace{-0.2cm}
e^{ {\rm i} \rho^{\rm R}_{} }_{} \left\{ |E|^{2}_{} s^{\rm R}_{13} c^{\rm R}_{13}
c^{\rm R}_{12} e^{{\rm i}\delta^{\rm R}_{}}_{}
- \frac{1}{2} \ s^{\rm R}_{12} c^{\rm R}_{13} \left( |C|^{2}_{}
- |D|^{2}_{} \right) + {\rm i} \ s^{\rm R}_{12} c^{\rm R}_{13}
{\rm Im} \left( z^{\ast}_{1} z^{}_{2} \right)
 \right.
\nonumber
\\
\hspace{-0.2cm}&&\hspace{-0.2cm}
- \sqrt{2} \ {\rm i} \ s^{\rm R2}_{13} c^{\rm R}_{12} {\rm Im} \left[ E^{\ast}_{}
\left( z^{}_{1} - z^{}_{2} \right) \right]  - \frac{1}{2}
e^{{\rm i} \delta^{\rm R}_{} }_{} s^{\rm R}_{13} c^{\rm R}_{13} c^{\rm R}_{12} |
z^{}_{1} - z^{}_{2}|^{2}_{}
\nonumber
\\
\hspace{-0.2cm}&&\hspace{-0.24cm}
\left. + \frac{1}{\sqrt{2}} \
c^{\rm R}_{12} E^{\ast}_{} \left( z^{}_{1} - z^{}_{2} \right)
- \frac{1}{\sqrt{2}} \ e^{{\rm i} \delta^{\rm R}_{} }_{} s^{\rm R}_{13}
s^{\rm R}_{12} E \left( z^{\ast}_{1} + z^{\ast}_{2} \right) \right\} \;,
\end{eqnarray}
as well as
\begin{eqnarray}
\left(\widetilde{M}^{\ast}_{\rm D}\right)^{}_{\tau1}
\left(\widetilde{M}^{}_{\rm D}\right)^{}_{\tau2}
\hspace{-0.2cm}&=&\hspace{-0.2cm}
e^{{\rm i}\left( \rho^{\rm R}_{} - \sigma^{\rm R}_{} \right)}_{}
\left\{ |E|^{2}_{} s^{\rm R}_{12} c^{\rm R}_{12} c^{\rm R2}_{13}
- \frac{1}{2} |z^{}_{1} + z^{}_{2}|^{2}_{} s^{\rm R}_{12} c^{\rm R}_{12}
+ \frac{1}{2} |z^{}_{1} - z^{}_{2}|^{2}_{} s^{\rm R2}_{13} s^{\rm R}_{12}
c^{\rm R}_{12} \right.
\nonumber
\\
\hspace{-0.2cm}&&\hspace{-0.2cm}
- \sqrt{2} \ s^{\rm R2}_{12} c^{\rm R}_{13} {\rm Re} \left[ E^{\ast}_{}
\left( z^{}_{1} + z^{}_{2} \right) \right] + \sqrt{2} \ {\rm i} \
e^{{\rm i}\delta^{\rm R}_{}}_{} s^{\rm R}_{13} c^{\rm R}_{13} s^{\rm R}_{12}
c^{\rm R}_{12} {\rm Im} \left[ E^{\ast}_{} \left( z^{}_{1} - z^{}_{2}
\right) \right]
\nonumber
\\
\hspace{-0.2cm}&&\hspace{-0.2cm}
- {\rm i} \ e^{{\rm i}\delta^{\rm R}_{}}_{} s^{\rm R}_{13}
\cos{2\theta^{\rm R}_{12}} {\rm Im} \left( z^{\ast}_{1} z^{}_{2} \right)
 + \frac{1}{2} s^{\rm R}_{13} e^{{\rm i}\delta^{\rm R}_{}}_{}
\left( |C|^{2}_{} - |D|^{2}_{} \right)
\nonumber
\\
\hspace{-0.2cm}&&\hspace{-0.24cm}
\left. + \frac{1}{\sqrt{2}} \ c^{\rm R}_{13} E \left( z^{\ast}_{1} + z^{\ast}_{2}
\right)  \right\} \;,
\nonumber
\\
\left(\widetilde{M}^{\ast}_{\rm D}\right)^{}_{\tau1}
\left(\widetilde{M}^{}_{\rm D}\right)^{}_{\tau3}
\hspace{-0.2cm}&=&\hspace{-0.2cm}
e^{ {\rm i} \rho^{\rm R}_{} }_{} \left\{ |E|^{2}_{} s^{\rm R}_{13} c^{\rm R}_{13}
c^{\rm R}_{12} e^{ {\rm i}\delta^{\rm R}_{}}_{} + \frac{1}{2} s^{\rm R}_{12}
c^{\rm R}_{13} \left( |C|^{2}_{} - |D|^{2}_{} \right) + {\rm i} \ s^{\rm R}_{12}
c^{\rm R}_{13} {\rm Im} \left( z^{\ast}_{1} z^{}_{2} \right) \right.
\nonumber
\\
\hspace{-0.2cm}&&\hspace{-0.2cm}
-\sqrt{2} \ {\rm i} \ s^{\rm R2}_{13} c^{\rm R}_{12} {\rm Im} \left[ E^{\ast}_{}
\left( z^{}_{1} - z^{}_{2} \right) \right] - \frac{1}{2}
e^{{\rm i} \delta^{\rm R}_{} }_{} s^{\rm R}_{13} c^{\rm R}_{13}
c^{\rm R}_{12} | z^{}_{1} - z^{}_{2}|^{2}_{}
\nonumber
\\
\hspace{-0.2cm}&&\hspace{-0.2cm}
\left.  - \frac{1}{\sqrt{2}} \ c^{\rm R}_{12} E \left( z^{\ast}_{1} - z^{\ast}_{2}
\right) - \frac{1}{\sqrt{2}} \ e^{{\rm i} \delta^{\rm R}_{} }_{} s^{\rm R}_{13}
s^{\rm R}_{12} E^{\ast}_{} \left( z^{}_{1} + z^{}_{2} \right) \right\} \;,
\end{eqnarray}
where $\eta = 1$ (or $-{\rm i}$) for $\phi^{\rm R}_{1} = 0$ (or $\pi/2$), and
\begin{eqnarray}
z^{}_{1} = \eta^{\ast}_{} C e^{-{\rm i}\phi^{\rm R}_{2}}_{} \;, \quad
z^{}_{2} = \eta^{}_{} D e^{{\rm i}\phi^{\rm R}_{2}}_{} \;.
\end{eqnarray}
With the help of Eq. (4.13) and Eqs. (4.22)---(4.25), we further obtain
\begin{eqnarray}
{\rm Im} \left[ \left(\widetilde{M}^{\ast}_{\rm D}\right)^{}_{e1}
\left(\widetilde{M}^{}_{\rm D}\right)^{}_{e2}
\left(\widetilde{M}^{\dagger}_{\rm D} \widetilde{M}^{}_{\rm D}
\right)^{}_{12} \right] \hspace{-0.2cm}&=&\hspace{-0.2cm} 0 \;,
\nonumber
\\
{\rm Im} \left[ \left(\widetilde{M}^{\ast}_{\rm D}\right)^{}_{e1}
\left(\widetilde{M}^{}_{\rm D}\right)^{}_{e3}
\left(\widetilde{M}^{\dagger}_{\rm D} \widetilde{M}^{}_{\rm D}
\right)^{}_{13} \right] \hspace{-0.2cm}&=&\hspace{-0.2cm} 0 \;,
\nonumber
\\
{\rm Im} \left[ \left(\widetilde{M}^{\ast}_{\rm D}\right)^{}_{\mu1}
\left(\widetilde{M}^{}_{\rm D}\right)^{}_{\mu2}
\left(\widetilde{M}^{\dagger}_{\rm D} \widetilde{M}^{}_{\rm D}
\right)^{}_{12} \right] \hspace{-0.2cm}&=&\hspace{-0.2cm} \eta_1 \left|
\left(\widetilde{M}^{\dagger}_{\rm D} \widetilde{M}^{}_{\rm D}
\right)^{}_{12} \right| {\rm Im} \left[ \frac{1}{\sqrt{2}}
c^{\rm R}_{13} E^{\ast}_{} \left( z^{}_{1} + z^{}_{2} \right) \right.
\nonumber
\\
\hspace{-0.2cm}&&\hspace{-0.24cm}
\left. - \frac{1}{2} s^{\rm R}_{13} e^{{\rm i}\delta^{\rm R}_{}}_{}
\left( |C|^{2}_{} - |D|^{2}_{} \right) \right]  \;,
\nonumber
\\
{\rm Im} \left[ \left(\widetilde{M}^{\ast}_{\rm D}\right)^{}_{\mu1}
\left(\widetilde{M}^{}_{\rm D}\right)^{}_{\mu3}
\left(\widetilde{M}^{\dagger}_{\rm D} \widetilde{M}^{}_{\rm D}
\right)^{}_{13} \right] \hspace{-0.2cm}&=&\hspace{-0.2cm} \eta_2 \left|
\left(\widetilde{M}^{\dagger}_{\rm D} \widetilde{M}^{}_{\rm D}
\right)^{}_{13} \right| {\rm Re} \left[ \frac{1}{\sqrt{2}} \
c^{\rm R}_{12} E^{\ast}_{} \left( z^{}_{1} - z^{}_{2} \right) \right.
\nonumber
\\
\hspace{-0.2cm}&&\hspace{-0.24cm}
\left. - \frac{1}{\sqrt{2}} \ e^{{\rm i}\delta^{\rm R}_{}}_{} s^{\rm R}_{13}
s^{\rm R}_{12} E \left( z^{\ast}_{1} + z^{\ast}_{2} \right)
- \frac{1}{2} s^{\rm R}_{12} c^{\rm R}_{13} \left( |C|^{2}_{} -
|D|^{2}_{} \right) \right]  \;,
\nonumber
\\
{\rm Im} \left[ \left(\widetilde{M}^{\ast}_{\rm D}\right)^{}_{\tau1}
\left(\widetilde{M}^{}_{\rm D}\right)^{}_{\tau2}
\left(\widetilde{M}^{\dagger}_{\rm D} \widetilde{M}^{}_{\rm D}
\right)^{}_{12} \right] \hspace{-0.2cm}&=&\hspace{-0.2cm} \eta_1 \left|
\left(\widetilde{M}^{\dagger}_{\rm D} \widetilde{M}^{}_{\rm D}
\right)^{}_{12} \right| {\rm Im} \left[ \frac{1}{\sqrt{2}} \
c^{\rm R}_{13} E \left( z^{\ast}_{1} + z^{\ast}_{2} \right) \right.
\nonumber
\\
\hspace{-0.2cm}&&\hspace{-0.24cm}
\left. + \frac{1}{2} s^{\rm R}_{13} e^{{\rm i}\delta^{\rm R}_{}}_{}
\left( |C|^{2}_{} - |D|^{2}_{} \right) \right]  \;,
\nonumber
\\
{\rm Im} \left[ \left(\widetilde{M}^{\ast}_{\rm D}\right)^{}_{\tau1}
\left(\widetilde{M}^{}_{\rm D}\right)^{}_{\tau3}
\left(\widetilde{M}^{\dagger}_{\rm D} \widetilde{M}^{}_{\rm D}
\right)^{}_{13} \right] \hspace{-0.2cm}&=&\hspace{-0.2cm} \eta_2 \left|
\left(\widetilde{M}^{\dagger}_{\rm D} \widetilde{M}^{}_{\rm D}
\right)^{}_{13} \right| {\rm Re}  \left[ - \frac{1}{\sqrt{2}} \
c^{\rm R}_{12} E \left( z^{\ast}_{1} - z^{\ast}_{2} \right) \right.
\nonumber
\\
\hspace{-0.2cm}&&\hspace{-0.2cm}
- \frac{1}{\sqrt{2}} \ e^{{\rm i}\delta^{\rm R}_{}}_{} s^{\rm R}_{13}
s^{\rm R}_{12} E^{\ast}_{} \left( z^{}_{1} + z^{}_{2} \right)
\nonumber
\\
\hspace{-0.2cm}&&\hspace{-0.24cm}
\left. + \frac{1}{2} s^{\rm R}_{12} c^{\rm R}_{13} \left( |C|^{2}_{}
- |D|^{2}_{} \right) \right] \;,
\end{eqnarray}
and
\begin{eqnarray}
{\rm Im} \left[ \left(\widetilde{M}^{\ast}_{\rm D}\right)^{}_{e1}
\left(\widetilde{M}^{}_{\rm D}\right)^{}_{e2}
\left(\widetilde{M}^{\dagger}_{\rm D} \widetilde{M}^{}_{\rm D}
\right)^{\ast}_{12} \right] \hspace{-0.2cm}&=&\hspace{-0.2cm} 0 \;,
\nonumber
\\
{\rm Im} \left[ \left(\widetilde{M}^{\ast}_{\rm D}\right)^{}_{e1}
\left(\widetilde{M}^{}_{\rm D}\right)^{}_{e3}
\left(\widetilde{M}^{\dagger}_{\rm D} \widetilde{M}^{}_{\rm D}
\right)^{\ast}_{13} \right] \hspace{-0.2cm}&=&\hspace{-0.2cm} 0 \;,
\nonumber
\\
{\rm Im} \left[ \left(\widetilde{M}^{\ast}_{\rm D}\right)^{}_{\mu1}
\left(\widetilde{M}^{}_{\rm D}\right)^{}_{\mu2}
\left(\widetilde{M}^{\dagger}_{\rm D} \widetilde{M}^{}_{\rm D}
\right)^{\ast}_{12} \right] \hspace{-0.2cm}&=&\hspace{-0.2cm}
\eta_3 \left| \left(\widetilde{M}^{\dagger}_{\rm D} \widetilde{M}^{}_{\rm D}
\right)^{}_{12} \right| {\rm Im} \left[ \frac{1}{\sqrt{2}} \
c^{\rm R}_{13} E^{\ast}_{} \left( z^{}_{1} + z^{}_{2} \right) \right.
\nonumber
\\
\hspace{-0.2cm}&&\hspace{-0.24cm} \left. - \frac{1}{2} s^{\rm R}_{13}
e^{{\rm i}\delta^{\rm R}_{}}_{} \left( |C|^{2}_{} - |D|^{2}_{} \right)
\right] \;,
\nonumber
\\
{\rm Im} \left[ \left(\widetilde{M}^{\ast}_{\rm D}\right)^{}_{\mu1}
\left(\widetilde{M}^{}_{\rm D}\right)^{}_{\mu3}
\left(\widetilde{M}^{\dagger}_{\rm D} \widetilde{M}^{}_{\rm D}
\right)^{\ast}_{13} \right] \hspace{-0.2cm}&=&\hspace{-0.2cm}
\eta_4 \left| \left(\widetilde{M}^{\dagger}_{\rm D} \widetilde{M}^{}_{\rm D}
\right)^{}_{13} \right| {\rm Re} \left[ \frac{1}{\sqrt{2}} \
c^{\rm R}_{12} E^{\ast}_{} \left( z^{}_{1} - z^{}_{2} \right)\right.
\nonumber
\\
\hspace{-0.2cm}&&\hspace{-0.2cm}
\left.
- \frac{1}{\sqrt{2}} \ e^{{\rm i}\delta^{\rm R}_{}}_{} s^{\rm R}_{13}
s^{\rm R}_{12} E \left( z^{\ast}_{1} + z^{\ast}_{2} \right)
- \frac{1}{2} s^{\rm R}_{12} c^{\rm R}_{13} \left( |C|^{2}_{}
- |D|^{2}_{} \right) \right] \;,
\nonumber
\\
{\rm Im} \left[ \left(\widetilde{M}^{\ast}_{\rm D}\right)^{}_{\tau1}
\left(\widetilde{M}^{}_{\rm D}\right)^{}_{\tau2}
\left(\widetilde{M}^{\dagger}_{\rm D} \widetilde{M}^{}_{\rm D}
\right)^{\ast}_{12} \right] \hspace{-0.2cm}&=&\hspace{-0.2cm}
\eta_3 \left| \left(\widetilde{M}^{\dagger}_{\rm D} \widetilde{M}^{}_{\rm D}
\right)^{}_{12} \right| {\rm Im} \left[ \frac{1}{\sqrt{2}} \
c^{\rm R}_{13} E \left( z^{\ast}_{1} + z^{\ast}_{2} \right) \right.
\nonumber
\\
\hspace{-0.2cm}&&\hspace{-0.2cm}
\left. + \frac{1}{2} s^{\rm R}_{13} e^{{\rm i}\delta^{\rm R}_{}}_{}
\left( |C|^{2}_{} - |D|^{2}_{} \right) \right]  \;,
\nonumber
\\
{\rm Im} \left[ \left(\widetilde{M}^{\ast}_{\rm D}\right)^{}_{\tau1}
\left(\widetilde{M}^{}_{\rm D}\right)^{}_{\tau3}
\left(\widetilde{M}^{\dagger}_{\rm D} \widetilde{M}^{}_{\rm D}
\right)^{\ast}_{13} \right] \hspace{-0.2cm}&=&\hspace{-0.2cm}
\eta_4 \left| \left(\widetilde{M}^{\dagger}_{\rm D} \widetilde{M}^{}_{\rm D}
\right)^{}_{13} \right| {\rm Re} \left[ - \frac{1}{\sqrt{2}} \
c^{\rm R}_{12} E \left( z^{\ast}_{1} - z^{\ast}_{2} \right) \right.
\nonumber
\\
\hspace{-0.2cm}&&\hspace{-0.2cm}
 - \frac{1}{\sqrt{2}} \ e^{{\rm i}\delta^{\rm R}_{}}_{} s^{\rm R}_{13}
s^{\rm R}_{12} E^{\ast}_{} \left( z^{}_{1} + z^{}_{2} \right)
\nonumber
\\
\hspace{-0.2cm}&&\hspace{-0.2cm}
\left. + \frac{1}{2} s^{\rm R}_{12} c^{\rm R}_{13} \left( |C|^{2}_{}
- |D|^{2}_{} \right) \right] \;,
\end{eqnarray}
where $\eta^{}_i = \pm 1$ (for $i=1,2,3,4$). Combining
Eqs. (4.26) and (4.27), we arrive at
\begin{eqnarray}
{\rm Im} \left[ \left(\widetilde{M}^{\ast}_{\rm D}\right)^{}_{\mu1}
\left(\widetilde{M}^{}_{\rm D}\right)^{}_{\mu2}
\left(\widetilde{M}^{\dagger}_{\rm D} \widetilde{M}^{}_{\rm D}
\right)^{}_{12} \right] \hspace{-0.2cm}&=&\hspace{-0.2cm} - {\rm Im}
\left[ \left(\widetilde{M}^{\ast}_{\rm D}\right)^{}_{\tau1}
\left(\widetilde{M}^{}_{\rm D}\right)^{}_{\tau2}
\left(\widetilde{M}^{\dagger}_{\rm D} \widetilde{M}^{}_{\rm D}
\right)^{}_{12} \right] \neq 0 \;,
\nonumber
\\
{\rm Im} \left[ \left(\widetilde{M}^{\ast}_{\rm D}\right)^{}_{\mu1}
\left(\widetilde{M}^{}_{\rm D}\right)^{}_{\mu3}
\left(\widetilde{M}^{\dagger}_{\rm D} \widetilde{M}^{}_{\rm D}
\right)^{}_{13} \right] \hspace{-0.2cm}&=&\hspace{-0.2cm} - {\rm Im}
\left[ \left(\widetilde{M}^{\ast}_{\rm D}\right)^{}_{\tau1}
\left(\widetilde{M}^{}_{\rm D}\right)^{}_{\tau3}
\left(\widetilde{M}^{\dagger}_{\rm D} \widetilde{M}^{}_{\rm D}
\right)^{}_{13} \right] \neq 0 \;,
\nonumber
\\
{\rm Im} \left[ \left(\widetilde{M}^{\ast}_{\rm D}\right)^{}_{\mu1}
\left(\widetilde{M}^{}_{\rm D}\right)^{}_{\mu2}
\left(\widetilde{M}^{\dagger}_{\rm D} \widetilde{M}^{}_{\rm D}
\right)^{\ast}_{12} \right] \hspace{-0.2cm}&=&\hspace{-0.2cm} - {\rm Im}
\left[ \left(\widetilde{M}^{\ast}_{\rm D}\right)^{}_{\tau1}
\left(\widetilde{M}^{}_{\rm D}\right)^{}_{\tau2}
\left(\widetilde{M}^{\dagger}_{\rm D} \widetilde{M}^{}_{\rm D}
\right)^{\ast}_{12} \right] \neq 0 \;,
\nonumber
\\
{\rm Im} \left[ \left(\widetilde{M}^{\ast}_{\rm D}\right)^{}_{\mu1}
\left(\widetilde{M}^{}_{\rm D}\right)^{}_{\mu3}
\left(\widetilde{M}^{\dagger}_{\rm D} \widetilde{M}^{}_{\rm D}
\right)^{\ast}_{13} \right] \hspace{-0.2cm}&=&\hspace{-0.2cm} - {\rm Im}
\left[ \left(\widetilde{M}^{\ast}_{\rm D}\right)^{}_{\tau1}
\left(\widetilde{M}^{}_{\rm D}\right)^{}_{\tau3}
\left(\widetilde{M}^{\dagger}_{\rm D} \widetilde{M}^{}_{\rm D}
\right)^{\ast}_{13} \right] \neq 0 \; . \hspace{1cm}
\end{eqnarray}
As a result,
\begin{eqnarray}
\epsilon^{}_{1e} = 0 \;, \quad \epsilon^{}_{1\mu} = - \epsilon^{}_{1\tau} \neq 0
\;, \quad \epsilon^{}_{1} = \sum^{}_{\alpha} \epsilon^{}_{1\alpha} = 0 \;.
\end{eqnarray}
It is therefore possible to realize $\mu$- or $\tau$-flavored leptogenesis
in this case, at least in principle. Since a realistic example of this kind
needs to include proper $S^{}_3$ reflection symmetry breaking effects, it
will be studied elsewhere.

Finally, if the mass hierarchies of three heavy Majorana neutrinos in categories
$A^{}_2$ and $A^{}_3$ are the same as that in category $A^{}_1$, then one can
get
\begin{eqnarray}
\left(\widetilde{M}^{A^{}_{2}\ast}_{\rm D}\right)^{}_{\alpha1}
\left(\widetilde{M}^{A^{}_{2}}_{\rm D}\right)^{}_{\alpha j}
\hspace{-0.2cm}&=&\hspace{-0.2cm}
\left(\widetilde{M}^{A^{}_{1}\ast}_{\rm D}\right)^{}_{\beta1}
\left(\widetilde{M}^{A^{}_{1}}_{\rm D}\right)^{}_{\beta j} \;,
\nonumber
\\
\left(\widetilde{M}^{A^{}_{3}\ast}_{\rm D}\right)^{}_{\gamma1}
\left(\widetilde{M}^{A^{}_{3}}_{\rm D}\right)^{}_{\gamma j}
\hspace{-0.2cm}&=&\hspace{-0.2cm}
\left(\widetilde{M}^{A^{}_{1}\ast}_{\rm D}\right)^{}_{\lambda1}
\left(\widetilde{M}^{A^{}_{1}}_{\rm D}\right)^{}_{\lambda j} \;,
\end{eqnarray}
and therefore
\begin{eqnarray}
\epsilon^{A^{}_{2}}_{1\alpha} = \epsilon^{A^{}_{1}}_{1\beta} \;, \quad
\epsilon^{A^{}_{3}}_{1\gamma} = \epsilon^{A^{}_{1}}_{1\lambda} \;,
\end{eqnarray}
where $\alpha\beta = e\mu$, $\mu\tau$ and $\tau e$;
$\gamma\lambda = e\tau$, $\mu e$ and $\tau \mu$; and $j= 2$ or $3$.
As a result,
\begin{eqnarray}
\epsilon^{A^{}_{2}}_{1\tau} \hspace{-0.2cm}&=&\hspace{-0.2cm} 0 \;, \quad
\epsilon^{A^{}_{2}}_{1e} = - \epsilon^{A^{}_{2}}_{1\mu} \neq 0 \;, \quad
\epsilon^{A^{}_{2}}_{1} = \sum^{}_{\alpha}
\epsilon^{A^{}_{2}}_{1\alpha} = 0 \; ;
\nonumber
\\
\epsilon^{A^{}_{3}}_{1\mu} \hspace{-0.2cm}&=&\hspace{-0.2cm} 0 \;, \quad
\epsilon^{A^{}_{3}}_{1\tau} = - \epsilon^{A^{}_{3}}_{1e} \neq 0 \;, \quad
\epsilon^{A^{}_{3}}_{1} = \sum^{}_{\alpha}
\epsilon^{A^{}_{3}}_{1\alpha} = 0 \; ,
\end{eqnarray}
for categories $A^{}_2$ and $A^{}_3$.

\setcounter{footnote}{0}
\setcounter{table}{0}
\setcounter{equation}{0}
\setcounter{figure}{0}
\section{Some further discussions}

In this work we have made a new attempt to specify the flavor structures
associated with the canonical seesaw mechanism, so as to promote its
predictability and testability. What we have done is to require the
relevant neutrino mass terms to be invariant under the $S^{}_3$
reflection transformations of both left- and right-handed neutrino fields.
This treatment allows us to constrain the Dirac mass matrix
$M^{}_{\rm D}$ and the right-handed neutrino mass matrix $M^{}_{\rm R}$
to some extent, and the effective light Majorana neutrino mass matrix
$M^{}_\nu$ is in turn constrained through the seesaw relation. We
find that the structures of $M^{}_{\rm D}$, $M^{}_{\rm R}$ and
$M^{}_\nu$ can be classified into 22 categories,
among which some structures respect the well-known $\mu$-$\tau$
symmetry and (or) flavor democracy. In particular, we find that the texture
of $M^{}_\nu$ may be either the same as or similar to that of $M^{}_{\rm R}$,
and this property reflects a seesaw mirroring
relationship between light and heavy Majorana neutrinos. To be specific,
we have calculated the light neutrino masses and flavor mixing parameters
for all the textures of $M^{}_\nu$, and examined whether the CP-violating
asymmetries in decays of the lightest heavy Majorana neutrino are
vanishing or not in the $S^{}_3$ reflection symmetry limit. Our
calculations show that only the flavored leptogenesis mechanism is possible
to work for categories $A^{}_1$, $A^{}_2$ and $A^{}_3$ listed in Table 2.1.

One might wonder whether some different neutrino mixing patterns and
related leptogenesis can be obtained in our approach if $S^{}_3$ symmetry
group is extended to $S^{}_4$ or $A^{}_4$. The answer is affirmative.
To illustrate, let us briefly discuss the situation associated with
$A^{}_4$ group in our framework. It is well known that $A^{}_4$ group is
defined as the even permutation of four objects and
has twelve elements being divided into four classes. So $A^{}_4$ group has
four irreducible representations --- three inequivalent
one-dimensional representations ($\underline{\bf{1}}$,
$\underline{\bf{1}}^{\prime}$ and $\underline{\bf{1}}^{\prime\prime}$) and
one three-dimensional representation ($\underline{\bf{3}}$). Now that
we work in the basis where $M^{}_l$ is diagonal, it is more interesting for
us to consider the three-dimensional unitary representation of $A^{}_4$ group,
which has been used in Refs. \cite{Altarelli,Altarelli:2005yx} rather
than Refs. \cite{Ma:2001dn,Babu:2005se}. In this representation the two
generators of $A^{}_4$, denoted as $S$ and $T$, are given by
\begin{eqnarray}
S = \frac{1}{3}\left(\begin{matrix} -1 & 2 & 2 \cr 2 & -1 & 2 \cr 2 & 2 &
-1   \end{matrix}\right) \;, \quad
T = \left(\begin{matrix} 1 & 0 & 0 \cr 0 & \omega^2 & 0 \cr 0 & 0 & \omega
\end{matrix}\right) \;,
\end{eqnarray}
where $\omega = \exp{({\rm i} 2\pi/3)}$. Then all the elements of $A^{}_4$
can be presented by the following twelve $3\times3$ matrices: $\bf{1}$, $S$,
$T$, $ST$, $TS$, $T^2$, $ST^2$, $T^2S$, $STS$, $TST$, $TST^2$ and $T^2ST$.

In a way similar to the $S{}_3$ reflection transformations, we may require the
neutrino mass term in Eq. (2.1) to keep invariant under the transformations
made in Eq. (2.2) with $S^{}_{(\rm L)}$ or $S^{}_{(\rm R)}$ being an
arbitrary element of the given subset of $A^{}_4$ group. In this case we
are left with the same form of the constraints on $M^{}_{\rm D}$ and
$M^{}_{\rm R}$ as obtained in Eq. (2.5). One may systematically categorize
all the possible structures of neutrino mass matrices as we have done in
Table 2.1 for $S^{}_3$ group, but for $A^{}_4$ group it seems unnecessary to do
so because in most cases the $A^{}_4$-induced constraints are so strong
that the resultant textures of neutrino mass matrices are trivial
and disinteresting. If a case with the given subset having more than one
element is considered, for example, it will be unable to result in any CP
violation in both light and heavy neutrino sectors. Some cases with only one
element may also lead to trivial results, and those more interesting cases
usually involve many unknown parameters. Here we only show a simple example
of this kind which allows us to obtain the textures of neutrino mass matrices
different from those listed in Table 2.1, together with a nonzero CP-violating
asymmetry $\epsilon^{}_1$ in the lightest heavy Majorana neutrino decays.
It is the case where the subset only contains element $S$, and in this
case $M^{}_{\rm D}$ and $M^{}_{\rm R}$ constrained by Eq. (2.5) are
\begin{eqnarray}
M^{}_{\rm D} \hspace{-0.2cm}&=&\hspace{-0.2cm}
\left(\begin{matrix} 2A^{}_{\rm r} & 2B^{}_{\rm r} & 2B^{}_{\rm r}
\cr 2B^{}_{\rm r} & A^{}_{\rm r}+B^{}_{\rm r} & A^{}_{\rm r}+B^{}_{\rm r}
\cr 2B^{}_{\rm r} & A^{}_{\rm r}+B^{}_{\rm r} & A^{}_{\rm r}+B^{}_{\rm r}
\end{matrix}\right) + {\rm i} C^{}_{\rm r} \left(\begin{matrix} 2 & 2 & 2
\cr -1 & -1 & -1 \cr -1 & -1 & -1 \end{matrix}\right) \;,
\nonumber \\
M^{}_{\rm R} \hspace{-0.2cm}&=&\hspace{-0.2cm}
\left(\begin{matrix} a^{}_{\rm r} & b^{}_{\rm r} & e^{}_{\rm r}
\cr b^{}_{\rm r} & d^{}_{\rm r} & a^{}_{\rm r} + e^{}_{\rm r} - d^{}_{\rm r}
\cr e^{}_{\rm r} & a^{}_{\rm r} + e^{}_{\rm r} - d^{}_{\rm r} & b^{}_{\rm r} -
e^{}_{\rm r} + d^{}_{\rm r} \end{matrix}\right) \; ,
\end{eqnarray}
where $ {\rm Re} \left[\langle M^{}_{\rm D}
\rangle^{}_{12} \right] = {\rm Re} \left[\langle M^{}_{\rm D} \rangle^{}_{13}
\right] = {\rm Re} \left[\langle M^{}_{\rm D} \rangle^{}_{21} \right] $, ${\rm Re}
\left[\langle M^{}_{\rm D} \rangle^{}_{22} \right] = \left( {\rm Re} \left[\langle
M^{}_{\rm D} \rangle^{}_{11} \right] + {\rm Re} \left[\langle M^{}_{\rm D}
\rangle^{}_{12} \right] \right)/2$, ${\rm Im} \left[\langle M^{}_{\rm D}
\rangle^{}_{11} \right] = {\rm Im} \left[\langle M^{}_{\rm D} \rangle^{}_{12}
\right] = {\rm Im} \left[\langle M^{}_{\rm D} \rangle^{}_{13} \right]$, ${\rm Im}
\left[\langle M^{}_{\rm D} \rangle^{}_{21} \right] = - {\rm Im} \left[\langle
M^{}_{\rm D} \rangle^{}_{11} \right] /2$ and ${\rm Im}
\left[ M^{}_{\rm R} \right] = 0$ have been assumed to reduce the number
of free parameters. The light Majorana neutrino mass matrix
$M^{}_\nu$ turns out to have the form
\begin{eqnarray}
M^{}_{\nu} = \left(\begin{matrix} 2a^{\prime}_{\rm r} - b^{\prime}_{\rm r} &
b^{\prime}_{\rm r} & b^{\prime}_{\rm r} \cr b^{\prime}_{\rm r} & a^{\prime}_{\rm r}
& a^{\prime}_{\rm r} \cr b^{\prime}_{\rm r} & a^{\prime}_{\rm r} & a^{\prime}_{\rm r}
\end{matrix}\right) + {\rm i} e^{\prime}_{\rm r} \left(\begin{matrix} 4 & 1 & 1
\cr 1 & -2 & -2 \cr 1 & -2 & -2 \end{matrix}\right) \;,
\end{eqnarray}
and the explicit relations between the parameters of $M^{}_\nu$ and those of
$M^{}_{\rm D}$ and $M^{}_{\rm R}$ can easily be derived.
We see that this example leads to a texture of $M^{}_{\nu}$
which respects the $\mu$-$\tau$ permutation symmetry.
Diagonalizing $M^{}_{\nu}$ allows us to obtain the masses of three light
neutrinos and their flavor mixing parameters. For the inverted
neutrino mass hierarchy
\footnote{The normal hierarchy case will not be discussed here, because
it yields $\theta^{}_{12} = \pi/2$ and thus disfavored.},
the results are
\begin{eqnarray}
m^{}_1 = \frac{1}{2} \left( -3b^{\prime}_{\rm r} + \Delta \right) \;,\quad
m^{}_2 = \frac{1}{2} \left( 3b^{\prime}_{\rm r} + \Delta \right) \;,\quad
m^{}_3 =0 \;
\end{eqnarray}
and
\begin{eqnarray}
V = \left(\begin{matrix} \displaystyle \pm\frac{2}{\sqrt{6}}
\frac{-4a^{\prime}_{\rm r} + b^{\prime}_{\rm r} - \Delta - 6{\rm i}
e^{\prime}_{\rm r}}{t} & \displaystyle \pm \frac{1}{\sqrt{3}}
\frac{4a^{\prime}_{\rm r} - b^{\prime}_{\rm r} + \Delta + 12{\rm i}
e^{\prime}_{\rm r}}{t} & 0 \cr \displaystyle \pm \frac{1}{\sqrt{6}}
\frac{4a^{\prime}_{\rm r} - b^{\prime}_{\rm r} + \Delta - 12{\rm i}
e^{\prime}_{\rm r}}{t} & \displaystyle \pm \frac{1}{\sqrt{3}}
\frac{4a^{\prime}_{\rm r} - b^{\prime}_{\rm r} + \Delta - 6{\rm i}
e^{\prime}_{\rm r}}{t} & \displaystyle -\frac{1}{\sqrt{2}} \cr
\displaystyle \pm \frac{1}{\sqrt{6}} \frac{4a^{\prime}_{\rm r} -
b^{\prime}_{\rm r} + \Delta - 12{\rm i} e^{\prime}_{\rm r}}{t} &
\displaystyle \pm \frac{1}{\sqrt{3}} \frac{4a^{\prime}_{\rm r} -
b^{\prime}_{\rm r} + \Delta - 6{\rm i} e^{\prime}_{\rm r}}{t} &
\displaystyle \frac{1}{\sqrt{2}} \end{matrix}\right) \;,
\end{eqnarray}
where $\Delta = \sqrt{(-4a^{\prime}_{\rm r} + b^{\prime}_{\rm r})^2 +
72e^{\prime 2}_{\rm r}}$ and $t= \sqrt{(4a^{\prime}_{\rm r} - b^{\prime}_{\rm r}
+ \Delta)^2 + 72 e^{\prime 2}_{\rm r}}$, and the ``$\pm$" signs correspond
to the sign of $e^{\prime}_{\rm r}$. It is obvious that $V$ can be regarded
as a variation of the tri-bimaximal flavor mixing pattern, and the
equalities $|V^{}_{\mu i}| = |V^{}_{\tau i}|$ hold (for $i=1,2,3$). To be
explicit,
\begin{eqnarray}
&&\theta^{}_{13} = 0 \;,\quad \theta^{}_{23} = \frac{\pi}{4} \;,\quad
\theta^{}_{12} = \arctan{\left[ \frac{1}{\sqrt{2}} \sqrt{\frac{ \left(
4a^{\prime}_{\rm r} - b^{\prime}_{\rm r} + \Delta \right)^2 + 144
e^{\prime 2}_{\rm r}}{ \left( 4a^{\prime}_{\rm r} - b^{\prime}_{\rm r}
+ \Delta \right)^2 + 36 e^{\prime 2}_{\rm r}}} \right]}\;,
\nonumber
\\
&&\delta \in \left[ 0, 2\pi \right) \;,\quad \rho = - \varphi^{}_2 \;
{\rm or} \; - \varphi^{}_2 + \pi \;,\quad \sigma = - \varphi^{}_1 - \pi \;
{\rm or} \; - \varphi^{}_1 \;,
\nonumber
\\
&&\phi^{}_e = \varphi^{}_1 + \varphi^{}_2 + \pi \;,\quad
\phi^{}_{\mu} = \pi \;,\quad \phi^{}_{\tau} = 0 \;.
\end{eqnarray}
where $\tan{\varphi^{}_1} = 6e^{\prime}_{\rm r}/
\left( 4a^{\prime}_{\rm r} - b^{\prime}_{\rm r} + \Delta \right)$
and $\tan{\varphi^{}_2} = 12e^{\prime}_{\rm r}/
\left( 4a^{\prime}_{\rm r} - b^{\prime}_{\rm r} + \Delta \right)$.

Given Eq. (5.2), a straightforward calculation leads us to the conclusion
$\epsilon^{}_1 \neq 0$ in the basis where $M^{}_{\rm R}$ is diagonal
(i.e., $M^{}_{\rm R} = \widehat{M}^{}_N = {\rm Diag}\{ d^{}_{\rm r}-
e^{}_{\rm r}-\Delta^{\prime}, d^{}_{\rm r}-e^{}_{\rm r} + \Delta^{\prime},
a^{}_{\rm r} + b^{}_{\rm r} + e^{}_{\rm r} \}$ with $\Delta^{\prime} =
\sqrt{ a^{2}_{\rm r} -a^{}_{\rm r}b^{}_{\rm r} +b^{2}_{\rm r} -2a^{}_{\rm r}
d^{}_{\rm r} +b^{}_{\rm r}d^{}_{\rm r} + d^{2}_{\rm r} +a^{}_{\rm r}
e^{}_{\rm r} -2b^{}_{\rm r}e^{}_{\rm r} - d^{}_{\rm r}e^{}_{\rm r} +
e^{2}_{\rm r}}$). It is therefore possible to realize both unflavored
and flavored leptogenesis in this case.

Finally, let us make another remark. Although the $S^{}_3$ reflection
symmetry helps a lot in determining the flavor structures in the seesaw
mechanism, it must be broken so as to make the relevant phenomenological
results fit current experimental data to a
good or acceptable degree of accuracy. A further work along this line of thought
will be done somewhere else. All in all, we expect that our structural
classification and discussions in the $S^{}_3$ reflection symmetry limit
will be useful for phenomenological studies of neutrino mass generation,
lepton flavor mixing, CP violation and leptogenesis when specific symmetry
breaking effects and more accurate experimental data are taken into account.
The same idea and similar analyses can be extended and applied to some other seesaw
mechanisms.

\vspace{0.5cm}

We would like to thank Shun Zhou and Jing-yu Zhu for useful discussions. This work is
supported in part by the National Natural Science Foundation of China under
Grant no. 11775231.

\end{document}